\documentclass{aa}  
\usepackage{graphicx}
\usepackage{txfonts}

\begin{document}

   \title{The VLTI/MIDI\thanks{Based on observations collected at the European Organisation for Astronomical Research in the Southern Hemisphere under ESO programmes 073.D-0711, 076.D-0620, 077.D-0294, 078.D-0122, 080.D-0801, 081.D-0021, 083.D-0234, 086.D-0737, 086.D-899, 187.D-0924, 089.D-0562, 090.D-410, 091.C-0468, 091.D-0344} view on the inner mass loss of
evolved stars from the Herschel MESS sample}

 \author{C. Paladini\inst{1,2}\and 
D. Klotz\inst{2}\and 
S. Sacuto\inst{2,3}\and  
E. Lagadec\inst{4}\and 
M. Wittkowski\inst{5}\and
A. Richichi\inst{6}\and 
J. Hron\inst{2}\and 
A. Jorissen\inst{1}\and 
M.~A.~T. Groenewegen\inst{7}\and 
F. Kerschbaum\inst{2}\and 
T. Verhoelst\inst{8}\and  
G. Rau\inst{2}\and
H. Olofsson\inst{9}\and
R. Zhao-Geisler\inst{10}\and
A. Matter\inst{4}}

  \institute{Institut d'Astronomie et d'Astrophysique, Universit\'e libre de Bruxelles, CP 226, Boulevard du Triomphe, 1050 Brussels, Belgium; \email{claudia.paladini@ulb.ac.be}
    \and
    University of Vienna, Department of Astrophysics, T\"urkenschanzstrasse 17, 1180 Wien, Austria
    \and
    University of Uppsala, Department of Physics and Astronomy, Division of Astronomy and Space Physics, Box 516, 75120, Uppsala, 
    Sweden\and
    Laboratoire Lagrange, Universit\'e C\^{o}te d'Azur, Observatoire de la C\^{o}te d'Azur, CNRS, 
Blvd de l'Observatoire, CS 34229, 06304 Nice cedex 4, France\and
  ESO, Karl-Schwarzschild-Str. 2, 85748 Garching bei M\"unchen, Germany\and
   National Astronomical Research Institute of Thailand, 191 Siriphanich Bldg., Huay Kaew Rd., Suthep, Muang, 50200 Chiang Mai,
    Thailand\and
    Koninklijke Sterrenwacht van Belgi\"e, Ringlaan 3, 1180 Brussel, Belgium\and
    Belgian Institute for Space Aeronomy (BIRA-IASB), Ringlaan-3-Avenue Circulaire, B-1180 Brussels, Belgium\and
    Onsala Space Observatory, Dept. of Earth and Space Sciences, Chalmers University of Technology, 43992 Onsala, Sweden\and
    National Taiwan Normal University, Department of Earth Sciences, 88 Sec. 4, Ting-Chou Rd, Wenshan District, Taipei, 11677, Taiwan, ROC
    \\
             }
                
      \date{Received; accepted}

 
  \abstract
   {The mass-loss process from evolved stars is a key
ingredient for our understanding of many fields of astrophysics, including
stellar evolution and the chemical enrichment of the interstellar medium (ISM) via
stellar yields. Nevertheless, many questions are still unsolved, one of which is the
geometry of the mass-loss process. }
   {Taking advantage of the results from the Herschel Mass loss of Evolved StarS (MESS)
programme, we initiated a coordinated effort to characterise the geometry of mass
loss from evolved red giants at various spatial scales.}
   {For this purpose we used the MID-infrared interferometric Instrument (MIDI) to resolve the inner envelope
of 14 asymptotic giant branch stars (AGBs) in the MESS sample. In this contribution we present an overview of the interferometric data
collected within the frame of our Large Programme, and we also add archive data for completeness.
We studied the geometry of the inner atmosphere by comparing the observations with predictions from different geometric models.
}
   {Asymmetries are detected for the following five stars: R~Leo, RT~Vir, $\pi^1~$Gruis, omi~Ori, and R~Crt.
   All the objects are O-rich or S-type, suggesting that asymmetries in the $N$ band are more common among stars with such chemistry.
   We speculate that this fact is related to the characteristics of the dust grains. 
   Except for one star, no interferometric variability is detected, i.e. the changes in size of the shells of non-mira stars correspond to changes of the visibility of less than 10\%. 
   The observed spectral variability confirms previous findings from the literature.
The detection of dust in our sample follows the location of the AGBs in the IRAS colour-colour diagram: more dust is detected around oxygen-rich stars in region II and in the 
carbon stars in region VII. The SiC dust feature does not appear in the visibility spectrum of the U~Ant and S~Sct, which are two carbon stars with detached shells. 
This finding has implications for the theory of SiC dust formation.}
   {}

\titlerunning{The VLTI/MIDI view on the inner mass loss of evolved stars}
   \maketitle
%

\section{Introduction}

Most of the material processed during the lifetime of low- to intermediate-mass stars is returned to the interstellar medium (ISM)
during the asymptotic giant branch (AGB) stage. This material is crucial for the chemical evolution of
galaxies \citep[perhaps even at high redshift;][]{valiante2009}, and it contributes to building new generations of stars and planets.

The general picture that explains the mass-loss process assumes that stellar pulsation triggers shock waves in the atmosphere. These shocks lift
the gas above the stellar surface, creating dense cool layers where dust may form.  Depending on the chemistry of the star, theory predicts that 
the radiation pressure on dust or the scattering on micron-size dust grains drive the stellar material away \citep{hoefner1997, woitke2006,hoefner2008}.
An important aspect of the mass-loss process that is poorly understood is its geometry, i.e. the deviation of the density distribution from spherical symmetry on different spatial scales.
The assumption about the geometry of the circumstellar environment affects calculations of mass-loss rates and other fundamental parameters \citep{ohnaka2008b}.
In recent years, several observing campaigns were carried out with the purpose of investigating the geometry of the 
envelope of AGB stars (references below). Observations suggest that the wind mechanism may depend on the initial mass of the objects 
and on the evolution along the AGB \citep{habing2003}. Despite first evidence
in favour of overall spherical symmetry, some observations \citep[][e.g. V~Hya]{knapp1997} show very complicated geometry at various spatial scales,
and no consensus on its origin has been reached so far \citep{habing2003}.
Understanding how the mass-loss shapes the envelope of AGB stars is crucial also for the progeny. 
Although a binary companion is currently the most accepted explanation, other mechanisms such as rotation velocity and
magnetic fields might still play a role \citep{demarco2009}.
Investigating the morphology of the atmosphere of AGB stars at different spatial scales and evolutionary stages (early-AGB and thermal-pulse AGB) 
helps to clarify the picture in the follow-up stages.

By scanning the envelope of an AGB star from the inside to the outside one can distinguish the following:

\emph{Inner circumstellar envelope (CSE)}. At milliarcsecond scales (1-2.5 stellar radii), close to the photosphere of the stars, asymmetries
are frequently detected with lunar occultations \citep{richichi1995, meyer1995} and optical interferometry 
\citep{ragland2006, lebouquin2009, pluzhnik2009, chiavassa2010, wittkowski2011, cruzalebes2013b, vanbelle2013, mayer2014, vanbelle2013, cruzalebes2015}. 
The asymmetric structures are often ascribed to convective patterns, but other interpretations are also invoked (mainly the effect of stellar rotation and binarity).
It is observed that asymmetries in the brightness distribution are more frequent 
for Miras \citep[][i.e. towards the end of the AGB life]{cruzalebes2015} and irregular variables \citep{ragland2006};
asymmetries are more frequent in C-rich stars than in the O-rich stars.

\emph{Intermediate CSE.} Between 2 and 10 stellar radii, 
asymmetries, and clumpiness are also observed for several objects \citep{weigelt1998,tuthill2000, weigelt2002, leao2006, tatebe2006, chandler2007, paladini2012, sacuto2013}.
In a very few cases the asymmetries have a clear pattern resembling a spiral or a disc. These cases
are usually related to the presence of a  hidden binary companion \citep{mauron2006, deroo2007, ohnaka2008a, maercker2012, mayer2013, decin2014, kervella2014, ramstedt2014, lykou2015, kervella2015}.
On the other hand, many other authors detect time variability but no clear signatures of asymmetries \citep{danchi1994, 
ohnaka2005, wittkowski2007, karovicova2011, sacuto2011, zhaogeisler2011, zhaogeisler2012, karovicova2013}. 
To confuse the picture even more, 
SiO maser observations show evidence for clumpy isotropic mass loss in the atmosphere of O-rich AGB stars,
while H$_2$O and OH masers (at R~$>10~R_{\star}$) probing the intermediate-outer part of AGB stars are less conclusive regarding the geometry.

\emph{Outer CSE.}  
Submillimiter observations of CO line profiles obtained towards M stars may deviate significantly 
from those expected from a spherical envelope \citep{knapp1998, winters2003, klotz2012a}.
Imaging  in the CO radio emission lines of carbon stars revealed spherically symmetric thin detached shells 
\citep{olofsson2000}, probably originating during the thermal pulses known to occur during the AGB phase. 
Images provided by the Herschel/PACS instrument within the frame of the 
Herschel Mass loss of Evolved StarS guaranteed time key programme \citep[MESS;][]{groenewegen2011} 
showed that the morphology of the outer atmosphere (R~$>1000~R_{\star}$) of AGBs
differs depending on various factors \citep[for example interaction between wind and the interstellar medium, or wind-wind interaction;][]{cox2012}. 

 
 Altogether it is clear that one has to probe all spatial scales to understand the physics of these complex outflows. 
 While previous studies with the aim of detecting asymmetries suffered from a lack of ${(u,v)}$ coverage (cfr. optical/infrared interferometry) and/or instruments with sufficient sensitivities, the advent of new facilities 
with improved resolution and sensitivity like Herschel, ALMA, and VLTI offers the unique chance to understand the mass-loss and dust-formation processes,
and generally speaking, the life cycle of dust and gas in the Universe.

In September 2010, we proposed  a Large Programme (LP) to the European Southern Observatory (ESO) to complement the Herschel observations with observations 
using the mid-infrared instruments MIDI on the Very Large Telescope Interferometer (VLTI)
and VISIR on the Very Large Telescope (VLT). 
The aims of the study are i) to establish whether asymmetries of the outer CSE originate in the dust-forming region or whether they are only
due to interaction with the ISM; ii) to evaluate at which height, the mass-loss process becomes manifestly non-spherical; and 
iii) to understand how the geometry of the atmosphere changes at the different evolutionary stages (M-S-C stars, and from almost dust-free to very dusty objects) within the AGB sequence. 

In this paper, we present the programme and data of the interferometric (MIDI) campaign interpreted with geometric models. 
All the ESO archive data available for the targets are incorporated in the analysis to give a complete overview. 
This first work will be followed by a series of papers including a detailed interpretation of the MIDI data in terms of model atmospheres.
The VISIR observing campaign was severely affected by bad weather conditions; a new observing proposal was recently accepted. 

The selection of the targets is described in Sect.~\ref{par:targets}, while the strategy for observations is given in
Sect.~\ref{par:strategy}. The data reduction is reported in Sect.~\ref{par:datareduction}. We also used archive data that are introduced in Sect.\ref{par:additional}.
The programme used for
the interpretation of the interferometric data is described in Sect.~\ref{par:tool}.
Results are presented in Sect.~\ref{par:results}. The discussion, conclusions, and outlooks are given in Sects.~\ref{Sect:discussion} and \ref{par:conclusions}, respectively.
The detailed description of every single target can be found in Appendix~\ref{singlestars.appendix}, and the journal of the observations is provided in Appendix~\ref{journal.appendix}.

\section{Target selection and data}
\label{data.sect}

\subsection{Target selection}
\label{par:targets}
\begin{figure}
   \includegraphics*[width=0.45\textwidth,bb = 70 369 547 701]{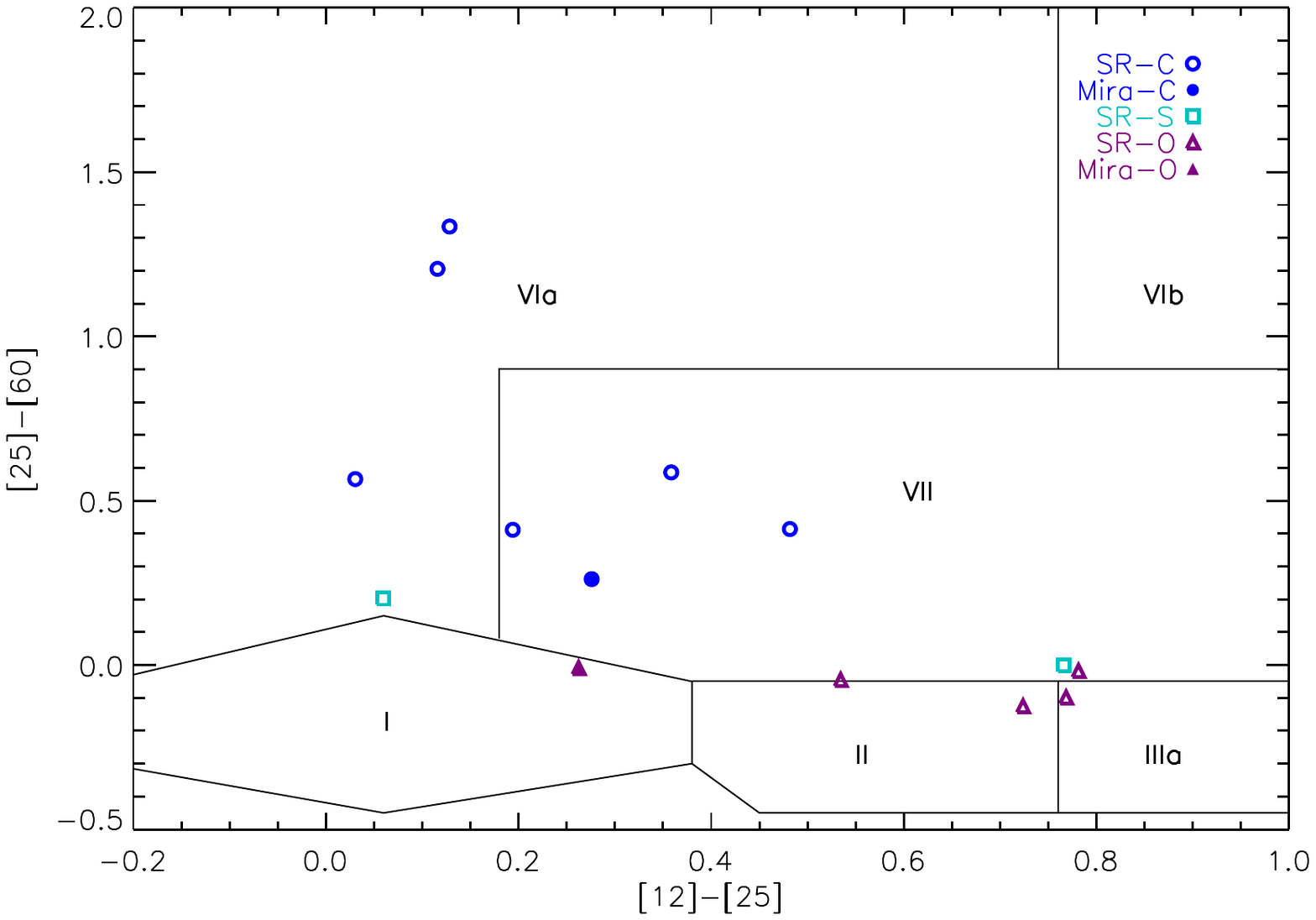}
   \caption{ \label{fig:iras} Targets of our sample shown in the IRAS two-colour diagram of \cite{vanderveen1988}.}
   \end{figure}

A sub-sample of the AGB stars observed within the Herschel key programme MESS \citep{groenewegen2011, cox2012}
was selected on the basis of the following criteria: declination accessible from the southern hemisphere where the ESO telescopes are located, 
brightness within the limits of the instruments, and  a range of chemistries and variability types. 
The IRAS two-colour diagram \citep{vanderveen1988} was used as a reference 
with the purpose of sampling AGBs with different shell properties (Fig.~\ref{fig:iras}). 
Every region includes targets with well-defined infrared characteristics: variability, and IRAS Low Resolution Spectrometer (LRS) classification.

The mildly variable O-rich AGB stars without extended circumstellar shells are expected to populate region I.
In region II the objects are surrounded by young O-rich shells, while in region IIIa the shells are more evolved and mass-loss rates are higher. 
Most of the C-rich objects with relatively cold dust are located in region VIa  (so-called detached-shell objects), 
while the objects located in region VIb have hot oxygen-rich dust close to the star
and cold dust at larger distances. 
The variable stars with evolved C-rich shells and infrared carbon stars are in region VII. 
The panels not covered within the observing sample are populated by objects with optically thick
envelopes having no visual counterparts, and by planetary nebulae.  
The list of targets is presented in Table~\ref{table:1} together with the location in the IRAS colour-colour diagram, the morphological class 
identified by \cite{cox2012} from the Herschel/PACS images, and general characteristics such as spectral type, variability class, IRAS 12$~\mu$m flux, period, distance, and
mass-loss rate.

\begin{table*}
\caption{Target list.}
\label{table:1}
\centering
\small{
\begin{tabular}{lllllllll}
\hline
\hline
Target & IRAS colour-colour region & Herschel/PACS$^{(a)}$ & Spectral Type & Variability & $F_{12}$ & Period & Distance$^{(b)}$ & Mass-loss rate\\
           &                                       & morphology              &               &             &  [Jy]   & [d]           &  [pc]    & $[M_{\sun} yr^{-1}]$\\
\hline
$\theta$~Aps&II (O-rich shells)& Fermata        &M6.5III    & SRb &734.30  & $119^{(c)}$  & $113^{+7}_{-6}$ & $0.4 \times 10^{-7} ~ ^{(d)}$ \\
R~Crt            &II (O-rich shells)& Eye               &M7III       & SRb &637.90  & $160^{(c)}$  & $261^{+86}_{-52}$ & $8 \times 10^{-7}~^{(d)}$\\
R~Leo          &I                         & Fermata        &M8IIIe     & M     &2161.00& $310^{(e)}$  & $110^{+17}_{-11}$ & $9.4 \times 10^{-8}~^{(f)}$\\
T~Mic           &II (O-rich shells)& Fermata        &M7III       & SRb &493.80  & $347^{(c)}$  & $200^{+57}_{-37}$  & $8 \times 10^{-8}~^{(d)}$\\
RT~Vir          &IIIa                    & Fermata        &M8III       & SR   &462.20  & $375^{(g)}$   & $136^{+17}_{-14}$ & $5 \times 10^{-7}~^{(d)}$\\
\hline                                                                                                                                     
$\pi^1$~Gru &II/IIIa/VII            &Irregular         &S5+G0V & SRb &908.50   & $198^{(h)}$ & $153^{+23}_{-18}$ & $4.6 \times 10^{-7}~^{(i)}$\\
omi~Ori   &VIa/I                   &Irregular         &S+WD    & SR   &85.35     & $ \ldots $     & $200^{+33}_{-25}$ & $< 0.4 \times 10^{-7}~^{(j)}$\\
\hline                                                                                                                                     
U~Ant       &VIa                        &Ring              &N:var      & Lb     &167.50  & $ \ldots $      & $270^{+45}_{-34}$ & $2 \times 10^{-8}~^{(k)}$\\
R~Lep       &VII                        &Point Source & CIIe       & M      &379.50  & $427^{(c)}$  & $470^{+301}_{-122}$ & $1.6 \times 10^{-6}~^{(k)}$\\
Y~Pav       &VII                        &Point Source & CII         & SRb  &72.38    & $233^{(c)}$  & $400^{+126}_{-77}$ & $4 \times 10^{-7}~^{(l)}$\\
TX~Psc      &VIa                      &Fermata        & CII         & Lb     &162.90  & $\ldots $       & $275^{+33}_{-27}$ & $3.2 \times 10^{-7}~^{(k)}$\\
S~Sct       &VIa                        &Ring              & CII         & SR    &65.31    & $148^{(c)}$  & $386^{+109}_{-70}$ & $5.6 \times 10^{-6}~^{(k)}$\\
AQ~Sgr      &VII                       &Fermata        & CII         & SR    &56.64    & $199^{(c)}$  & $333^{+96}_{-61}$ & $7.7 \times 10^{-7}~^{(k)}$\\
X~TrA       &VII                        &Ring               & C           & Lb     &201.00  & $ \ldots $      & $360^{+68}_{-49}$ & $1.8 \times 10^{-7}~^{(k)}$\\
\hline
\end{tabular}
\tablebib{
(a)~\cite{cox2012};
(b)~\citet{vanleeuwen2007}; (c)~\citet{samus2009}; (d)~\citet{olofsson2002}; (e)~\citet{whitelock2000};  (f)~\citet{knapp1998};
(g)~\citet{imai1997}; (h)~\citet{tabur2009}; (i)~\citet{jorissen1998};
(j)~\citet{groenewegen1998}; (k)~\citet{Bergeat2005}; (l)~\citet{winters2003}.}
}
\end{table*}

\subsection{Observing strategy}
\label{par:strategy}
In this work we concentrate on possible asymmetries developing in the dust-forming region.
\cite{danchi1994} and more recently \cite{norris2012} observed that dust is forming between 3 and 5 stellar radii, and even closer to the star
in the case of Mira variables, as \cite{draine1981} already predicted.
As a consequence, 
one needs to have an angular resolution of $\sim 20$ mas to observe the \emph{locus} of dust formation for AGBs
within 1~kpc. 
For this reason, the MIDI instrument \citep{leinert2003} 
installed at the ESO Very Large Telescope Interferometer (VLTI) on Cerro Paranal (Chile) until March 2015 was chosen. The MIDI instrument is a two-beam combiner
interferometer observing in the $N$ band (8-13~$\mu$m). For every observation, the instrument delivers one visibility spectrum (sometime shortened to visibility in the text),
one total flux spectrum, and one differential phase spectrum (phase difference between different spectral channels; Sect.~\ref{Sect:diffpha}).
Depending on the correlated magnitude of the target, either the HIGH-SENS (where the correlated and total flux are measured one after the other) or SCI-PHOT (where the correlated and total flux are measured simultaneously) mode was chosen. 
All the LP data have spectral resolution $R= \lambda/\Delta\lambda \sim 30$.

To optimise the observations for studying the geometry the following strategy was used.
The choice for the baseline configurations was split in two different categories, based on the Herschel data available: 
configurations for non-spherical \citep[such asthe bow shock in the case of TX~Psc;][]{jorissen2011}, and for spherical objects \citep[such as the detached-shell object U~Ant;][]{kerschbaum2010a}.
 For the non-spherical cases we selected one baseline oriented in the direction of the asymmetry and one baseline perpendicular to the latter.  
A third baseline, with orientation in between the two, was selected to put constraints on the possible elongation.
Baselines with random orientation were selected 
for the symmetric stars, as well as in the cases where PACS images
were not yet available. Thus, we are able to constrain any possible deviation from sphericity.
We also tried to sample the same spatial frequencies by choosing the same baseline lengths for all points.
The baseline length also has to be selected carefully for resolving the dust formation zone.
Given the fact that not all the targets of the sample had a measured photospheric diameter, 
the latter was estimated through the $(V-K)$ relation of \cite{vanbelle1999}. 
Following the results of \cite{danchi1994}, we thus assumed that the diameter in the $N$ band is approximately three times the photospheric diameter,
and the baseline was selected accordingly.

For planning the observations, we used ASPRO and ASPRO2 developed by the Jean-Marie Mariotti Center \texttt{Aspro}
service \footnote{Available at \tt{http://www.jmmc.fr/aspro}}.
The observations were carried out between 2011 April 23 and 2012 July 01 on the 1.8\,m Auxilliary Telescopes (ATs). 
We used the recommended observation sequence CAL-SCI-CAL. The following selection criteria for calibrator stars were applied: brightness (difference between calibrator and science target
 of $\pm1$\,mag), position (RA and Dec as close as possible to the science target), size (the calibrator should be as small as possible), and spectral type
 (if possible the calibrator should have a spectral type earlier than M0). The list of calibrators and their main characteristics are
 presented in Table~\ref{table:2}. 
 The journal of the MIDI observations is available as online material (Appendix~\ref{journal.appendix}).

\begin{table*}
\caption{Calibrator list.}
\label{table:2}
\centering
{\small
\begin{tabular}{llllllll}
\hline
\hline
HD & Spectral type\tablefootmark{a} & F$_{12}\tablefootmark{(a)}$ & $\theta\tablefootmark{(b)}$ & used for \\
   &               & [Jy]        & [mas]        &          \\
\hline
18884           &       M1.5IIIa        &       234.7     &     $12.28\pm0.05$  &       omi~Ori, R~Lep\\
20720               &      M3/M4III        &       162.70  &        $10.14\pm0.04$      &       R~Lep\\
25025           &       M1IIIb          &       109.6 &  $8.74\pm0.09$  &       R~Leo\\
29139           &       K5III           &       699.7   &       $20.398\pm0.087$        &       R~Leo, U~Ant\\
32887           &       K4III           &        56.82  &       $5.90\pm0.06$    &       R~Lep \\
39425           &       K1IIICN+1       &        28.0   &        $3.752\pm0.017$        &       omi~Ori\\
48915           &       A1V                     &       143.1   &        $6.08\pm0.03$   &R~Leo, R~Lep, U~Ant\\
50778           &       K4III           &        24.6   &        $3.904\pm0.015$        &       omi~Ori\\
61421           &       F5IV-V          &        79.1   &        $5.25\pm0.21$  &       R~Leo\\
81797           &       K3II-III        &       157.6   &        $9.142\pm0.045$        &       $\theta$ Aps, R Crt, RT Vir, R Leo\\
112142          &       M3III           &        47.0   &        $5.90\pm0.7$   &       R~Crt, RT~Vir, AQ~Sgr\\
120323          &       M4.5III         &       255.4   &       $13.25\pm0.06$  &       R~Crt, RT~Vir, R~Leo, U~Ant, AQ~Sgr\\
123139          &       K0III           &       56.9    &        $5.33\pm0.057$ &       R~Crt, RT~Vir, R~Leo, AQ~Sgr\\
129456          &       K5III           &       21.4    &        $3.37\pm0.014$ &       T~Mic\\
133216          &       M3/M4III        &       200.7   &       $11.154\pm0.046$        &       RT~Vir\\
150798          &       K2II-III        &       144.0   &        $8.76\pm0.12$  &       $\theta$~Aps, R~Crt R~Leo, Y~Pav, X~TrA\\
151249          &       K5III           &        52.18  &        $5.515\pm0.179$   &       Y~Pav\\ 
152786          &       K3III           &       82.1    &        $8.02\pm3.23$  &       $\theta$~Aps, R~Crt R~Leo, AQ~Sgr\\
165135          &       K1III           &       23.4    &        $3.47\pm0.015$ &       T~Mic\\
167618          &       M3.5III         &       213.7   &       $11.665\pm0.043$        &       $\theta$~Aps, T~Mic\\
168454          &       K3III           &       62.17   &       $5.874\pm0.026$   &       S~Sct, AQ~Sgr\\
169916          &       K0IV            &       31.2    &        $3.995\pm0.019$        &       T~Mic, Y~Pav, S~Sct, AQ~Sgr\\
177716          &       K1III           &       26.0    &       $3.78\pm0.21$   &       T~Mic\\
206778          &       K2Ib            &       103.9   &        $7.59\pm0.046$ &       $\theta$~Aps, R~Leo, $\pi^1$~Gru\\
211416          &       K3III           &       59.3    &        $5.92\pm0.28$   &       AQ~Sgr\\
224935          &       M3III           &       86.90   &        $7.25\pm0.03$   &       $\pi^1$~Gru\\
\hline
\hline
\end{tabular}
\tablefoot{
\tablefoottext{a}{\tt{http://simbad.u-strasbg.fr/simbad/}; } \\
\tablefoottext{b}{\tt{http://www.eso.org/observing/dfo/quality/MIDI/qc/calibrators\_obs.html}}
}
}
\end{table*}

\smallskip


\subsection{Observations and data reduction}
\label{par:datareduction}
The MIDI data were reduced using the data reduction pipeline MIA+EWS\footnote{\tt{http://www.strw.leidenuniv.nl/$\sim$jaffe/ews/MIA+EWS-\\Manual/index.html}}
\citep{jaffe04, ratzka05, leinert04}. A detailed description of the data quality tests that were applied during the data reduction can be found
 in \citet{klotz2012b}. Data were reduced with all calibrators observed in the same night (if possible within $\pm2$\,hours) and with the same
 baseline configuration as the science target. The final calibrated visibilities are then the mean of all the
 visibilities, differential phases, and fluxes reduced with suitable calibrators. The error is derived from the standard deviation of that series. If
 the former error is lower than $\pm10$\% or only one calibrator was available during the night, a multiplicative error of $\pm10$\% was used
 \citep{chesneau07}. 
 
 The extraction of the differential phase was carried out following the standard procedure of EWS \citep{jaffe2004}. 
 The differential phase was corrected for the changing  index of refraction of air by subtracting a linear slope, so that the mean phase over the $N$ band is zero. We did not correct for higher order effects owing to water vapour content (PWV). Instead, we employed a very simplistic approach, by calibrating the differential phases using several calibrators taken over the night and deriving our error estimate on the phase from the scatter of these multiple calibrations. The instrumental phase measured for the calibrators during the different nights is usually stable 
 and of the order of $\pm 5 \degr$.
Any uncertainty on the differential phase caused by the lack of correction for the dispersion effects due to PWV is therefore transferred to the errors on our phases. 

A word of caution must be issued concerning the MIDI spectra. The water vapor content in the Earth atmosphere can change conspicuously on a timescale of half an hour without any changes in seeing or coherence time.
So, it is possible that the water vapour content and thus the transmission of the
Earth atmosphere changes between science and calibrator. Whereas the calibration of the interferometric visibility is not affected,
the fluxes (i.e. the MIDI spectra) are affected. 
To limit such effects, fluxes were only derived if the airmass difference to the science target was smaller than 0.2 and the calibrator was observed within $\pm~2$~hours from the science target. 
Only calibrators of spectral type earlier than M0 were selected for the flux calibration. Henceforth the possibility that the science spectra is
contaminated by possible dust around the calibrator is minimised \citep{chesneau07}.
Adding the LP to the archive data (Sect.~\ref{par:additional}), we collected a total of 201 visibility points; $60\%$ of these data were of good quality and were used in this work. 

%

\subsection{Additional observations and variability check}
\label{par:additional}
Archive MIDI observations were available for TX~Psc, AQ~Sgr, U~Ant, T~Mic, R~Crt, R~Leo, RT\,Vir, $\pi^{1}$~Gru, omi~Ori, and R~Lep.
Some of these data were observed in {\tt GRISM} mode (spectral resolution $R = 230$). 
These high resolution archive observations were convolved
to a spectral resolution $R = 30$ before any comparison with the LP data.
We also noticed that most of these observations were carried out with different baselines but at the same position angle. 
Such data sets allow us to probe the atmosphere at different spatial scales, i.e. these data sets are optimal for tomography studies.
Moreover these observations carry information about intra-cyle and cycle-to-cycle interferometric and spectroscopic variability.
When possible we assigned a variability phase calculated from the visual light curve  to every MIDI observation.
For this purpose light curves were collected from the American Association of Variable Star Observers (AAVSO),
the All Sky Automated Survey (ASAS), and the Association Fran\c caise des Observateurs d'\'Etoiles Variables (AFOEV).
The phase is determined from the light curve using the following relation:
$$
\phi = \frac{(t-{\rm{T}_0})}{P} - int \left(\frac{(t-{\rm{T}_0})}{P}\right),
$$ 
where $t$ stands for the date of the MIDI observation(s) expressed in Julian date, T$_0$ is the phase-zero point that
was selected as the maximum light closest in time to the first MIDI observation, and P is the period of variability already listed in Table~\ref{table:1}. 
Visual phases are assigned to the stars $\theta$~Aps, R~Crt, R~Leo, T~Mic, RT~Vir, R~Lep, Y~Pav, S~Sct, and AQ~Sgr.
The values are listed in Appendix~\ref{journal.appendix}, and errors are assumed to be of the order of 10\% of the period.

\begin{figure*}[!ht]
\centering
\includegraphics[height=0.95\textheight,bb=55 363 271 682]{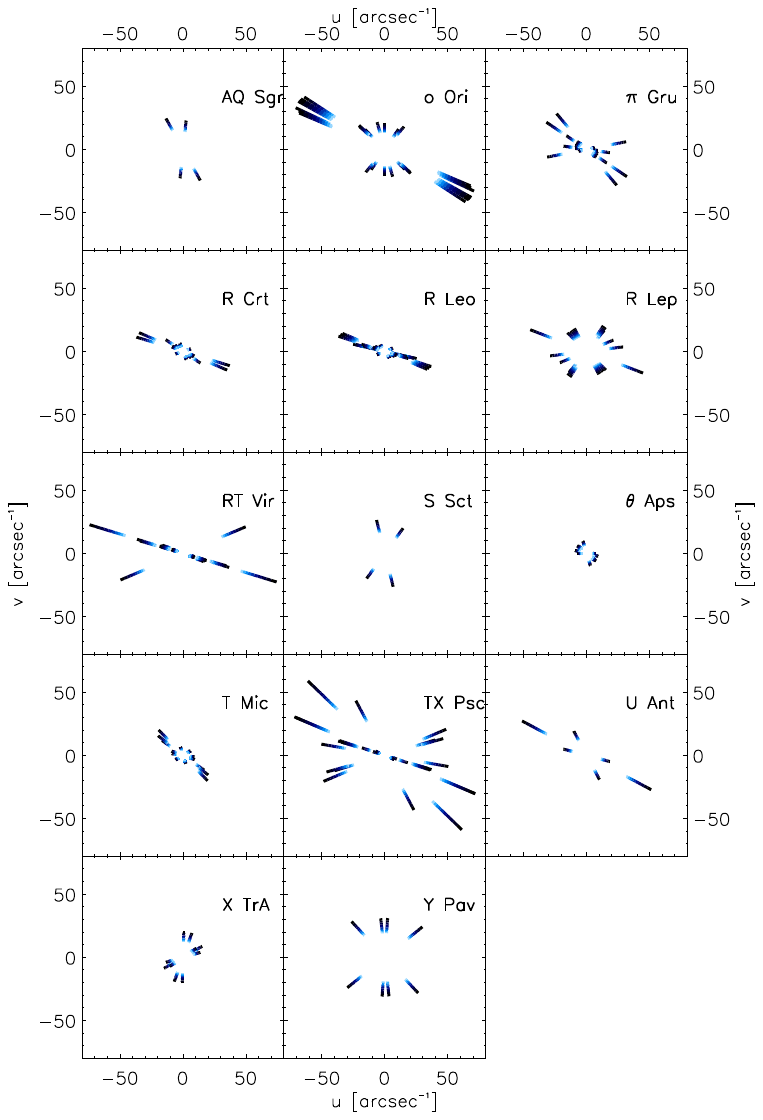}
\caption{$(u,v)$ coverages obtained for all the targets. The different wavelengths from $8-13~\mu$m are colour-coded (dark to light, respectively).}
\label{uvcoverage.fig}
\end{figure*}

The analysis of spectroscopic variability is performed by comparing MIDI spectra obtained at different visual phase, and also by comparing the MIDI spectra
to available ISO and/or IRAS spectra.
The interferometric variability was studied by comparing (when available) sets of visibilities at similar baseline lengths and position angles, observed at different dates. If no interferometric variability is detected, one can assume that the data can be combined for the geometric fit.
The $(u,v)$-coverages obtained for all the data of the LP, including the archive data
is shown in Fig.~\ref{uvcoverage.fig}.

\section{Geometric fitting}
\label{par:tool}

\begin{table*}
{\small
\caption{Reduced $\chi^{2}_{\rm{red}}$ from GEM-FIND fitting.}
\label{tab:gem-find}      
\centering
\vspace{0.5cm}
\begin{tabular}{lllllllll}
\hline
\hline
Target       & Baseline & Year & N$_{OB}$       &  UD & Gauss & Ell. UD & Ell.Gauss. & UD+Gauss   \\
             & [m]      &&&&&&&\\
\hline
$\theta$\,Aps & 10--17  & 2011/12       & 6     &{\bf 0.78}     &       0.69    &       0.41    &0.38&\ldots    \\

R\,Crt        & 10--16  & 2012          & 5     &       10.17   &       2.53    &       1.18      &  {\bf 1.04}   &       \ldots          \\
              &  {\bf 10--64}   & 2009/11/12    & 8     &       46.27   &       28.31   &       18.18   &       17.74   &{\bf   0.69}   \\

R\,Leo        & 11--16  & 2012          & 4     &{\bf   0.93}   &       0.78    &       0.09    &       0.11    &       \ldots          \\
              & 11--64  & 2006/07       & 18    &       38.55   &       14.73   &       8.03    &       10.84   &{\bf   7.17}   \\

T\,Mic        & 11--16  & 2011          & 4     &{\bf   0.28}   &       0.27    &       0.09    &        0.11    &       \ldots  \\
              & 11--46  & 2004/11       & 7     &       1.16    &{\bf   0.96}   &       0.49    &       0.46    &       0.24    \\              

RT\,Vir       & 13--15  & 2012          & 2     &1.31           &{\bf   1.13}   &       \ldots  &       \ldots          &       \ldots          \\
              & 12--128 & 2008/09/11/12 & 14    &       49.10   &       25.13   &       22.80   &       20.51   &{\bf   3.04}   \\
$\pi^1$\,Gru   & 10-15  & 2011          & 3     & 4.83  & 3.44  & {\bf 0.57} & 0.42 & \ldots\\
             & 10-62    & 2006/11       & 11    & 23.23 & 23.56 & 21.59 & 21.49 & {\bf 0.73}      \\

$o^1$\,Ori      & 32--46  & 2011          & 7     & 1.30  & 1.30  & {\bf 0.79} & 0.78  & \ldots \\
             & 32-129  & 2005/11          & 14    & 2.04  & 2.01  & 1.90 &    1.85  & {\bf 1.43} \\
U\,Ant        & 30      & 2012          & 1     & \ldots& \ldots & \ldots & \ldots & \ldots\\             
              &30-95    & 2008-2012     & 3     & {\bf 1.20} &  1.23 & 0.65 &  0.63 &\ldots  \\

R\,Lep        &  34-40  & 2012          & 6     & 1.59 & {\bf 0.93}  & 1.32 &  0.90   & \ldots\\
              & 34-79   & 2011/12       & 10    & 2.38 & 3.65  & 1.39 &  3.55  & {\bf 0.96}      \\

Y~Pav         & 50-63   & 2011          & 4     &  0.77  & {\bf 0.78} & 0.44 &  0.44 \\
TX\,Psc       & 60-140  & 2011          & 6     &{\bf 0.99} & 1.19 &  0.90 & 1.09 & \ldots \\
              & 11-140  &2004-2011      &15     &{\bf 1.28} & 1.32  & 1.29 & 1.32 & 1.34 \\
S\,Sct        & 40-45   & 2011          & 2     & 2.63 &  {\bf 2.61} & \ldots & \ldots & \ldots                       \\
X\,TrA        & 21-34   & 2011          & 5 & {\bf 0.95} & 0.89 & 0.31 &  0.30 & \ldots                  \\
AQ\,Sgr & 37-42 & 2011 & 2 & {\bf 2.39} & 2.47 & \ldots& \ldots& \ldots \\
\hline
\hline
\end{tabular}
\tablefoot{The range in baseline length is given as 'Baseline'. The number of observations used for the
 fitting is given as N$_\mathrm{OB}$. The first row corresponds to the data of the LP.
 The second row corresponds to fits were the LP and archive data were merged. If 
only data points with similar position angles were available, the elliptical models were not fitted.
The $\chi^2_{\rm{red}}$ of the model best fitting the data is highlighted in boldface. 
The resulting best-fitting parameters are given in Table~\ref{table:diameters}, whereas the dates of the observations are given in Appendix~\ref{journal.appendix}.}
}
\end{table*}

A model-independent way to identify departure from spherical symmetry of the CSE is by comparing visibilities taken at similar 
baseline and different position angle. We performed this check where the data set allowed it.
As a second approach we employ the software GEM-FIND \citep[GEometrical Model Fitting for INterferometric Data;][]{klotz2012b} to
 interpret our observations. This software fits geometrical models to interferometric visibility data, where different
 spherically symmetric, centro-symmetric, and asymmetric models are available. The different parameters of the models can be either wavelength dependent
 (e.g.\,diameter, flux ratio of two components) or wavelength independent (e.g. inclination or axis ratio of a disk)
 as GEM-FIND fits each wavelength point separately. This gives us the possibility to study the dependence of the
 model parameters on, for example\,molecular and dust features. The output of GEM-FIND is a $\chi^2_{\rm{red}}$, the best-fitting parameters, 
 and wavelength-dispersed visibilities and differential phases. The errors of the best-fitting parameters given by GEM-FIND are the
 1$\sigma$ statistical errors derived from the covariance matrix (calculated within the MPFIT\footnote{\tt{http://purl.com/net/mpfit}} IDL routines implemented in GEM-FIND). 
It is known that the models in the Fourier space are not linear, therefore the errors are not Gaussian distributed. 
We tested the validity of the approach with Monte Carlo simulations \citep{klotz2012b}. We find that 1$\sigma$ errors from the Monte Carlo simulations are
 comparable to those derived from the covariance matrix. Therefore, in the following, errors are computed from the covariance matrix.
%

For the study of the geometry of the circumstellar environment, the following models were used to fit the data: circular
 uniform disk (UD, representing an approximation of the stellar disk), circular Gaussian distribution (Gauss, approximation to an object with a molecular or dusty environment and limb darkening), elliptical uniform disk (Ell.~UD, such as in the UD case with non-central symmetric brightness distribution), elliptical Gaussian
 distribution (Ell.~Gauss, such as in the Gauss case with non-central symmetric brightness distribution). The latter two models were applied only for the objects with more than two position angles available.
 In the case where a sufficient number of observations sampling different spatial frequencies were available, 
 a spherical two-component model (a circular UD plus circular Gaussian, where the two components typically represent the photosphere and an optically thin dust and/or molecular component) was used additionally.
 In this latter case the diameter of the UD was fixed to a value corresponding to the  $\theta_{(V-K)}$ diameter \citep{vanbelle1999} to simulate the central star
 (or to the observed $K-$band value, when available). Only the Gaussian envelope was fitted.
The fit with GEM-FIND was performed in two stages: first only the LP data are fitted, and afterwards the LP data are merged with the archive data and a new fit is performed.
The reasoning behind this strategy is that the LP data are chosen to sample the same spatial frequencies and different position angles,
therefore they are more suitable for detecting possible elongations owing to a non-central symmetric distribution.
The fit with all data (LP + archive) is performed for completeness, and it allows us to study the stratification of the stars.

\section{Results}
\label{par:results}
This section summarises the general findings of the LP.
Detailed discussion for the single targets are given in Appendix~\ref{singlestars.appendix}.

\subsection{Visibility versus wavelength}
\label{visibilityvs.wavelength.sect}
\begin{figure*}[!t]
        \centering
        \includegraphics[bb=68 460 502 710]{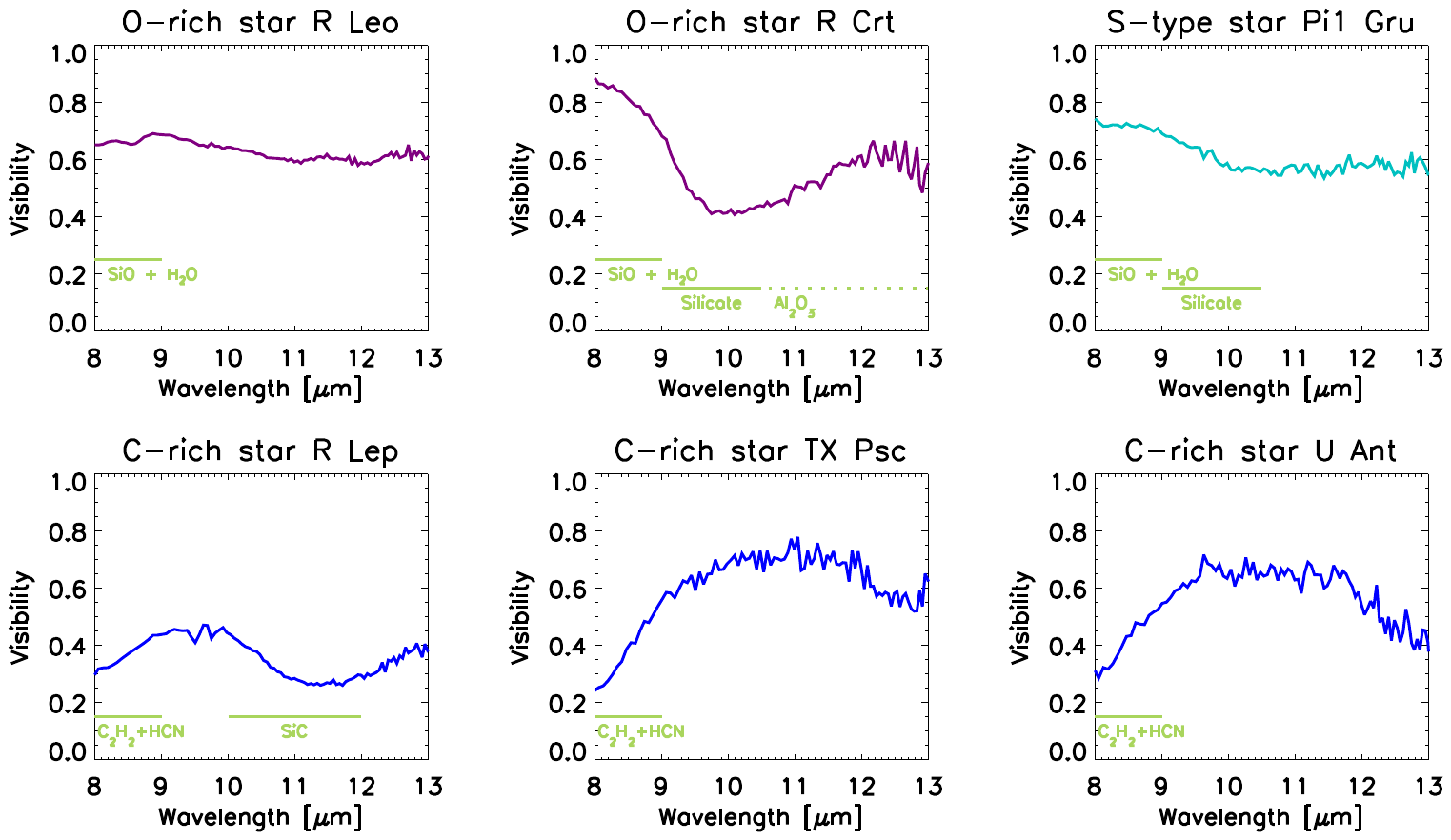}
\caption{\label{fig:shape} Some examples of visibility spectra for given baselines.
The complete sample can be found in Appendix~\ref{singlestars.appendix}.
Molecular and dust features are highlighted for stars with different chemistry: R~Leo and R~Crt are M-type stars, $\pi^1$~Gru is an S-type object,
while R~Lep and TX~Psc are C-type stars. The lower right panel shows an example of the visibility spectra of the carbon star U~Ant (cf. Sect.~\ref{visibilityvs.wavelength.sect}).
The typical error bars are of the order of $10\%$.} 

        \end{figure*}

A visual inspection of the visibility spectrum reveals certain spectral features characterising the chemical composition of the CSE.
For this study, we inspect the visibilities in the range between $0.1\la V \la 0.9$. 
In the case of $ V \ge 0.9$, it is not possible to distinguish details
of the spectral signature, or even distinguish the visibilities from those of a point source owing to the typical errors.
At $V\la 0.1$ the relation between visibility and spatial frequency may not be univocal (as the visibility function may consist of several lobes).
Fig.~\ref{fig:shape} shows the spectrally dispersed visibility curves for stars with different chemistry.

\emph{M-type and S-type stars.} The most prominent molecular feature of oxygen-rich stars in the $N$ band is SiO around 8~$\mu$m. 
In some cases, this is followed by silicates and Al$_2$O$_3$ dust.
For stars not showing a pronounced silicate feature such as R~Leo (upper left panel of Fig.~\ref{fig:shape}), T~Mic, and the S-type stars omi~Ori and $\pi^1$~Gru
(upper right panel of Fig.~\ref{fig:shape}), the visibility is rather flat with a small bump at short wavelengths. The diameter increases slightly at longer wavelengths.
For stars showing the dust and molecular features (R~Crt shown in the central upper panel of Fig.~\ref{fig:shape}; RT~Vir, and $\theta$~Aps) 
the visibility has a peak in the 8-9~$\mu$m
region,  a decrease of between 9-11~$\mu$m, and a subsequent increase.

\emph{C-type stars.} The molecules contributing to the carbon stars opacity in the $N$ band are mainly C$_2$H$_2$ and HCN. Concerning the dust,
evolved carbon-rich objects show SiC dust at 11.3~$\mu$m and amorphous carbon dust (featureless). 
Examples of visibilities of stars with SiC dust were shown by e.g. \cite{ohnaka2007}, \cite{sacuto2011}, \cite{paladini2012}, and \cite{rau2015}.
The lower left panel of Fig.~\ref{fig:shape} shows the visibility of the C-rich Mira R~Lep with the typical drop
at 11.3~$\mu$m because SiC. Y~Pav, X~Tra, and AQ~Sgr show similar visibility curves.
The visibilities of carbon-rich stars \emph{without} SiC have a typical 
bow shape. The visibility gets lower (i.e. the star is larger)
between 8-9~$\mu$m and after 12.5~$\mu$m where the molecular opacity is higher. 
Fig.~\ref{fig:shape} shows an example of this kind of star, i.e. TX~Psc \citep[also presented in][]{klotz2013}.

\emph{Other stars.} The IRAS and MIDI spectra of U~Ant show the signature of SiC dust feature, while there is no trace of such a feature in the (spatially resolved)
interferometric observations (lower right panel of Fig.~\ref{fig:shape}).
This could be the result of resolving out part of the total
emission, only revealing the emission of the spatial scale to which the interferometer is sensitive (at the employed baselines). 

The case of S~Sct is slightly different. There is no trace of SiC in the (spatially resolved)
interferometric observations nor in the MIDI spectrum. 
The dust feature is observed in the ISO spectrum recorded 14 years before the LP observations. 
The IRAS spectrum of S~Sct obtained $\sim30$ years before is very noisy but seems to agree with the MIDI spectrum (Fig.~\ref{ssct-spec}).
The latter was derived by averaging data taken on two different days, therefore one should be able to rule out 
a problem with the calibration. The S~Sct observations suggest a temporal variability in the stratification of SiC.

To our knowledge it is the first time that such findings are reported for AGB stars. 
  Section~\ref{Sect:discussion} provides a detailed discussion of these findings.

\subsection{Differential phase}\label{Sect:diffpha}

\begin{figure*}[!htbp]
\begin{center}
\resizebox{\hsize}{!}{
\includegraphics[width=\hsize,angle=0,bb=51 427 503 681]{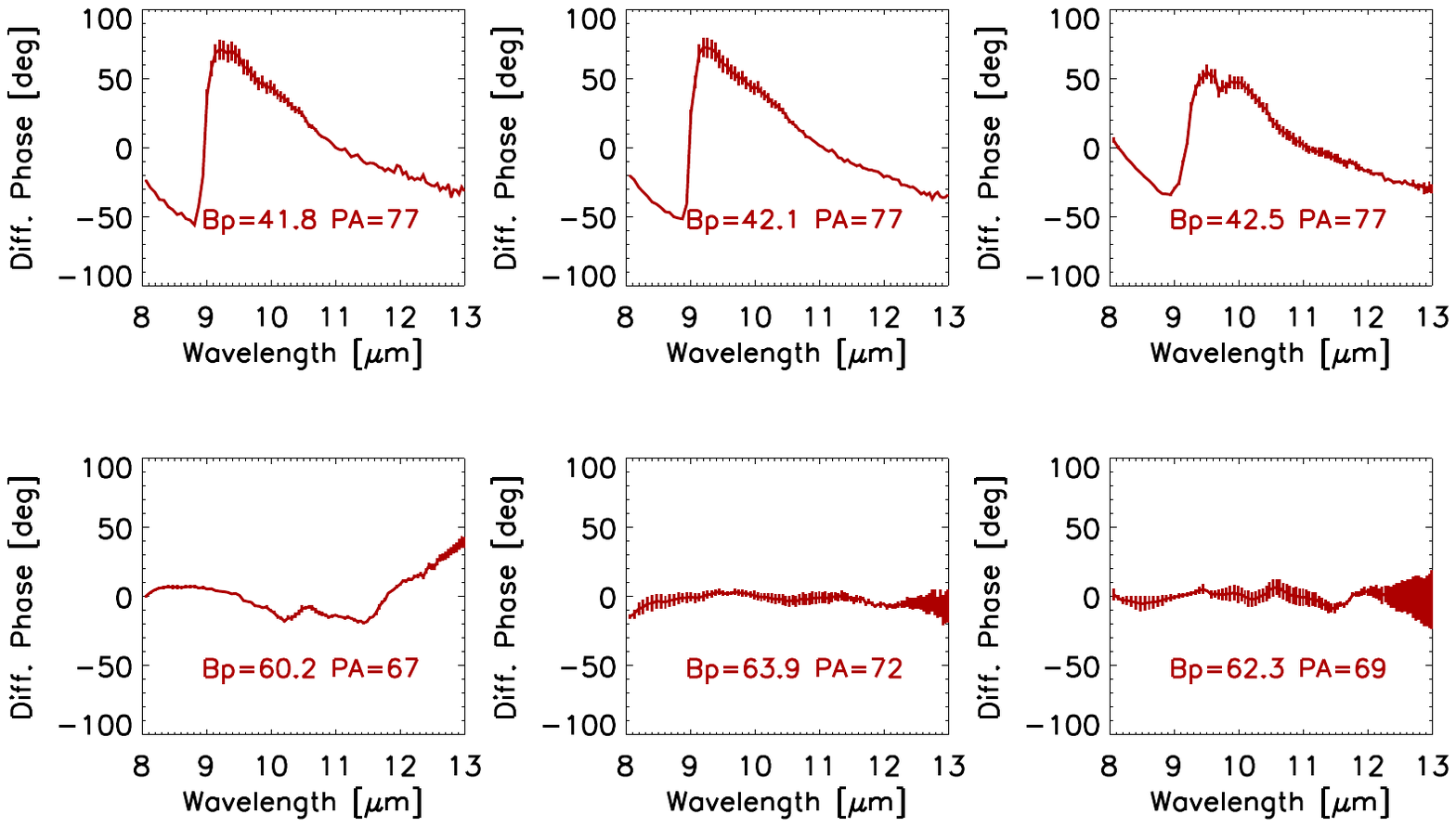}}
\caption{Non-zero differential phases measured by MIDI for R~Leo ordered in increasing projected baseline (B$_{\rm{p}}$). }
\label{diffpha-rleo.fig}
\end{center}
\end{figure*}
Except for the cases where the differential phase is equal to $180^{\circ}$, and/or it is accompanied by a 
null value in the visibility, a non-zero differential phase measured by MIDI implies an asymmetric brightness distribution.
The latter can be explained by two effects \citep{tristram2014}:
first, the object is composed of two sources (for example, the photosphere and resolved dust component) with different spectral distribution through the $N$ band; and, second, the object is composed of two objects with a spatial distribution more resolved at a certain wavelength than at another across the $N$band. 
In nature we usually observe a mixture of these two effects, which are very difficult to
distinguish, unless one has enough information for detailed modelling or to attempt an image reconstruction.
Non-zero differential phase was observed in a few AGB stars. Usually this is interpreted 
as a typical signature of a disc \citep{kervella2014, ohnaka2008a, deroo2007}
or the signature of a clump \citep{sacuto2013, paladini2012}. 
Since modelling very few differential phases
gives highly non-unique solutions, no attempt to interpret the differential phase is carried out here.

\begin{figure}[!htbp]
\begin{center}
\resizebox{\hsize}{!}{
\includegraphics[angle=0,bb=51 567 201 681]{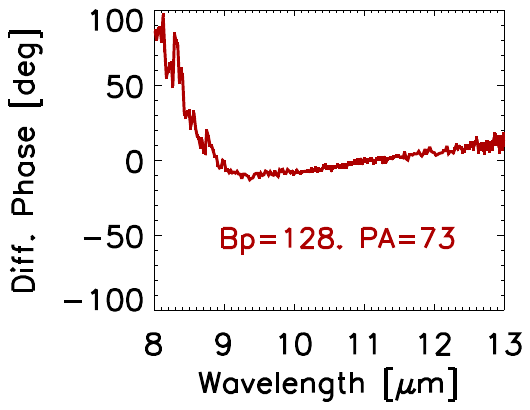}}
\caption{Same as Fig.~\ref{diffpha-rleo.fig} for RT~Vir. }
\label{diffpha-rtvir.fig}
\end{center}
\end{figure}

We report non-zero differential phase only for two objects: R~Leo and RT~Vir (Figs.~\ref{diffpha-rleo.fig} and \ref{diffpha-rtvir.fig}).
The morphology of the differential phases of R~Leo can be classified in two groups, according to the projected baseline used for the observations. 
The three upper panels of Fig.~\ref{diffpha-rleo.fig} show a jump of the differential phase around 9~$\mu$m. 
These data are acquired at different times (see Sect.~\ref{Sect:Spectroscopic and interferometric variability}), at the same position angles, and very similar projected baselines.
The lower panels show a much more complex behaviour with features at $\sim~10$ and $\sim~11.8 ~\mu$m.

The differential phase of RT~Vir shown in Fig.~\ref{diffpha-rtvir.fig} 
is characterised by a jump between 8 and 9~$\mu$m, followed by a monotonic increase. 
For an interpretation of the RT~Vir differential phase with a geometric model, we refer to \cite{sacuto2013}.

All the non-zero differential phases occur at visibility spectra below 10\%.

\subsection{Spectroscopic and interferometric variability}
\label{Sect:Spectroscopic and interferometric variability}
As already mentioned in Sect.~\ref{par:additional}, we used archive spectroscopic and interferometric observations 
to study the $N$-band variability. 

The spectroscopic variability typically corresponds to the variation in the colour (temperature) 
or a specific variation in certain line strength.
On the other hand, the interferometric variability is usually connected with a change in the morphology of the object, in particular the spatial scale of the $N$-band emission region.
Of course the real picture is more complex, and interferometric variability might also be due to
brightness variation. When the stellar atmosphere is spatially resolved, and there are multiple components in the FOV of the interferometer 
(i.e. photosphere plus extended molecular/dust layer, or photosphere plus clumpy structures), it is possible to observe 
variation in the visibility (at the same spatial frequency) because of a change in the flux ratio between the two components.

\begin{figure*}[!htbp]
\begin{center}
\resizebox{\hsize}{!}{
\includegraphics[width=0.8\hsize,angle=0, bb = 43 381 525 729]{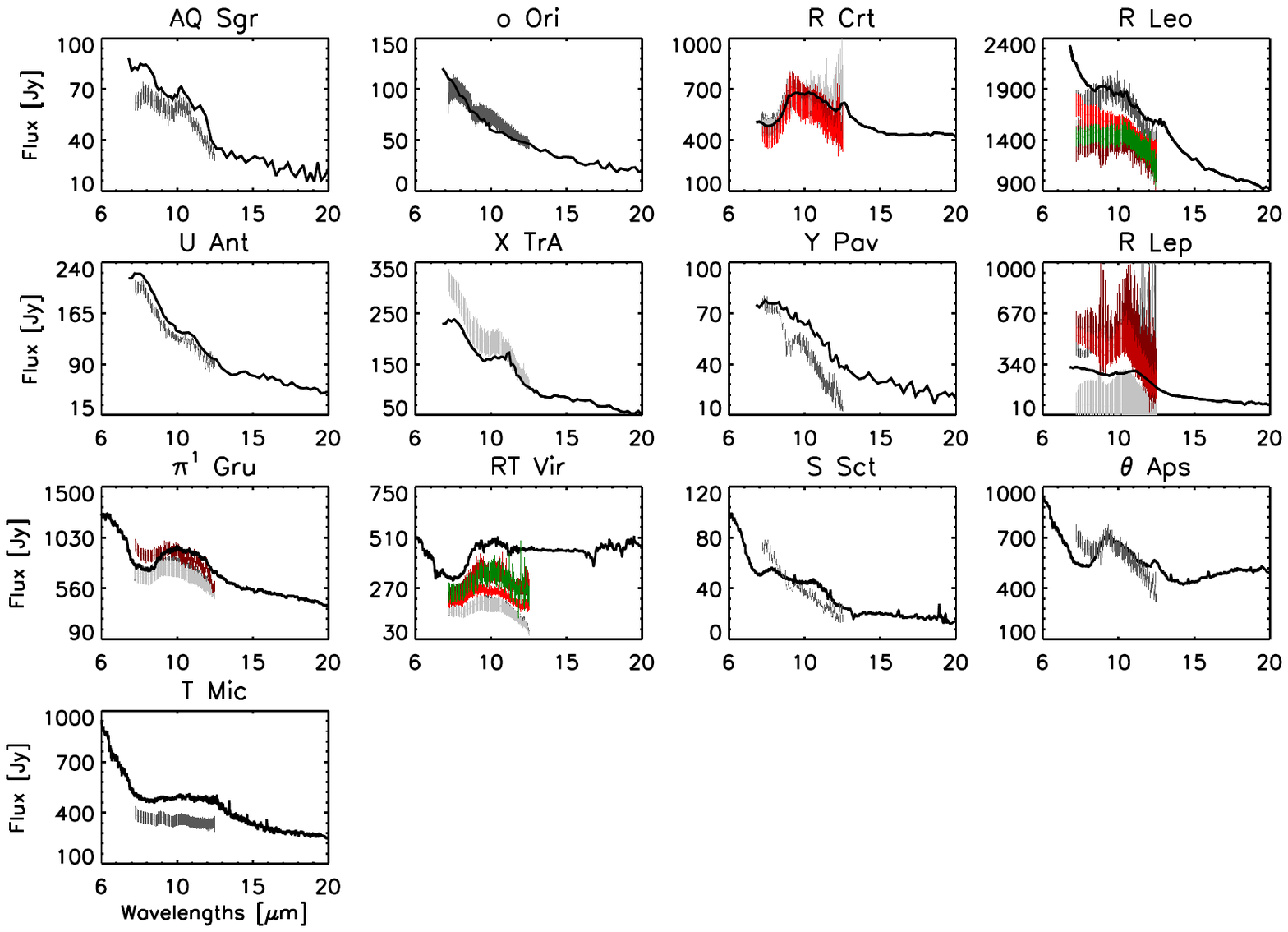}}
\caption{IRAS or ISO spectra (black lines) compared with the MIDI spectroscopic observations (shaded areas).  }
\label{spec.fig}
\end{center}
\end{figure*}

In Fig.~\ref{spec.fig} we compared the level of the MIDI spectra with ISO and IRAS spectra (when the ISO observations were not available).
The ISO and IRAS observations are taken approximately 30 years apart from the MIDI observations. Such a comparison can in principle provide information around
long-time variability due to dust formation and/or mass-loss variation. On the other hand, this can be caused by the FOV difference between the various telescopes. 
The FOV of the MIDI observations is $\sim2.3\arcsec \times 1.6\arcsec$ , and it is smaller than the FOV of IRAS and ISO, $45\arcsec\times45\arcsec$, and $33\arcsec\times20\arcsec$, respectively). Only a detailed modelling is able to distinguish between these effects. Such  modelling is beyond our scope, and we 
simply report cases of suspected variability and leave the modelling to a future investigation \citep{rau2017}.

We observe that 3 stars out of 13 have a mid-infrared flux very similar to the IRAS flux (U~Ant, omi~Ori, and R~Crt).
AQ~Sgr, Y~Pav, RT~Vir, and T~Mic have a flux level below that observed by ISO/IRAS. 
The shape of the spectrum is usually consistent, with an exception made for Y~Pav where a calibration problem cannot be excluded. 
This hypothesis is also supported by the fact that in the Y~Pav spectrum, one can still see the telluric ozone feature at 9.7~$\mu$m. 

For R~Crt, R~Leo, R~Lep, and RT~Vir, we had several MIDI spectra observed  at different visual phases. 
By plotting the flux a various wavelengths (8,10, and 12~$\mu$m) versus visual phase, we study the intra-cycle and cycle-to-cycle variability of the star
 (Figs.~\ref{flux-phase-rcrt},~\ref{rleo-phase-flux},~\ref{rtvirflux},~and \ref{flux-phase-rlep}).
R Crt shows no significant variability (Fig.~\ref{flux-phase-rcrt}). R~Lep and R~Leo show variations.
RT~Vir is by far the star with the best coverage in phase. The variation of the flux over the pulsation period resembles a sinusoid (Fig.~\ref{rtvirflux}).
The flux variation within the cycle corresponds to an amplitude of variability of 0.48 mag at 8~$\mu$m, and 0.75 mag at 10 and 12~$\mu$m. 

The interferometric variability of the visibility spectrum was studied only for $\theta$~Aps, R~Leo, RT~Vir, and R~Lep. 
The intra-cycle observations of the carbon-rich mira R~Lep are taken at very similar visual phases (0.01 difference), therefore it is not a surprise if no interferometric variability is detected  (Fig.~\ref{rlep-interfvariab.fig}, left panel).  However, a cycle-to-cycle variation is observed in the level of the visibility spectrum (Fig.~\ref{rlep-interfvariab.fig}, right panel). 
The variation is more pronounced in the molecular dominated region between 9 and 10~$\mu$m. We do not observe variation between 11 and 12~$\mu$m, where SiC is located. 
The visibility level is higher before the visual phase minimum ($\phi_{\rm{V}} = 1.43$), corresponding to a smaller diameter.

Neither $\theta$~Aps and RT~Vir show any evidence of variability in the visibility spectrum.
Two sets of data are available to check the interferometric variability for R~Leo. The first set includes three observations taken with the short baseline configurations. The visibility level is 
$\sim0.6$; the observations are taken at similar visual phases, but one of them was observed six cycles before. No variability is observed for this set of data (Fig.~\ref{rleo-interf-var}, left panel).
The second set of data also includes three data points, but the first two were averaged because they were taken within two consecutive days (with very similar  PA and projected baseline).
The observations are shown in the right panel of Fig.~\ref{rleo-interf-var} and the difference between the visual phases is 0.13.
It is obvious that in this case we observe a variation in the visibility level from one visual phase to the other. However these observations are associated
with a differential phase signature (Fig.~\ref{diffpha-rleo.fig}, upper row). The differential phase also changes from one set of observations to the other.


\subsection{The geometric fitting results}\label{par:The geometric fitting results}

{\emph{Large Programme data only}}. 
As a first step the geometric models are only fitted to the LP data.
As described in Sect.~\ref{par:strategy}, the LP observations sample very similar spatial frequencies (i.e. the same part of the star)
at different position angles. In Table~\ref{tab:gem-find}, for each star, we present in the first row the results of the fit on the large-programme data only.
The model with the $\chi^2_{\rm{red}}$ closest to 1 is considered as the one best fitting the data, and it is highlighted in bold in Table~\ref{tab:gem-find}.
The elliptical models have been tested for 10 objects out of 14 because in some cases not enough data points were available.
One oxygen-rich star (R~Crt) out of the 4 tested with elliptical models is asymmetric. 
Both S-type objects ($\pi^1$~Gru and omi~Ori) also show indications of ellipticity from the GEM-FIND fit. 
Out of the sample of 4 carbon-rich objects that were tested, none turned out to be asymmetric.
It has to be stressed out that an elliptical solution does not necessarily imply that the environment has a truly elliptical shape.
It  only means that the CSE is non-central symmetric. More complex geometries than ellipses cannot be excluded.

\begin{figure*}[htbp]
\begin{center}
\resizebox{\hsize}{!}{
\includegraphics[width=\hsize,angle=0,bb=70 369 547 701]{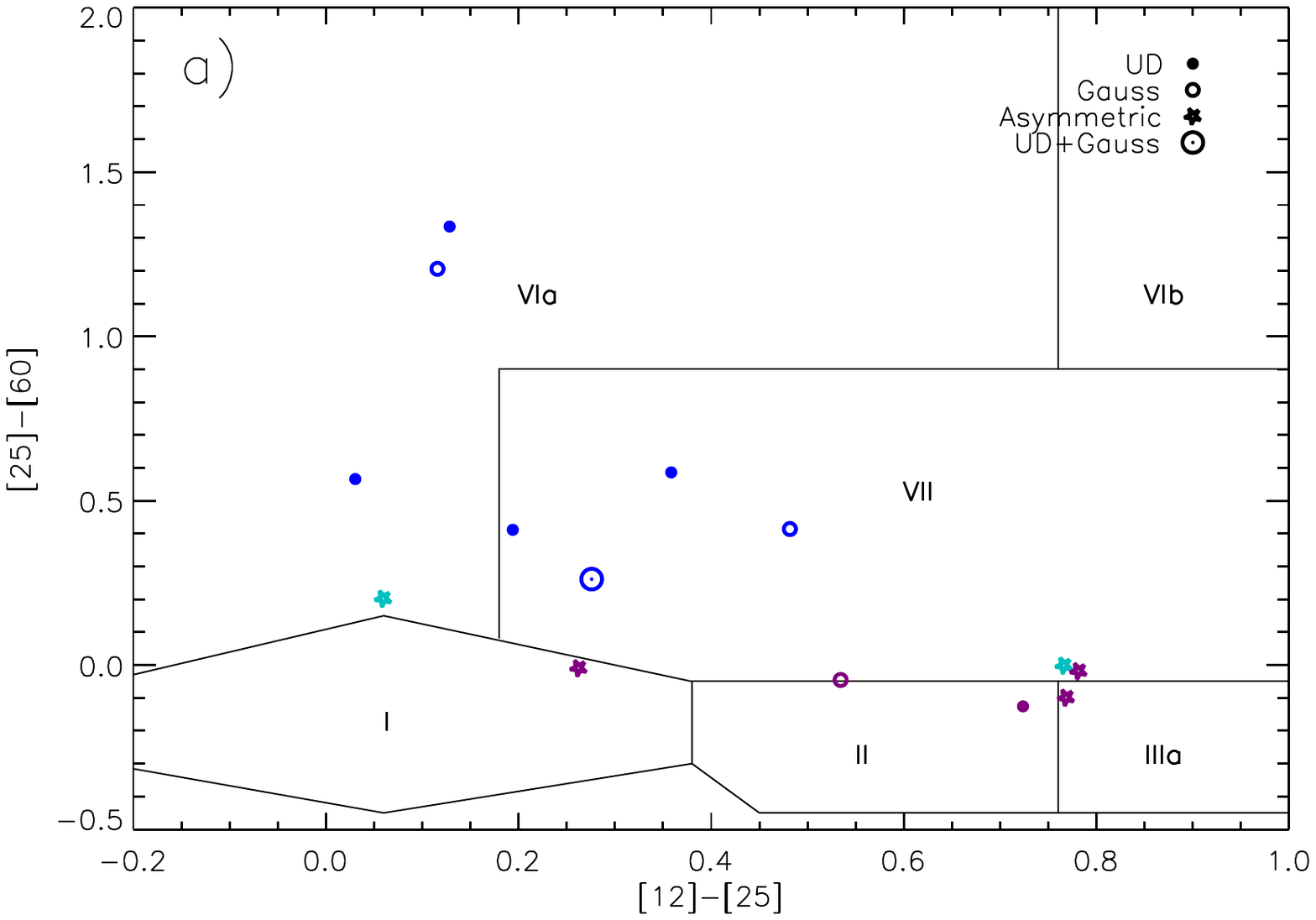}
\includegraphics[width=\hsize,angle=0,bb=70 369 547 701]{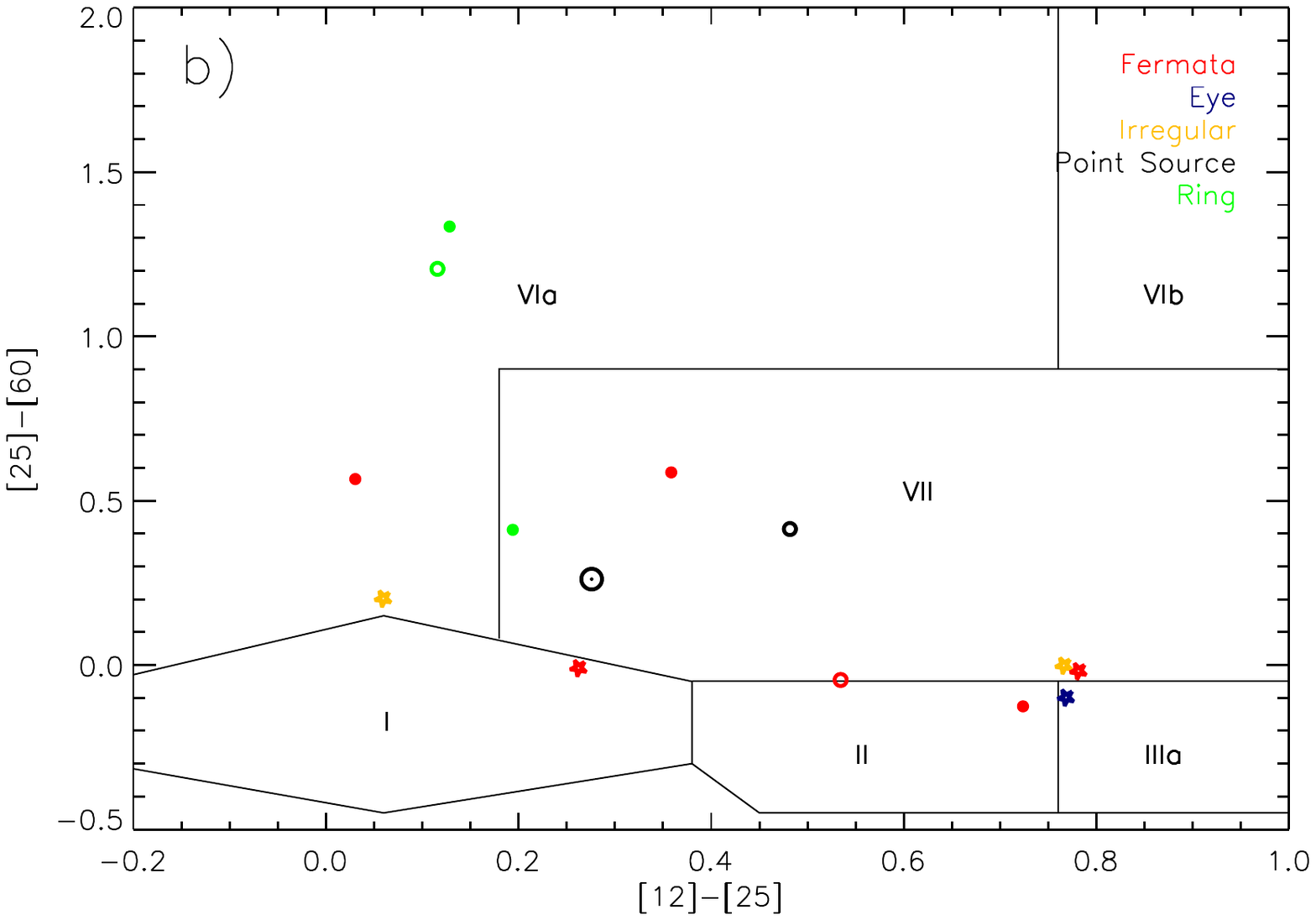}}
\caption{\emph{Panel a):} our sample of stars in the IRAS colour-colour diagram. The symbols show the results of the GEM-FIND fit. Asymmetric solutions are a result of the fit or due to non-zero differential phase. \emph{Panel b):} same as Panel a) colour coded to highlight the Herschel/PACS morphology observed in \cite{cox2012}.}
\label{comparison}
\end{center}
\end{figure*}

\noindent{\emph{Large Programme \& archive data}}.
Nine stars of our sample have archive data, and two of these stars have non-zero differential phases.
There are, therefore, a total of  five targets showing evidence of asymmetric environment. All these asymmetric stars have O-rich chemistry and are located in the lower part of the IRAS colour-colour diagram, as shown in panel a) of Fig.~\ref{comparison}.

After excluding interferometric variability, we combined the data of the LP with those obtained from the archive. 
Besides R~Lep and TX~Psc, all the other stars have archive data that sample mostly the same position angle but at different spatial scales.
As already stated in Sect.~\ref{par:additional}, such data are optimal for studying the stratification of the star, but they are obviously less sensitive to asymmetries.
Exception made to TX~Psc, every time archive data are added to the fit, 
the composite (UD plus Gaussian) model turns out to be the best-fitting model (Table~\ref{tab:gem-find}).
This implies that the stars have an extended environment because of molecular and/or dust opacities.
We note that the additional archive data \emph{wash away} the elliptical solution for all the three objects mentioned at the beginning of this section.
This is a consequence of the fact that the additional data always have the same position angle. Because these additional data are more numerous than the LP data, they  drive the fit solution towards symmetric solutions; but this result certainly does not imply that the asymmetry detected by the dedicated LP data must be considered spurious.
Very likely the best-fitting model would be an elliptical UD+Gaussian, 
but this model has too many free parameters to be tested versus our data set.
By looking at the LP+archive fitting results of Table~\ref{tab:gem-find}, one notes the following:
six stars are best fit by the composite model UD+Gauss, i.e. six stars have an extended, optically thin, component. 
However, it is possible that by adding more visibility points, other objects
also increase in complexity and they are best fitted with a composite geometric model. 
This is very likely for X~Tra, AQ~Sgr, $\theta$~Aps, and Y~Pav where the coverage of the visibility spectrum does not extend below 0.6.
Given the fact that no dust feature is detected around S~Sct, we expect that the visibility spectrum are reproduced only with one component,
even by adding other visibility points.
So far, in our sample only TX~Psc has a visibility that goes down to V$\sim0.2$ without showing departure from uniform disc:
no dust envelope is detected.  
The carbon stars in region VIa, following the loop for the carbon stars, can be fit with single component models
and are supposed to be younger than those located in region VII. The only carbon mira of the sample  is fit with the composite model in region VII. 
All the objects in region VIa do not show SiC in the visibility. The geometry of the CSE increases in complexity from left to
right, where this complexity could be caused by dust or an increased
asymmetry of the CSE.
\smallskip

\subsection{The diameters}\label{The diameters}
In Sect.~\ref{par:strategy} the strategy for choosing the baseline that allows us to resolve the dust-forming region was explained.
One of the output of the GEM-FIND fitting is the diameter (or the full width half maximum for the Gaussian profile)
of the object in the fitted wavelength. 
If one assumes that the UD or Gaussian are a crude approximation of the stellar disk, then one can hazard 
a comparison between the diameter computed from the $V-K$ relation and that derived in the $N$ band via geometric fitting. Such a comparison is presented in Table~\ref{table:diameters}. 
The photospheric angular diameter (see Sect.~\ref{par:strategy}) calculated from the $V-K$ relation \citep{vanbelle1999} are shown in Col.~2, while in Col.~3 we list (when available) the measured $K-band$ values. Column~4 lists the best-fitting GEM-FIND model. 
Columns~5 and 6 report the value of the angular UD-diameters and/or FWHM resulting from the geometric fitting at 8~$\mu$m. Whereas the best-fitting model is the composite model (UD+Gauss; Table~\ref{tab:gem-find}), both $\theta$ and $FWHM$ values are given. In the latter case we do not indicate the error on the UD diameter to point out that in the composite model
we kept the value of the UD fixed to the $\theta_{V-K}$ or to the observed $K$-band value (see Sect.~\ref{par:tool}). Columns~7, 8, 9, and 10 are the same as Cols.~5 and 6 at 10 and 12~$\mu$m. The photospheric diameter and the 8~$\mu$m UD diameters (FWHM) are converted in linear sizes in Cols.~11 and 12 (Col.13), respectively. 
Finally, in Col.~14 we list the ratio between the 8~$\mu$m UD diameter (FWHM in the cases when Gaussian or the composite model is the best-fitting model) and
the photospheric diameter. 
The comparison between the ratios is not straightforward because of the mix of UD and FWHM sizes, but overall most of the diameters at 8~$\mu$m are of the order of two times
the photospheric diameter. The exceptions are the two S-type stars, the carbon Mira R~Lep, and AQ~Sgr, which have a much larger environment ($>2.4~D_{V-K}$). T~Mic and Y~Pav 
have a diameter ratio that is smaller than one, indicating that the environment is very similar to that observed in the near-infrared. Another explanation could be that the $\theta_{V-K}$ diameter of these two stars is overestimated.
Diameters at longer wavelength are even larger with some extreme cases for the more dusty stars such as R~Crt and RT~Vir.
The diameters $ratio$ obtained for the carbon stars are in general agreement with the ratio between the photospheric and 8-9~$\mu$m diameters predicted by 
dynamic model atmospheres \citep[][Fig.~6]{paladini2009}. The sizes here derived are intended as a guideline for the preparation of future interferometric imaging campaigns. Detailed radiative transfer helps constrain the molecular and dust statification, and will be performed in follow-up studies.

\begin{sidewaystable*}
\caption{Predicted, observed diameters, and geometric characteristics.}
\label{table:diameters}
\centering
\begin{tabular}{llllllllllllll}
\hline
\hline
Target      &$\theta_{V-K}$&$\theta_{K}$            &GEM-FIND       &$\theta_{UD, 8}$   & FWHM$_8$    & $\theta_{UD, 10}$  & FWHM$_{10}$      & $\theta_{UD, 12}$  & FWHM$_{12}$  & $D_{V-K}$ or $D_K$  &$D_{UD, 8}$& $D_{FWHM, 8}$ & ratio\\
            &  [mas]      &   [mas]               &               & [mas]           &    [mas]    &   [mas]          &[mas]            &[mas]             & [mas]       &    [AU]            & [AU]    & [AU]       &\\
\hline                                                   
$\theta$~Aps& 24    &18.1\tablefootmark{a}                  &UD             &40$\pm$5         &  \ldots     &85$\pm3$         &\ldots           &86$\pm5$          &\ldots                     &2.0$^{+0.1}_{-0.1}$    &4.6$^{+0.8}_{-0.8}$ &\ldots            &2.24 \\
R~Crt       & 19         &\ldots                  &UD+Gauss       &19               &38$\pm$8     &19               &162$\pm$28       &19                &89$\pm$7            &5.0$^{+1.6}_{-1.0}$    &\ldots           &9.9$^{+5.3}_{-4.1}$ &1.99 \\
R~Leo       & 25        &\ldots\tablefootmark{b} &UD+Gauss       &25               &49$\pm$1     &25               &50$\pm$1         &25                &57$\pm$1   &1.8$^{+0.4}_{-0.3}$    &\ldots           &3.5$^{+0.9}_{-0.6}$  &1.96 \\
T~Mic       & 25         &\ldots                  &Gauss          & \ldots          &36$\pm$7     &\ldots           &55$\pm$8         & \ldots           &64$\pm$9               &5.3$^{+1.4}_{-1.0}$    &4.4$^{+1.5}_{-1.0}$ &\ldots              &0.84 \\
RT~Vir      & 15         & 15\tablefootmark{c}                 &UD+Gauss       &15               &28$\pm$5     &15               &115$\pm$15       &15                &52$\pm$3          &2.0$^{+0.3}_{-0.2}$    &\ldots           &3.7$^{+1.1}_{-1.0}$  &1.84 \\
$\pi^1$~Gru & 28      &21.6\tablefootmark{d}   &UD+Gauss       &21.6             &131$\pm$52   &21.6             &172$\pm$46       &21.6&179$\pm$31    &3.5$^{+0.5}_{-0.4}$    &\ldots           &21.4$^{+11.5}_{-10.9}$&6.36 \\    
omi~Ori     & 11         &9.78\tablefootmark{e}   &UD+Gauss       &9.78             &214$\pm$21   &9.78             &214$\pm$22       &9.78 & 221$\pm22$   &1.9$^{+0.3}_{-0.2}$    &\ldots           &42.8$^{+11.3}_{-9.5}$&21.92\\
U~Ant       & 10         &\ldots                  &UD             &16$\pm$1         &\ldots       &12$\pm$1         &\ldots           &17$\pm1$          &\ldots                   &2.7$^{+0.5}_{-0.3}$    &4.2$^{0.8}_{0.6}$  &  \ldots               &1.56\\
R~Lep       &  9         &12.0\tablefootmark{f}   &UD+Gauss       &12               &29$\pm$1     &12               &33$\pm$2         &12                &44$\pm$2  &5.0$^{+3.6}_{-1.5}$    &\ldots           &12.0$^{+9.1}_{-3.9}$&2.42 \\
Y~Pav       &  7         &\ldots                  &Gauss          &\ldots           &5$\pm$1      &\ldots           &5$\pm$2          &\ldots            &12$\pm$1                 &2.8$^{+0.9}_{-0.5}$    &\ldots           &2.1$^{+1.1}_{-0.8}$ &0.76 \\
TX~Psc      & 11        &\ldots                  &UD             & 12$\pm$1        &\ldots       &10$\pm1$         &\ldots           &$12\pm1$          &\ldots                  &3.0$^{+0.4}_{-0.3}$    &3.4$^{+0.5}_{-0.4}$ &\ldots             &1.13 \\
S~Sct       &  7          &6.22\tablefootmark{g}   &Gauss          &\ldots           &12$\pm$1     &\ldots           &11$\pm$2         &\ldots            &13$\pm$3     &2.4$^{+0.7}_{-0.4}$    &\ldots           &4.5$^{+1.8}_{-1.3}$ &1.89 \\
AQ~Sgr      &  6        &6.13\tablefootmark{h}   &UD             & 17$\pm$3        &\ldots       &20$\pm$4         &\ldots           &33$\pm$3          &\ldots        &2.0$^{+0.6}_{-0.4}$    &5.5$^{+2.5}_{-2.0}$ &\ldots             &2.70 \\
X~Tra       & 11         &\ldots                  &UD             & 22$\pm$2        &\ldots       &23$\pm$4         &\ldots           &39$\pm$3          &\ldots                  &4.0$^{+0.7}_{-0.5}$    &7.9$^{+2.4}_{-2.0}$ &\ldots             &1.99 \\
\hline
\end{tabular}
\tablefoot{A detailed description of the various columns of the Table can be found in Sect.~\ref{The diameters}.
The diameter of carbon stars showing SiC at 11.3~$\mu$m is usually comparable to that measured at 8~$\mu$m or 12~$\mu$m, depending on the visibility level 
(see Appendix~\ref{singlestars.appendix}).
\tablefoottext{a} Adopted from \cite{dumm98} 
\tablefoottext{b} Several diameters ranging between 11 mas \citep{mennesson02} and 27 mas \citep{perrin99} have been published for R~Leo. Given the large range of numbers we decided to stick to the $\theta_{V-K}$ diameter for our calculations.  
\tablefoottext{c} Adopted from \cite{sacuto2008}, and recently confirmed by VLTI/PIONIER data (Paladini et al., prep.).
\tablefoottext{d} Adopted from \cite{sacuto2013}.
\tablefoottext{e} Adopted from \cite{cruzalebes2013}.
\tablefoottext{f} Adopted from \cite{richichi05}.
\tablefoottext{g} Recently measured with VLTI/PIONIER (Le Bouquin, private communications).
\tablefoottext{h} Adopted from \cite{richichi05}.
}
\end{sidewaystable*}

\section{Discussion}
\label{Sect:discussion}

\subsection{Silicon carbide dust}
Silicon carbide is a dust species observed around carbon stars. The typical signature in the spectrum is an emission feature
at $\sim$11.3~$\mu$m. The condensation temperature (and condensation radius) of SiC is matter of investigation, 
however there is increasing evidence that for galactic AGB stars SiC condenses at similar temperatures as (amorphous) carbon dust 
\citep{lagadec2007}.

As a matter of fact, up to now, all the carbon stars observed with MIDI that were showing SiC in the spectrum, 
always had a signature of such dust component in the visibility \citep{ohnaka2007, sacuto2011, paladini2012, zhaogeisler2012, rau2015}.  

\cite{vanboekel2004} showed the case of protoplanetary discs where the (normalised) correlated flux spectrum (or visibility spectrum) reveals certain chemical
features that are not detected in the total flux spectrum. The MIDI observations of U~Ant show a similar behaviour, although in this specific case the chemical feature
(SiC) is present in the total flux and not in the correlated one. 

Assuming that SiC forms in a uniform shell expanding because of a slow wind, then this shell will
be not visible if (i) it is smaller than the resolving power of the interferometer, or (ii) if the contrast
between the shell and the main source is very low. 
The maximum resolving power is given by the longest baseline, and in the case of the U~Ant observations it corresponds to
$\lambda/ (2\times B_{max}) = 11$~mas at 10~$\mu$m. Since 10~mas is the diameter of the
photosphere, it sounds more reasonable to believe that the SiC shell is optically thin.
Further constraints can be given with detailed radiative transfer modelling and future observations with baselines longer than 95~m.

The MIDI observations (correlated and uncorrelated flux) of S~Sct point to variability in the SiC abundance.  
We searched in the literature for possible variation in the mass-loss rate values to test if a recent strong stellar wind event might have pushed away or dissolved the 
SiC dust. No such event was reported within recent decades. A spectrum with VISIR is needed to confirm the lack of SiC in the uncorrelated MIDI flux and to eventually monitor the 
changes in abundance. 

\subsection{The variability}

Mid-IR flux changes of up to 30\% with less flux observed at the minimum visual phase are reported by \cite{karovicova2011, wittkowski2007}, and \cite{tevousjan2004}.
\cite{monnier1998} studied multi-epoch mid-infrared spectra of 
30 late-type stars and found that stars with a strong silicate feature exhibit spectral shape fluctuations, 
where a narrowing of the feature is detected near maximum light. They report that R~Leo shows unusually 
large spectral shape variations that may be attributed to a dynamic dust condensation zone. 
Although based only on four targets, our findings on the $N-$band spectroscopic variability largely confirm previous literature results.
The flux variation of RT~Vir, which is the star in our sample with the best temporal coverage, is comparable 
to the amplitude of $\sim0.8-1$~mag found by \cite{lebertre1993} for O-rich mira stars. 

Interferometric variability was so far observed in the carbon-rich miras V~Oph \citep{ohnaka2007} and R~For \citep{paladini2012}.
While the variability is interpreted as connected to pulsation for V~Oph,  the presence of an asymmetric structure plays a role for R~For.
The variability we observe in the case of R~Lep is just within the 3 sigma uncertainties, however it goes in the same direction 
 of the literature findings. The visibility level suggests that R~Lep is smaller close to the minimum visual phases, 
 similar to V~Oph \citep{ohnaka2007}, and model predictions at 10~$\mu$m showed in Fig.~7 of \cite{paladini2009}.
Model predictions for O-rich Mira stars presented in \cite{karovicova2011} show that the expected visibility changes with visual phase 
are wavelength dependent and in the range of $5-20\%$. During the pulsation phase, the largest difference within the $N$ band is predicted at around 10~$\mu$m.
\cite{karovicova2011} argue that these limited variations are because the sizes of the molecular and dust layers 
do not change significantly with visual phase. According to the authors, the changes are within their measurement uncertainties. 
The star observed by \cite{karovicova2011} is a Mira variable. 
As these objects are known to exhibit larger variability with visual phase, the size changes for our semi-regular stars (all except R Leo) 
are expected to be even smaller. In fact, the RT~Vir and R~Crt data suggest that interferometric variability for semi-regular O-rich stars should be less than $10\%$. 
The interferometric variability of R~Leo reported here is associated with a non-zero differential phase, as in the case of R~For \citep{paladini2012}, and is therefore very likely caused
by (variable) asymmetries.
Since low visibilities are more constraining for the size changes, 
future interferometric snapshot campaigns on temporal variability should concentrate on those projected baselines that are able to probe V$\le 0.2$.

\subsection{The geometry}

One may wonder why only two objects out of 14 show a non-zero differential phase.
Both objects are O-rich M8III giants, the coolest objects of the sample. 
The fact that the two stars appear in a completely different part of the IRAS colour-colour diagram could be due to different masses, hence mass loss. 
What the observations of these stars also have in common are the spatial scales probed.
In both cases the differential phase is detected while sampling visibilities below 0.2.
Looking at our whole sample, these spatial scales are also probed in two observations of the carbon Mira R~Lep. The differential phase is $< 10$\degr,
hence we conclude that R~Lep is spherical (at least at PA = 48\degr and 248\degr).
There have been only a few cases of non-zero differential phase in the literature detected with MIDI; these cases are 
R~For \citep{paladini2012}, BM Gem \citep{ohnaka2008a}, IRAS+18006-3213 \citep{deroo2007}, and L$_2$~Pup \citep{kervella2014}.
All these observations also had in common the fact of probing visibilities $< 0.2$. 
Thereupon we can assume that the lack of detection of non-zero differential phase does not necessarily exclude the existence of asymmetries in the brightness distribution 
in the other targets of our sample. Very likely asymmetric structures show up in the differential phase when probing high spatial frequencies, 
similar to what is reported in the near-IR \citep{ragland2006, cruzalebes2013}. 

If we include the objects with a differential phase signal, then $\sim57\%$ of the stars with O-rich chemistry (O-type plus S-type stars) have a CSE geometry that is 
consistent with being an asymmetric structure.
These results suggest that asymmetries in the dusty environment are more common among the oxygen-rich objects.
Near-infrared investigations of AGBs \citep{ragland2006, cruzalebes2015}, on the other hand, indicate that asymmetries are more common among carbon stars. 
This result can be explained with the large-grains scenario for O-rich stars \citep{hoefner2008}.  
Oxygen-rich dust grains are nearly transparent in the near-infrared, therefore not many brightness asymmetries are expected at those wavelengths. 
The dusty region of O-rich stars probed in the mid-IR
should be more "blobby" than for C-rich stars because there are stronger non-linear effects in radiative acceleration. 
This is valid \emph{if} the scattering on transparent grains scenario is correct; in fact scattering depends much more steeply on grain size (S. H\"ofner, private communications).
Bipolar objects on the post-AGB are mainly O-rich \citep{lagadec2011}. The latter well agrees with our results on the AGB.

The result obtained by fitting only the LP data, plus the results of the differential phase study,
indicate that asymmetries in the dust-forming region are concentrated in the lower part of the IRAS colour-colour diagram.
No object located in region VII so far showed asymmetries. More observations covering long baselines (high spatial frequencies), or (ideally) $N$-band images
obtained  with the next generation VLTI/MATISSE instrument \citep{lopez2006} are needed to confirm this finding.

\subsection{Comparison with the MESS results}
The aim of our programme is to observe the evolution of asymmetric structures through the atmosphere of a representative sample of AGBs.
For this purpose we selected stars imaged with the Herschel/PACS instrument and we complemented these 
observations of the large scales of the atmospheres with interferometric VLTI/MIDI observations of 
the inner spatial scales. 
Panel d) of Fig.~\ref{comparison} shows the stars of our sample in the IRAS colour-colour diagram. The objects are colour coded
to highlight their MESS classification. The symbols are the same as in panel b) and indicate the presence of a non-spherically symmetric environment.
Three objects out of the LP sample are classified by \cite{cox2012} as rings, six 
are fermata type, two irregulars, one eye, and two unresolved.

All the stars from the ring class are fitted with the one-component geometric model,
and no asymmetry is detected in the dust-forming region.
For two of these stars (S~Sct and U~Ant) no SiC dust component is spatially resolved.
The dust observed in their spectra might have been pushed away by episodic wind, or
it is optically thin.
As already mentioned in the previous section, for X~TrA we have only a few data
points and we cannot exclude that by adding more visibilities the picture will become more complex.
 
The carbon-rich stars TX~Psc and AQ~Sgr, plus most of the Herschel images of our oxygen-rich sample, are of fermata type, i.e. show an
interaction with the ISM. R~Crt is classified as an eye morphology by \cite{cox2012}.
However, the fact that R~Crt is the only oxygen-rich AGB star in the MESS sample that is
of eye shape, and the fact that the eye shape is not very pronounced may indicate that R~Crt may also be of fermata type \citep{cox2012}.
If we consider R~Crt among the possible fermata, three out of seven of the stars in this class show a direct detection
of an asymmetry (via differential phase) or at least a hint of a non-spherically symmetric environment (detected by fitting only the LP data).
This  indicates that 42$\%$ of the stars classified as fermata have an asymmetric dust-forming region, and they all have O-rich chemistry. 

The irregular class contains objects with diffuse irregular extended emission. 
Both the stars classified as irregular are S-type stars, and they can be fitted with two-components model and show non-central symmetric CSE in the LP data.

\section{Conclusions and outlook}
\label{par:conclusions}

In this paper, we present mid-IR interferometric and spectroscopic data observed with VLTI/MIDI for a sample of 14 AGB~stars.
The sample is based on the list of objects observed within the frame of the MESS 
programme with Herschel PACS \citep{groenewegen2011}. The aim of our study is to investigate the morphology of the
dusty environment of these objects at different spatial scales and to answer the questions:  i) Are the asymmetries of the outer CSE intrinsic to the mass-loss process, and are they only
due to interaction with the  ISM? ii) At which height does the mass-loss process become manifestly non-spherical? 
iii) How does the geometry of the atmosphere change at the different evolutionary stages (M-S-C stars, and from almost dust free to very dusty objects) within the AGB sequence?

The first question cannot be addressed using only the mid-infrared interferometric data presented here. These data, in fact, 
scan the stellar atmosphere between 1-2 and 10 stellar radii, and the intermediate spatial scales probed by VISIR are needed to 
have a connection with the Herschel/PACS images. Nevertheless, we report that all the objects
in our sample classified as symmetric rings by \cite{cox2012} so far have a symmetric dust-forming region. 
All the objects classified as irregular need a composite model (i.e. they have an extended environment), and the fit of the LP data indicates some asymmetric behaviour.
Approximately 42\% of the objects classified as fermata (i.e. show interaction with the ISM) also have an asymmetric dust-forming region.

To answer the second question, it is obvious that asymmetries in the brightness distribution appears already in the dust-forming region of AGB stars.
We directly detected asymmetries via differential phase measurement in two cases: RT~Vir \citep[already reported by][]{sacuto2013},
and R~Leo. However, the fact that we find the other objects to be symmetric does not preclude the presence of clumps or
small or very faint asymmetric structures. GEM-FIND fitting of the LP data shows that beside these two stars,
three more show a hint of non-central symmetric morphology. 
Concerning the third question, all the stars with asymmetric behaviour have O-rich chemistry.
Therefore asymmetric structures in the mid-infrared are more common among O-rich and S-type stars. On the contrary
literature suggests that C-rich stars are more asymmetric in the near-IR. We speculate that this result supports the large-grains scenario for O-rich stars. 

Another major finding of our programme is that silicon carbide dust is not detected in the correlated flux of S~Sct and U~Ant. 
U~Ant shows SiC in the total spectrum, and we speculate that SiC is optically thin. 
The case of S~Sct might be related to variability in the stratification of this material.

Finally, by analysing archive data, we studied the spectroscopic and interferometric variability.
Spectroscopic variability is reported in a few cases and the flux variations are in agreement with what was already observed in the literature
\citep{lebertre1993}. An exception is made for the carbon-rich Mira R~Lep case, which nevertheless requires confirmation through 
monitoring campaigns; interferometric variability (i.e. change in the shell size) is observed in the O-rich Mira R~Leo. In the latter case
the variability is associated with an asymmetry in the brightness profile of the star. 
The lack of detection for the other stars suggest that interferometric variability in non-mira stars is less that $\sim10\%$ of the visibility.



The next step of this study will be, as already mentioned, to bridge the gap between MIDI and Herschel with additional
single-dish images in the mid-infrared. These observations will be crucial for completing the picture of the morphology of the dusty
environment.
Finally, MIDI was recently decommissioned, but a second generation instrument VLTI/MATISSE will be available at VLTI
in 2018. This interferometer will combine the light from four telescopes, and it will observe in the $L$, $M$, and $N$ bands.
While MIDI, with only two telescopes, gave us a rough idea of the appearance of the stars within this sample,
MATISSE will provide images allowing  us not only to detect asymmetric structures, but also to unveil their nature (disk-like or clumps).
Because of the complex nature of the targets here studied,
a complete coverage of the visibility curve is mandatory. In the $N$ band, this will require in certain cases baselines longer than the 150 m (for stars such as omi~Ori and TX~Psc),
but also very short baselines. In this frame, an aperture masking experiment on the VISIR instrument matching the spectral resolution of MATISSE is required.

\begin{acknowledgements}
We thank the anonymous referee for comments that helped improving the paper.
This work is supported by the Austrian Science Fund FWF under the project AP23006, the Belgian Federal Science Policy Office via the PRODEX Programme of ESA,
the Belgian Fund for Scientific Research F.R.S.-FNRS, and the 
European Community's Seventh Framework Programme under Grant Agreement 312430.
This research has made use of the SIMBAD database, operated at CDSm Strasbourg, France, and 
the Jean-Marie Mariotti Center \texttt{Aspro} service. 
We acknowledge with thanks the variable star observations from the AAVSO International Database 
contributed by observers worldwide and used in this research.
We thank T. Lebzelter, M. Cesetti, and F. Bufano for helpful comments during the proposal preparation; 
the ESO Paranal team for supporting our VLTI/MIDI observations; 
G.~C.~Sloan for providing the IRAS and ISO spectra;
J.-B. Le Bouquin who kindly provided us reduced PIONIER data of S~Sct; 
S. H\"ofner and K. Tristram for useful discussion.

\end{acknowledgements}

\bibliographystyle{aa}
\bibliography{mybib}

\clearpage
\appendix
\section{MIDI results on individual stars}
\label{singlestars.appendix}
\subsection{$\theta$~Aps}
$\theta$\,Aps is a semi-regular variable star located at a distance of 113\,pc \citep{vanleeuwen2007}. In its ISO spectrum the star shows silicate emission that is typical for an oxygen-rich AGB star \citep[e.g.][]{fabian01} with features at 10, 13, 19.5, and 32\,$\mu$m. The mass-loss rate of the star was determined to be $0.4\times10^{-7}$\,M$_\odot$\,yr$^{-1}$ by \cite{olofsson02}. The Herschel image is of fermata-type with indications for a companion \citep{cox2012}.  \cite{mayer2013} find indications for a jet that is interfering with the wind-ISM bow shock.\\
\\
$\theta$\,Aps was observed in 2011 and 2012. Six out of seven data sets are of good quality. There are no archive data. 

\subsubsection{Variability}
The variability period for $\theta$\,Aps is given as $P$=119\,d \citep{samus2009}. We calculated visual phases of the observations adopting phase-zero point $T_0=2\,454\,622$~JD.

The MIDI spectra have similar visual phases. Thus, a check for intra-cycle variability is not possible. However, a check for cycle-to-cycle spectroscopic 
variations can be performed by comparing the MIDI with ISO and IRAS data (Fig.~\ref{spec.fig}). 
The flux level of the ISO and MIDI spectra  is the same at least in the range $\sim 9.8$\,--\,12.5$~\mu$m, i.e.\,no spectroscopic cycle-to-cycle variations are found. 
However, below $\sim 9.8~\mu$m, a clear change in flux level is observed.
 
\begin{figure}[htbp]
\begin{center}
\resizebox{\hsize}{!}{
\includegraphics[width=\hsize,angle=180,bb=535 74 766 313]{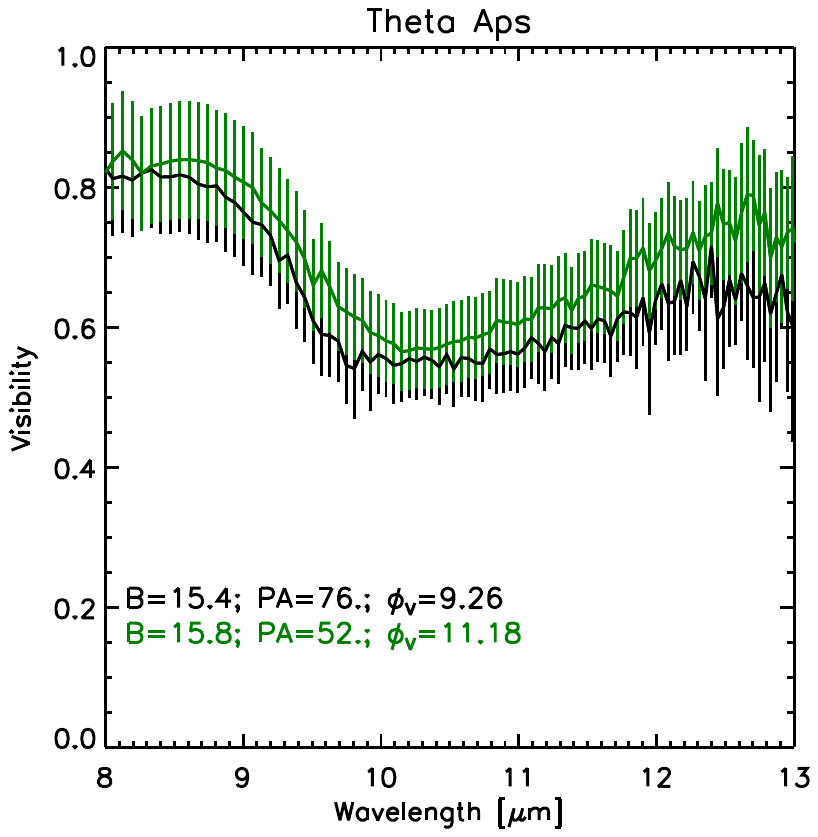}}
\caption{Interferometric variability check for $\theta$\,Aps. B is the projected baseline, PA is the position angle, and $\phi_{\rm{V}}$
is the visual phase.}
\label{tetaps-variability}
\end{center}
\end{figure}

To check for interferometric variability, a set with similar baseline lengths and position angles 
observed at different visual phases is shown in Fig.\,\ref{tetaps-variability}. No interferometric variability is detected.

\subsubsection{Morphology}
\begin{figure*}[htbp]
\begin{center}
\resizebox{\hsize}{!}{
\includegraphics[width=\hsize,angle=0,bb=52 583 407 716]{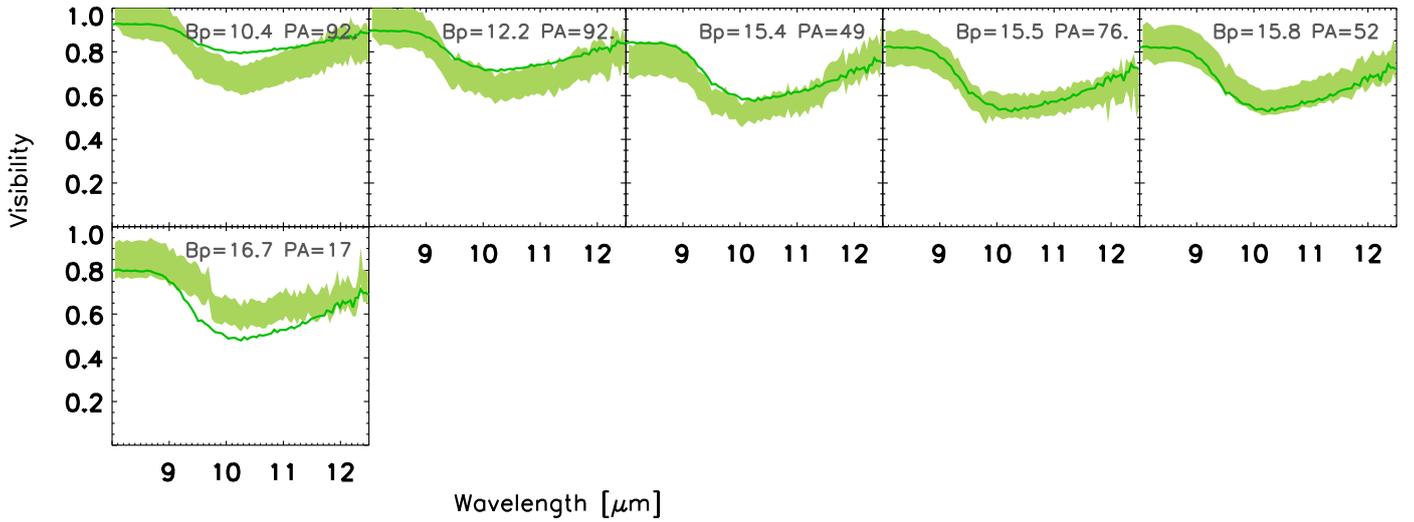}}
\caption{Best-fitting GEM-FIND model (solid line) for the MIDI visibilities of $\theta$\,Aps.}
\label{tetaps-vis-model}
\end{center}
\end{figure*}

Calibrated visibilities are shown in Fig.\,\ref{tetaps-vis-model}. No differential phase signature is detected, i.e.\,no asymmetries are observed in the brightness distribution. 

The $\chi^2_{\rm{red}}$ of the GEM-FIND fitting for the different models are given in Table\,\ref{tab:gem-find}. The increase of the UD diameter from 8\,--\,10\,$\mu$m is pronounced for $\theta$\,Aps, 
indicating that the star shows a strong silicate feature. This is in agreement with the ISO spectrum. 
The difference between the $K$-band \citep[18.1\,mas,][]{dumm98} and $N$-band diameter ($\sim$40--90\,mas) is evidence for dusty material in the circumstellar 
surrounding of the star. As $\theta$\,Aps has a low mass-loss rate, the circumstellar environment can 
be assumed to be optically thin, 
therefore one cannot exclude that observations at larger baselines would point to a two-component structure of the CSE of $\theta$\,Aps, as is the case for RT\,Vir and R\,Crt. 

\subsection{R~Crt}
R\,Crt is a semi-regular variable star. Its distance is given as 261\,pc \citep{vanleeuwen2007}. The star was part of many maser and CO-line studies. The CO-envelope seems to be consistent with a uniformly expanding envelope \citep{kahane94}. The Herschel image was interpreted to be of `eye'-shape \citep{cox2012}. This shape, 
however, is not well constrained (A. Mayer, private communication). The strength of the silicate feature is quite large \citep{begemann97}. R\,Crt also has a higher mass-loss rate ($8\times10^{-7}$\,M$_\odot$\,yr$^{-1}$) than the other O-rich sources.

R\,Crt was observed in 2009, 2011, and 2012, and 8 out of 12 data sets are of good quality. 

\subsubsection{Variability}
The variability period of R\,Crt is given as $P=160$\,d \citep{samus2009}. We calculated the visual phase of our observations assuming a phase-zero point $T_0=2\,454\,225$~JD. 
The observations were carried out at different visual phases within different cycles. This makes a check for cycle-to-cycle and intra-cycle variability necessary.
 
\begin{figure*}[htbp]
\begin{center}
\resizebox{\hsize}{!}{
\includegraphics[width=\hsize,angle=180,bb=48 86 769 311]{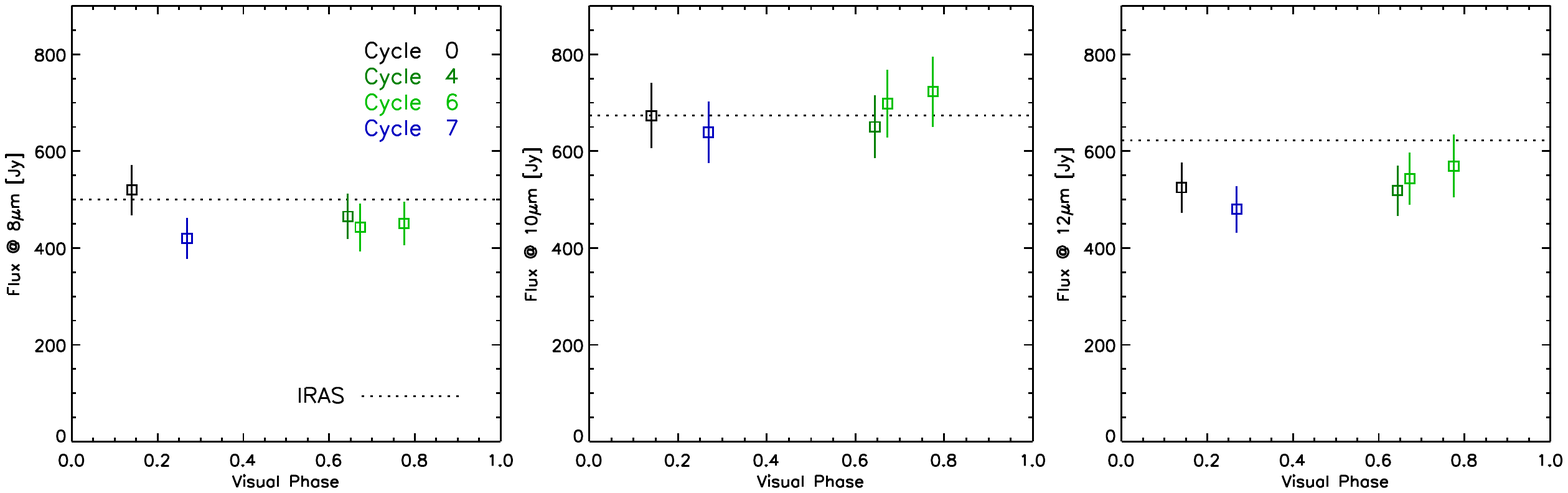}}
\caption{MIDI fluxes at 8, 10, and 12~$\mu$m observed for R~Crt at various visual phases. These fluxes are compared with the IRAS flux.}
\label{flux-phase-rcrt}
\end{center}
\end{figure*}

Possible variability effects can be checked in Fig.\,\ref{flux-phase-rcrt}, which gives the visual phase versus flux at 8, 10, and 12\,$\mu$m. Fluxes agree within the errors for 8 and 10\,$\mu$m. At 12\,$\mu$m there seems to be a variation in the flux level when comparing MIDI and IRAS. 

No sets with similar baseline lengths and position angles observed at different epochs are available. Therefore, no statement can be made concerning the interferometric variability. 

\subsubsection{Morphology}
\begin{figure*}[htbp]
\begin{center}
\resizebox{\hsize}{!}{
\includegraphics[width=\hsize,angle=0,bb=52 583 407 716]{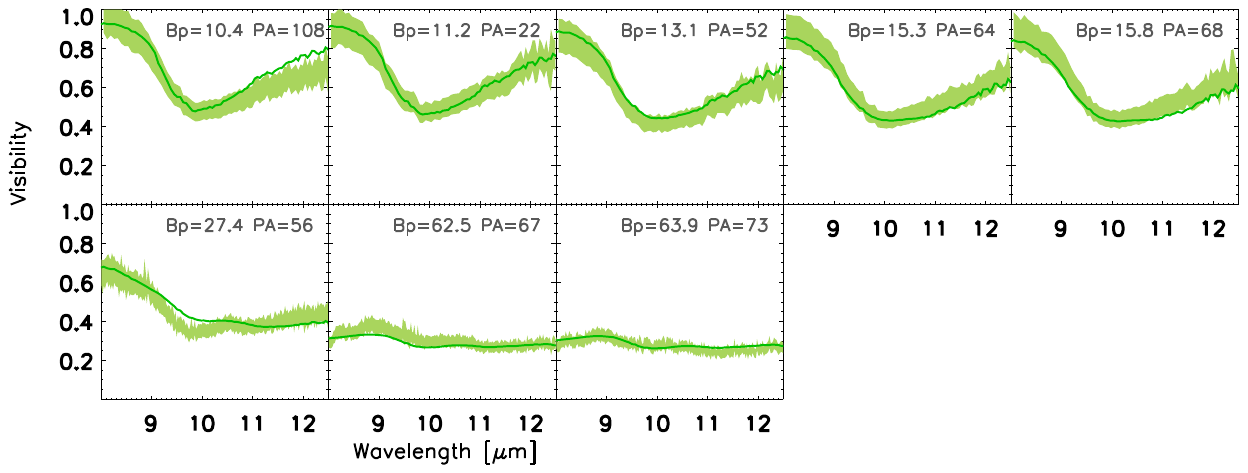}}
\caption{Best-fitting GEM-FIND model (solid line) for the MIDI visibilities of R\,Crt.}
\label{rcrt-vis-model}
\end{center}
\end{figure*}

Calibrated visibilities are shown in Fig.\,\ref{rcrt-vis-model}; the differential phase is always zero. 

As for $\theta$~Aps, the strong silicate feature can be seen in the visibility profile at $\sim$\,9.8\,$\mu$m for short baselines.
On the other hand, for R~Crt we also have at our disposal long baseline observations. 
The shape of the visibility for the longer baselines may still show some silicate feature, but much less pronounced than for the short features. 
Contributions from other molecular and dust species may shape the visibility profile (e.g.\,SiO, H$_2$O, and Al$_2$O$_3$).

The LP observations of R\,Crt are best described by an elliptical Gaussian profile with an axis ratio of 0.7 and an inclination angle of $\phi_\mathrm{incl}=157^\circ$. 

As can be seen in the second row of Table\,\ref{tab:gem-find}, all one-component models fail to fit the observations of the LP combined with the archive data. The best-fitting model is a two-component model (CircUD+CircGauss; Fig.~\ref{rcrt-vis-model}). This suggests that the environment of R\,Crt is optically thin, where the outer environment is dominated by silicate-rich dust. 

\subsection{R~Leo} 
R\,Leo is a well-studied O-rich AGB star of Mira type. Its distance is 110\,pc \citep{vanleeuwen2007}. 
Its magnitude in the $V$ band ranges from 11.3 to 4.4 \citep{kholopov98}. 
\cite{knapp1998} determined a mass-loss rate of $9.4\times10^{-8}$M$_\odot$~yr$^{-1}$. The Herschel image suggests that R\,Leo is of `fermata' type \citep{cox2012}.\\
        \emph{Asymmetries.} Evidence for asymmetries were found by \cite{ireland2004b} and \cite{burns98} from the closure 
        phases obtained with optical interferometry (650\,--\,1000\,nm). Non-zero closure phases are also reported in the 
        mid-infrared by \cite{tatebe08}. They suggest that the closure phase signal of the star comes from an asymmetry that is 
        located in the southern hemisphere of the star. \cite{wiesemayer09} claim that they detect a planet at a separation of 24\,mas 
        in their SiO maser data. They do not mention whether such a planet could be responsible for the closure phase signals observed 
        in the other wavelength ranges. \cite{perrin99} used near-infrared interferometry and show that their
        observations at low spatial frequency ($<40$\,arcsec$^{-1}$) are well represented by a UD model. On the other hand, 
        data at high spatial frequencies (40\,--\,80\,arcsec$^{-1}$) cannot be explained by 
        a UD or Gaussian intensity distribution, which may point to the presence of one or more extra structures. This is confirmed 
        by the observations of \cite{mennesson02}. \cite{perrin99} mention that the low spatial frequency observations within the first null 
        cannot resolve small structures such as spots and this may be the reason why a UD model fits those data well.
        \cite{monnier04}, however, 
        report that both the low and high frequency data (1\,--\,50\,arcsec$^{-1}$) in the near-infrared can be perfectly fitted with a UD model. 
        This may depend on the visual phase and is further discussed below under the heading Visual phase-dependent diameter.
        The CO line profile of R\,Leo shows an asymmetric shape with the red side being stronger \citep{knapp1998, teyssier06}.\\
        \emph{Ellipticity.} Reports of an elliptical shape of the CSE of R\,Leo are reported in the optical 
        \citep{lattanzio97} and in the mid-infrared \citep{tatebe08}. On the other hand, several other studies do not 
        find any signs of ellipticity \citep[e.g.\,][]{burns98, monnier04}.\\
        \emph{Wavelength-dependent diameter.} A dependence of the diameter on wavelength is reported in the optical by 
        \cite{hofmann2001}, where the size is twice as large in the TiO band-head as in the continuum. A size increase from the 
        $K$ band to the $L$ band of 20\,--\,30\% was reported by \cite{mennesson02} and \cite{schuller04}. The multiwavelength 
        study of \cite{woodruff09} in the near-infrared showed that there is a clear anti-correlation between the angular
        diameter and the features seen in the spectrum. \\
        \emph{Visual phase-dependent diameter.} \cite{chagnon02} interferometrically observed the star in the near-IR at two 
        different visual phases and find that one set can be modelled with a UD, the other not. This is confirmed by 
        \cite{fedele05} who find that their pre-maximum data can be explained by a UD, however, the post-maximum data cannot. 
        Such a variability in the size of R\,Leo is detected by several authors in the near-IR 
        \citep{ perrin99, chagnon02, mennesson02, fedele05, woodruff08, woodruff09}, with the star being largest at visual 
        minimum as already predicted by pulsation models; none of these models, however, can reproduce the pulsation amplitude of R~Leo \citep{ireland2004}.
        Such variability was also detected in the optical \citep{burns98} and mid-infrared \citep{tatebe2006, tatebe08}. \cite{mennesson02} mention that such changes may be caused by variations in the spatial extent and/or in the 
        opacity of the outer atmospheric layers. \\
        \\
R\,Leo was observed in 2006, 2007, and 2012. Out of 26 data sets, 17 are of good quality. 

\subsubsection{Variability}
The variability period of R\,Leo is $P=310$\,d \citep{whitelock2000} and we adopted phase-zero point $T_0=2\,453\,540$ JD to calculate the corresponding visual phase. 
The observations were carried out at different visual phases within different cycles. This makes a check for cycle-to-cycle and intra-cycle variability possible.
\begin{figure*}
\begin{center}
\resizebox{\hsize}{!}{
\includegraphics[width=\hsize,angle=180,bb=48 86 774 311]{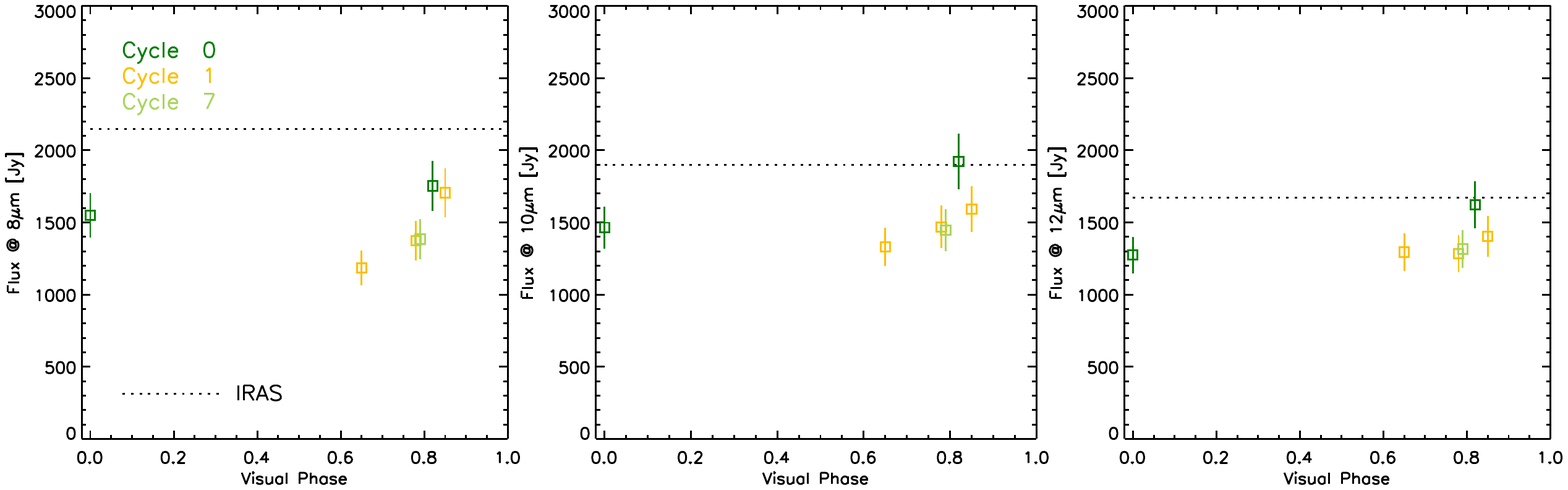}}
\caption{Visual phase vs MIDI flux at 8, 10, and 12~$\mu$m for R~Leo.}
\label{rleo-phase-flux}
\end{center}
\end{figure*}

\begin{figure*}[htbp]
\begin{center}
\resizebox{\hsize}{!}{
\includegraphics[width=\hsize,angle=180,bb=535 74 766 313]{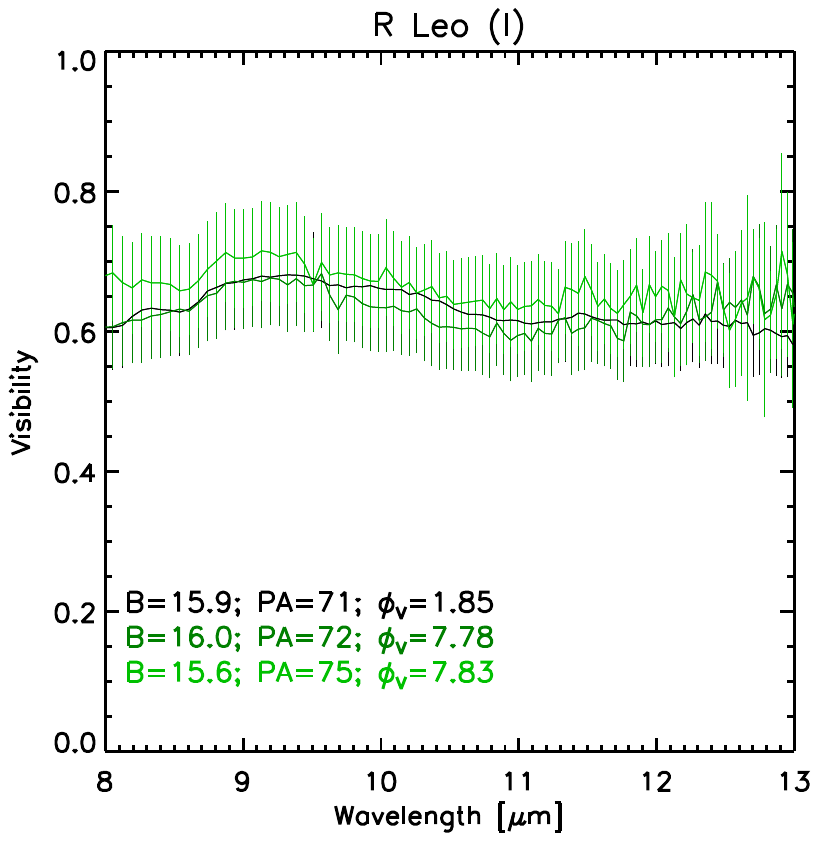}
\includegraphics[width=\hsize,angle=180,bb=535 74 766 313]{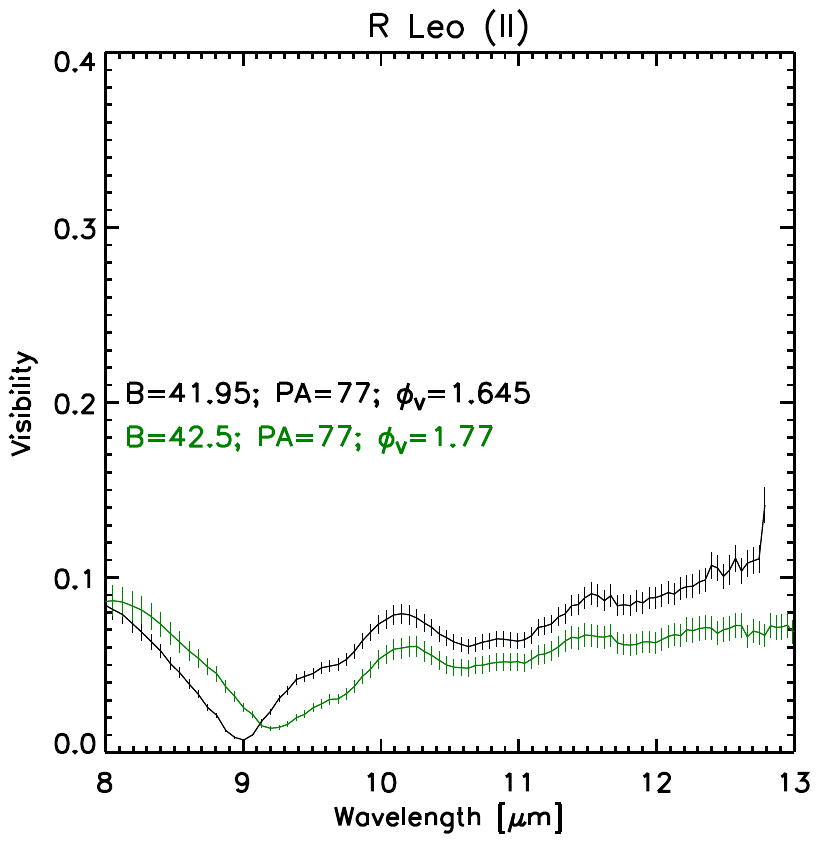}}
\caption{Interferometric variability check for R~Leo. The black line shown in the right panel is the result of the average of two observations. The latter were observed one day apart
with similar projected baseline and PA. }
\label{rleo-interf-var}
\end{center}
\end{figure*}
The variation of the flux with visual phase was detected in the mid-infrared by \cite{monnier99} and \cite{tatebe2006}, where the flux is 30\% lower at visual minimum. 
We confirm the trend of lower fluxes towards the minimum in Fig.\,\ref{rleo-phase-flux}, which shows visual phase versus flux at 8, 10, and 12\,$\mu$m.  

The interferometric observations with similar baselines and position angle shown in green in the left panel of Fig~\ref{rleo-interf-var} are taken at very similar visual phase, therefore it does not come as a surprise that
no intra-cycle variability is detected. A third observation (black line in the left panel of Fig.~\ref{rleo-interf-var}) is taken six cycles apart.
Even in this case no cycle-to-cycle interferometric variability is observed.
Changes in the visibility with visual phase are shown in the right panel, Fig~\ref{rleo-interf-var}. The black line here is the result of the average of two observations
taken at the same baseline and position angle, and one day apart. The variability in the visibility is accompanied with a non-zero differential phase.
Such variations were found in the mid-infrared by \cite{tatebe2006} and \cite{tatebe08}. This may be
an indication that for R\,Leo observations at different visual phase should not be combined.

\subsubsection{Morphology}
\begin{figure*}[htbp]
\begin{center}
\resizebox{\hsize}{!}{
\includegraphics[width=\hsize,angle=0,bb=52 471 407 716]{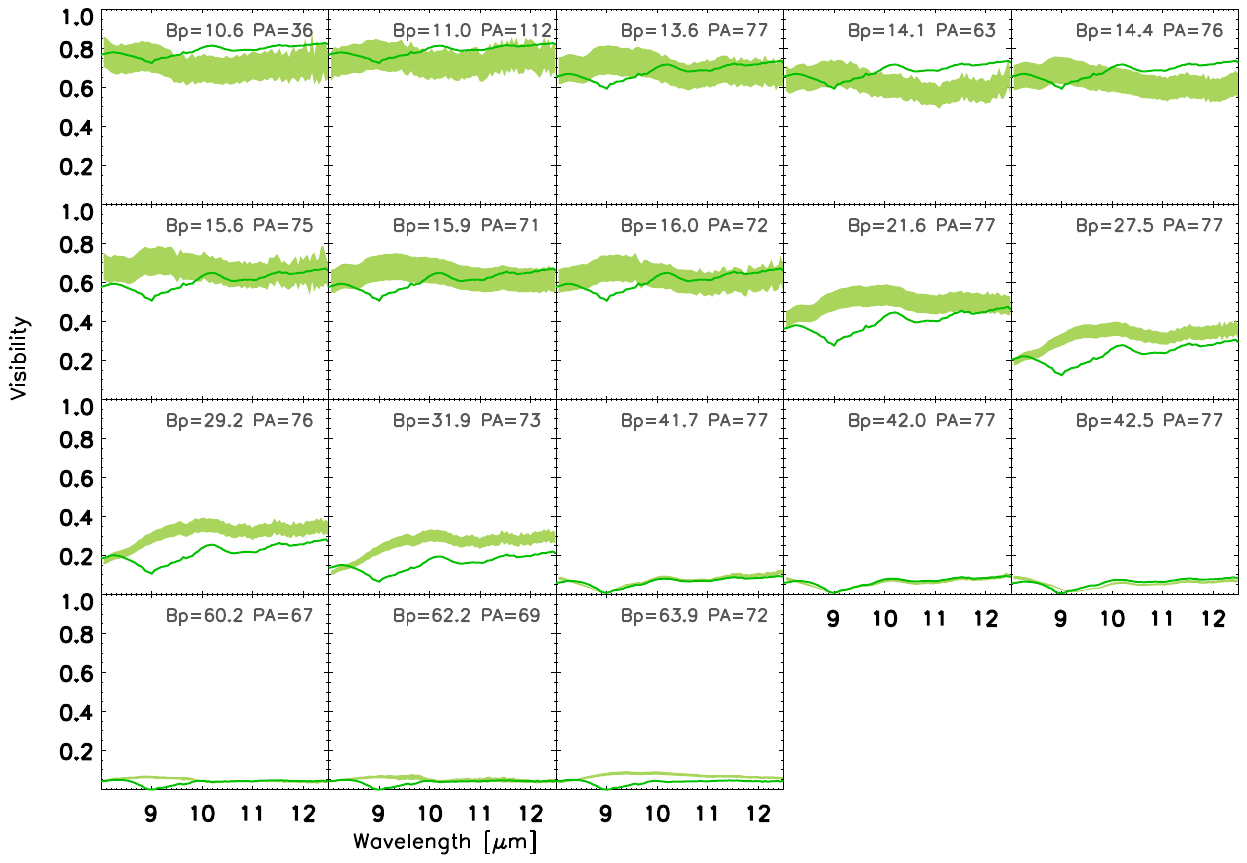}}
\caption{Best-fitting GEM-FIND model (solid line) for the MIDI visibilities of R\,Leo.}
\label{rleo-vis-model}
\end{center}
\end{figure*}

Calibrated visibilities are shown in Fig.\,\ref{rleo-vis-model}. 

The $\chi^2_{\rm{red}}$ of the GEM-FIND fitting for the LP data is given in the first row of Table\,\ref{tab:gem-find}. 
The morphology of R\,Leo seems to be very complex. When performing a GEM-FIND fit on the whole LP data sets, the best model is a UD. 
However, combining the LP and archive data, none of the GEM-FIND models were able to provide a good fit (confirmed by the high value of $\chi^2_{\rm{red}}$ for the best-fitting model). 
One explanation could be that the environment of R\,Leo is made of more than two components, i.e.\,multiple shells are observed with MIDI. 
Another possible explanation could be variability in the $N$ band, which was already reported by \cite{tatebe2006} and \cite{tatebe08} and confirmed in this work. 
Therefore, considering that there may be intra-cycle variability, we combined all data at similar visual phases to do the fitting. 
Also in this case the UD model does not provide a good fit to the data. As all the observations in the archive were observed at approximately the same position angle, 
they do not provide supplementary constraints to fit more complex asymmetric models. 

There is a differential phase signature present for observations with baselines $\sim$40 and $\sim$60\,m (Fig.~\ref{diffpha-rleo.fig}). 

\subsection{T~Mic} 
T\,Mic is a semi-regular variable star located at a distance of 200 pc \citep{loup1993}. This star was part of different photometric, spectroscopic, and CO-line-profile studies. 
Its mass-loss rate is estimated to be $8\times10^{-8}$\,M$_\odot$\,yr$^{-1}$ \citep{olofsson02}. 
The Herschel image is interpreted as of `fermata'-type \citep{cox2012}, i.e. shows signs of interaction with the ISM. 
T\,Mic was observed in 2004, 2010, and 2011. Out of 10 data sets, 7 are of good quality. 

\subsubsection{Variability}
We derived visual phases for the observations, however
the values need to be considered with caution. In fact light curves available from AAVSO, ASAS, DIRBE, and HIPPARCOS 
do not cover the epoch of the ISO or MIDI data.
The adopted origin for the visual phase determination is $T_0=2\,452\,832$~JD. 
The observations were carried out at different visual phases within different cycles. 
The two spectra obtained on 2004 July 30 and 31 were averaged.
The flux level of the ISO and MIDI spectra in Fig.\,\ref{spec.fig} is not the same, i.e.\,the level of $N$-band emission may have changed. 
No data set with similar baseline lengths and position angles observed at different epochs is available. Therefore, no statement can be made on the interferometric variability. 

\subsubsection{Morphology}
\begin{figure*}[htbp]
\begin{center}
\resizebox{\hsize}{!}{
\includegraphics[width=\textwidth,angle=0,bb=52 583 407 716]{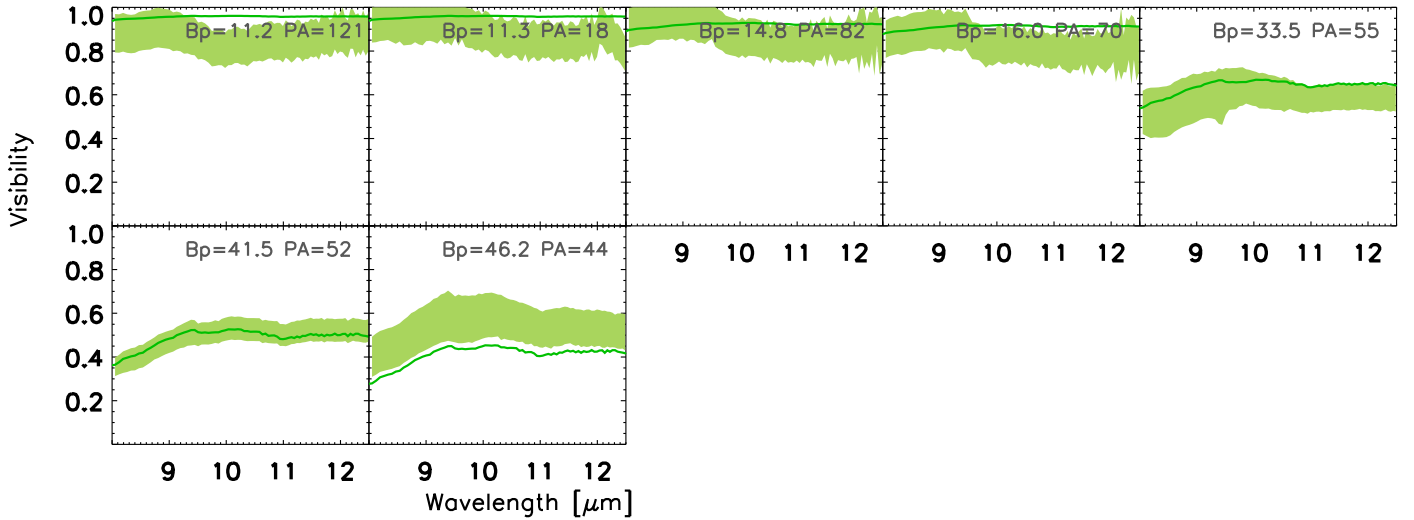}}
\caption{Best-fitting GEM-FIND model for the MIDI visibilities of T\,Mic.}
\label{tmic-vis-model}
\end{center}
\end{figure*}

Calibrated visibilities are shown in Fig.\,\ref{tmic-vis-model}. 

T\,Mic does not show a strong silicate feature. However, the slight drop that is observed in the visibility of baselines shorter than 16\,m at $\sim$\,9.8\,$\mu$m can be attributed to silicate dust. No differential phase signature is detected in T\,Mic, i.e.\,no asymmetries are detected with MIDI. 

The fitting of the LP data of 2011 with GEM-FIND reveals that T\,Mic can be well described with a circular UD model. The $\chi^2_{\rm{red}}$ of the GEM-FIND fitting for the different models is given in the first row in Table\,\ref{tab:gem-find}.  

If the LP data are combined with the archive data, the best-fitting model is a circular Gaussian model. 
The fit, however, does not describe the short baseline and 46 m baseline data well. 
This could mean that the close environment of T\,Mic is more complex and cannot be described with a one-component or two-component geometric models. The other possibility is that the short-baseline data (observed in 2011) cannot be combined with the longer baseline data (2004 and 2010) because of a variability effect. 

\subsection{RT~Vir}
RT\,Vir is a semi-regular variable star that is located at a distance $d=136$\,pc \citep{vanleeuwen2007}. RT\,Vir is a well-studied and, one of the brightest, water maser sources \citep{richards11}. Several mass-loss estimates are given, lying between $1.1\times10^{-7}\,M_\odot\,$yr$^{-1}$ \citep{knapp1998} and $5\times10^{-7}\,M_\odot$\,yr$^{-1}$ \citep{olofsson02}. 
\cite{sacuto2013} recently reported an asymmetry revealed through the MIDI differential
phase at 9 stellar radii (12 AU), and also spectroscopic cycle-to-cycle variability.
The Herschel image is interpreted as being of `fermata' type \citep{cox2012}.\\
\\
RT\,Vir was observed in 2008, 2009, 2011, and 2012. For our programme, we obtained six data sets, but only two turned out to be of good quality. The archive 
data, on the other hand, are of good quality. Twelve out of 13 data sets can be used.

\subsubsection{Variability}
The visual phase of the MIDI observations is derived using the light curve from ASAS \citep{pojmanski02}. The variability period for RT\,Vir is given as $P$=375\,d \citep{imai1997} and the phase-zero point is $T_0=2\,454\,854$~JD \citep[adopted from][]{sacuto2013}. 
The observations were carried out at different visual phases within different cycles.

The flux level of the ISO and MIDI spectra is not the same, i.e.\,the level of $N$-band emission has changed.
From Fig.~\ref{rtvirflux}, one can see that the flux is lowest at visual minimum.
{At a glance, it also seems that the flux follows a sinusoid with maximum flux shifted from the visual maximum. This kind of behaviour was already predicted by model atmosphere simulations \citep[][for near-infrared wavelengths]{ireland2004}, and observed by \cite{zhaogeisler2012}. 
Since these observations are carried at different cycles, the effect of cycle-to-cycle variation cannot be completely ruled out \citep{sacuto2013}. 

\begin{figure*}[htbp]
\begin{center}
\resizebox{\hsize}{!}{
\includegraphics[width=\textwidth,angle = 180, bb=48 86 769 311]{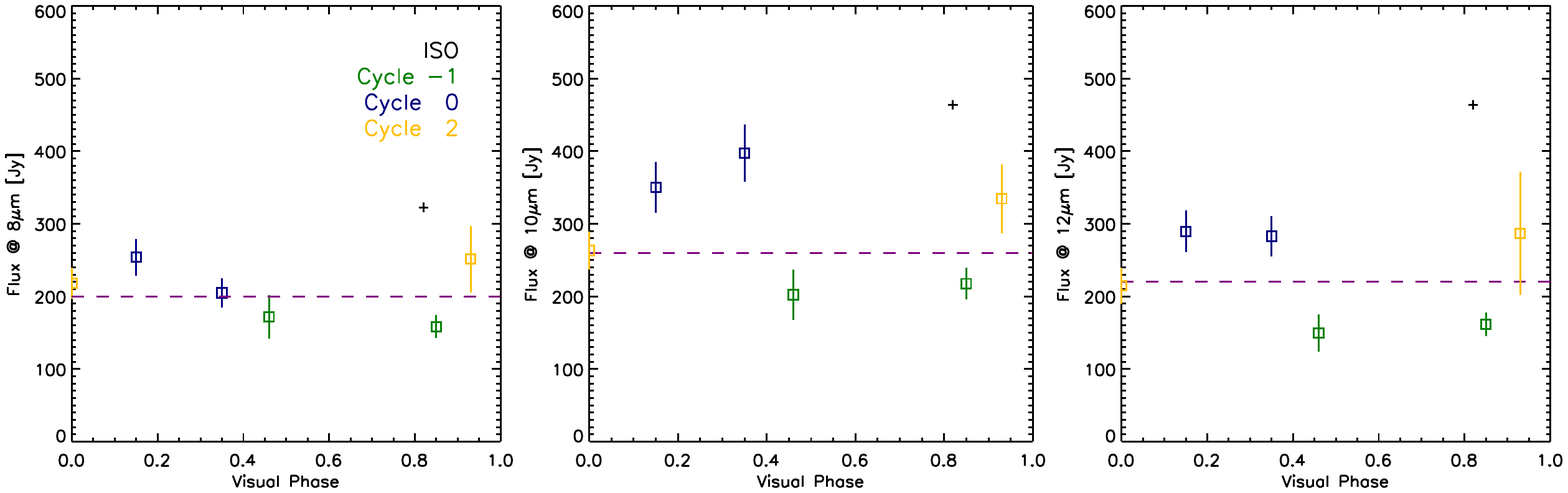}}
\caption{Visual phase vs. 8 (left), 10 (centre), and 12~$\mu$m (right)  flux for RT Vir. The colours refer to the cycles, while the cross indicates the
ISO flux. A cyan horizontal line was drawn to guide the eye and to make the sinusoidal-like variations clearer.}
\label{rtvirflux}
\end{center}
\end{figure*}

To check for interferometric variability, data sets with similar baseline lengths and position angles observed at different visual phases are shown in Fig.~\ref{rtvirvisvariab}. 
Although photometric variability is present, we do not report any interferometric variability, confirming the results of \cite{sacuto2013}.
It would be interesting to monitor the atmosphere with long baselines to check whether interferometric variability can be detected at high spatial frequencies, where the object also presentsan asymmetry (see following section).
As no interferometric variability is so far observed, all data sets can be combined in the further analysis.

\begin{figure}[!htbp]
\begin{center}
\resizebox{\hsize}{!}{
\includegraphics[width=\textwidth,angle = 180, bb=535 75 766 312]{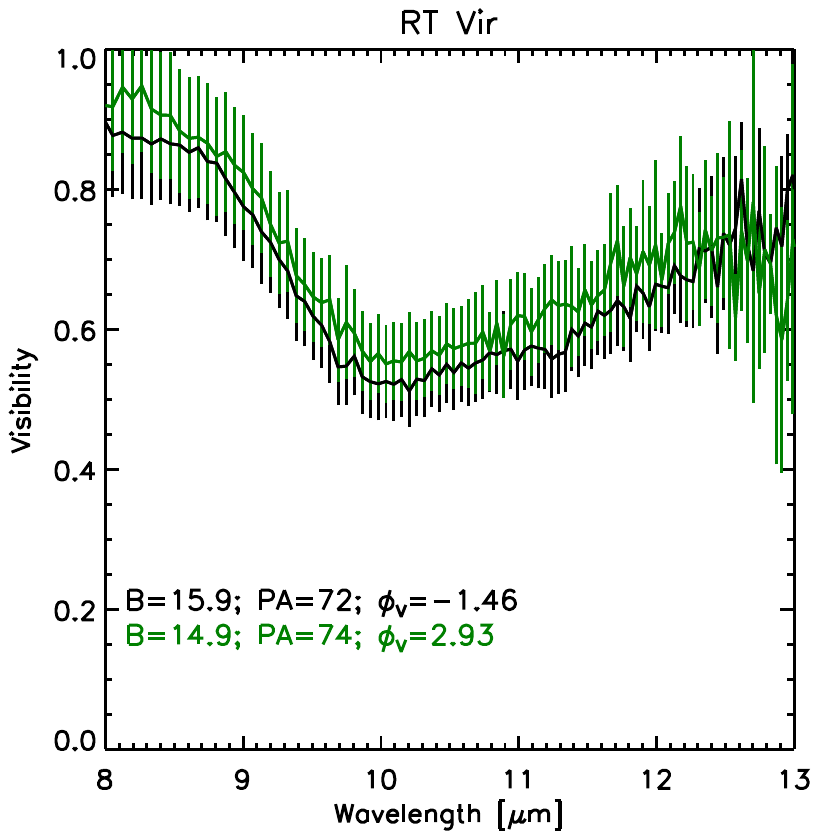}}
\caption{Interferometric variability check for RT~Vir.}
\label{rtvirvisvariab}
\end{center}
\end{figure}

\subsubsection{Morphology}
Calibrated visibilities are shown in Fig.~\ref{rtvir-vis-model}. The drop that is observed in the visibility can be attributed to amorphous silicate. For RT\,Vir, a non-zero differential phase is detected for the longest baselines (89 and 128\,m) by \cite{sacuto2013}. This is the signature of an asymmetry. \cite{sacuto2013} used GEM-FIND to fit different geometric models to a subset of the data. Not only one-component but also two- and three-component models are used in their work. The best-fitting model is a three-component model: a UD that describes the central star, a spherical Gaussian that represents the optically thin dust environment, and a Dirac function that represents the unresolved asymmetry (which could be a companion or a dust clump). This model is also able to reproduce the observed differential phase.
\begin{figure*}[htbp]
\begin{center}
\resizebox{\hsize}{!}{
\includegraphics[width=\textwidth,angle=0,bb=52 527 407 716]{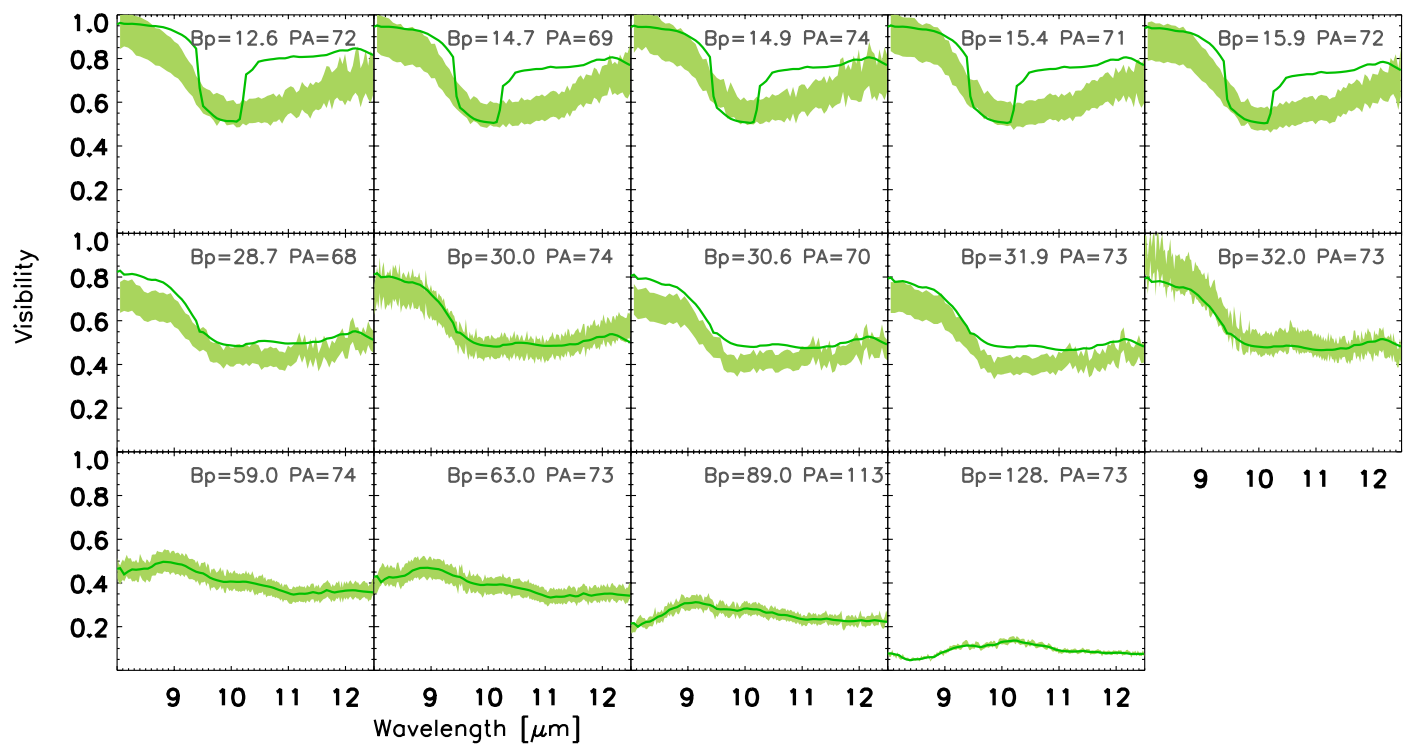}}
\caption{Best-fitting GEM-FIND model (solid line) for the MIDI visibilities of RT~Vir.}
\label{rtvir-vis-model}
\end{center}
\end{figure*}

Although \cite{sacuto2013} find the environment of RT\,Vir to be asymmetric, we start by fitting spherically symmetric models to the LP data that do not show any asymmetry. The LP data can be fitted best with a circular Gaussian profile with a FWHM of $\sim$60\,mas at 10\,$\mu$m (see first row in Table\,\ref{tab:gem-find}). As only two observations with the same position angle are available, no attempt to fit 
elliptical models was made. If we combine the LP data with the archive data, a one-component model is no longer able to describe the observations (see second row in Table\,\ref{tab:gem-find}). Therefore, we fit a spherical model, composed of a circular UD and a circular Gaussian, to the data. The best-fitting diameter of the circular UD is 16\,mas. The diameter derived by \cite{richichi05} in the $K$ band is $\theta_K$=12.4\,mas. Our model does not fit the short-baseline data well enough. The model of \cite{sacuto2013}, on the other hand, fits all the visibility and differential phase observations (as the model is asymmetric). The diameter of their UD is wavelength dependent and larger than the $K$-band diameter determined by \cite{richichi05}, which suggests that the environment is composed of more than one silicate-rich layer \citep{sacuto2013}.

\subsection{$\pi^1$~Gru}
$\pi^1$~Gru is a well-studied S-type star close to the tip of the AGB \citep{jorissen1993,vaneck2000}.
The star is known to have a G0V companion located $\approx 2.7\arcsec$ ($\approx 450$\,AU) away from the primary,
and possibly a third component much closer as suggested by \cite{Makarov2005, Frankowski2007} and more recently, \cite{mayer2014}.
CO line observations \citep{sahai1992, knapp1999, chiu2006} reveal an asymmetric, double-peaked structure and extended emission wings, which are interpreted as an expanding disk and a fast bipolar outflow oriented
perpendicular to the disk. The Herschel/PACS image shows an elliptical emission and a hook east of the star at a distance of
$38\arcsec$ \citep[Fig.~5,][]{mayer2014}. These kinds of hooks or arcs is most probably part of
an Archimedean spiral formed by the interaction with a companion \citep[ e.g. $o$~Cet;][]{mayer2011}.
Therefore the presence of the companion(s) affects the geometry of the atmosphere at different spatial scales.
The star was observed with MIDI in 2008 and 2011. We have a total of 11 data sets to fit, 4 of which are from the LP. 

\subsubsection{Variability}
$\pi^1$~Gru is a SRb variable with period 198 d. 
However, as already stated in Sect.~2.2 of \cite{sacuto2008}, because of the semi-regular nature of the object, it is very difficult to assign a visual phase
to a given observation.  
\cite{sacuto2008} give two explanations for the discrepancy between the level of the ISO and the 2006 MIDI spectra: 
first, MIDI is missing some of the flux that is seen by the larger ISO beam; and, second, there is a difference in the emission due to pulsation. The difference we observe between the 2011 LP data, ISO, and the 2006 data could be due to cycle-to-cycle variability.

\subsubsection{Morphology}

Concerning the shape of the visibility versus wavelength, we observe the typical shape of silicate dust with a bump between 8 and 9~$\mu$m. A drop is also observed in the visibilities
at wavelengths longer than 12~$\mu$m. The lower left panel of Fig.~\ref{pi1gru-vis-model} shows a slightly different shape in the visibility with respect to the other panels. 
This may be a residual of a non-perfect calibration \citep{sacuto2008}. 
A fit including only the LP data points to an elliptical UD morphology with axis ratio 0.2 and inclination $140\degr$.
The visibilities observed for $\pi^1$ Gru can be well fitted with a composite model UD+Gaussian, as shown in Fig.~\ref{pi1gru-vis-model}.

\begin{figure*}[!htbp]
\begin{center}
\resizebox{\hsize}{!}{
\includegraphics[width=\textwidth,angle=0,bb=52 527 407 716]{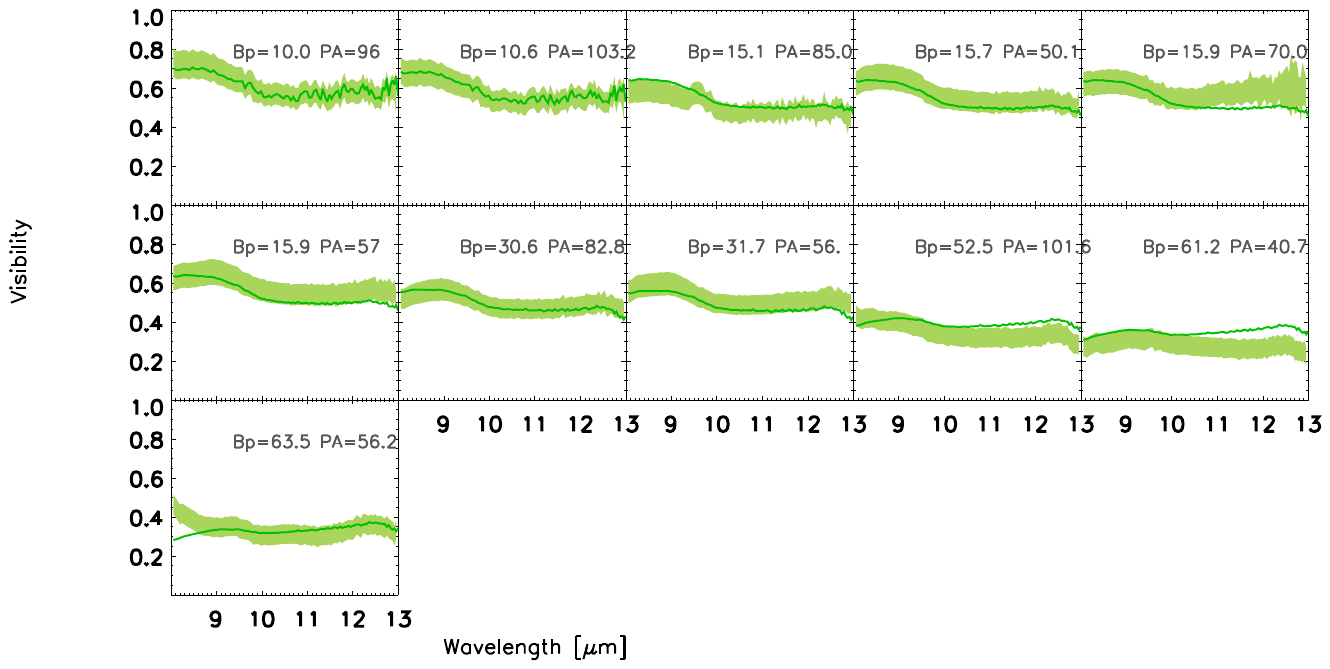}}
\caption{Best-fitting GEM-FIND model (solid line) for the MIDI visibilities of $\pi^1$ Gru.}
\label{pi1gru-vis-model}
\end{center}
\end{figure*}


\subsection{omi~Ori}
omi~Ori is one of the two S-type stars of our sample.
According to the recent parameter determination by \cite{cruzalebes2013b}, the star has a $K-$band angular diameter of 9.78 mas. 
\cite{cruzalebes2013} reports asymmetric structures detected (via a closure-phase signature) in the near-infrared. 
The star is known to have a white dwarf companion \citep{ake1988},
but the separation between the two stars, and more generally the orbit of the system, are not known.
The star was classified among the irregular morphologies in the Herschel/PACS images \citep{cox2012}.
omi~Ori was observed by MIDI in 2005 and 2011, and we collected a total of 14 data sets, 7 of which are archive data.

\subsubsection{Variability}
Classified as SRb, omi~Ori has a period of 30 days. We collected a light curve from ASAS, but the data are limited to 2010,
therefore we could not estimate the visual phase for the MIDI data of the LP.
A comparison of the IRAS spectrum with the MIDI spectrum from 2005 shows no evidence of variability.
The data of the LP were collected within a few days, therefore we do not need to worry about
interferometric intra-cycle variability. As there is no overlap between the baselines and position angles of the archive data and those from the LP,
it is not possible to issue any statement concerning cycle-to-cycle interferometric variability. As a consequence, one should consider
the results from the $\chi^2_{\rm{red}}$ combining all the data only as indicative.

\subsubsection{Morphology}
The visibility curve of omi~Ori does not show any sign of dust when plotted versus wavelength.
The feature observed around 9.58~$\mu$m is due to telluric ozone. 
The LP data sample the upper part of the visibility curve. Therefore a fit on these data 
cannot really distinguish between Gaussian and UD profiles.
As a confirmation, the $\chi^2_{\rm{red}}$ of the two models are very close to each other. 
GEM-FIND fits of the LP data point to an elliptical UD model with inclination $55\degr$ and axis ratio 0.4.
The $\chi^2_{\rm{red}}$ obtained by fitting all the data (archive+LP) points towards a composite model (UD+Gauss)
with a very extended Gaussian (FWHM $> 220$ mas) enshrouding the UD (see also Fig.~\ref{omiori-vis-model}). The flux ratio is $> 50$, meaning that the central source is the dominant contributor.

\begin{figure*}[!htbp]
\begin{center}
\resizebox{\hsize}{!}{
\includegraphics[width=\textwidth,angle=0,bb=52 527 410 717]{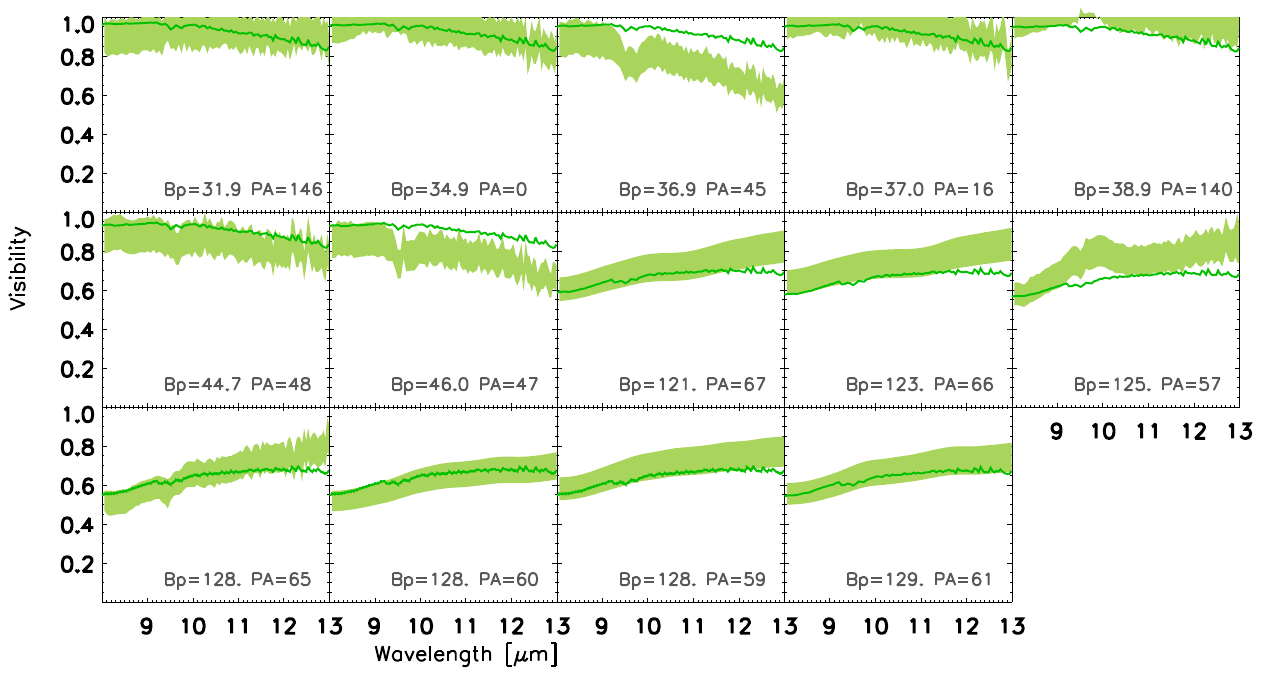}}
\caption{Best-fitting GEM-FIND model for the MIDI visibilities of omi~Ori.}
\label{omiori-vis-model}
\end{center}
\end{figure*}

\subsection{U~Ant}\label{uant.sect}
U~Ant is a nearby N-type carbon star. \cite{Knapp2003} reported a distance of 260~pc.
\cite{Bergeat2005} estimated its parameters as follows: $T_{\rm{eff}}=2810K$, the $C/O=1.44$, $\dot{M} = 2.0\times10^{-6}M_{\sun}$~yr$^{-1}$. 
More recently, \cite{mcdonald2012} estimated a much hotter temperature (3317~K) through spectral energy distribution (SED) fitting.
Five spherical detached shells were detected at long wavelength and large scales $\sim 25\arcsec, 37\arcsec, 43\arcsec, 50\arcsec, 3\arcmin$
\citep{Izumiura1996, Olofsson1996, Gonzales2001, Gonzales2003, Maercker2010}.
Herschel-PACS imaged in the FIR a shell with a distance from
the star of 42'' \citep{kerschbaum2010a}. 

The star was observed with MIDI in 2008 and 2012. For our programme, we obtained five observations, but only one of them
turned out to be of any use. The reason for this is mostly poor seeing conditions during the nights of observation.
Most of the archive data are also of mediocre quality. Only two points out of five from the archive is hereafter used.


\subsubsection{Variability}
Since U~Ant is an irregular variable of type Lb, it is not possible to determine at which visual phase the observations were taken.
Moreover, there are no $V$-band measurements available at the time of the MIDI observations. Therefore, a study of the interferometric variability
is not possible for this star.
We retrieved the IRAS spectrum from the archive and compared it with the MIDI spectrum 
in Fig.\ref{spec.fig}. 
The star is classified as SiC+ by \cite{sloan1998}. This class of objects is characterised by a spectrum with a weak 8.5-9~$\mu$m feature,
 a weak dust continuum, and a weak SiC feature that is observed in our spectra. The MIDI spectrum is within the error bars of the 
the IRAS spectrum.

\subsubsection{Morphology}

Three visibility points (Fig.~\ref{uant-vis}) are definitely too few to have an idea about the morphology of the star, especially if one considers that
the data are taken four years apart. Therefore, for this star, we can only derive sizes and study
the molecular and dust components. The diameter of the best-fitting FWHM varies between 7 and 10~mas as a function of wavelength.

\begin{figure*}[htbp]
\begin{center}
\resizebox{\hsize}{!}{
\includegraphics[width=\textwidth,angle=90,bb=446 436 558 739]{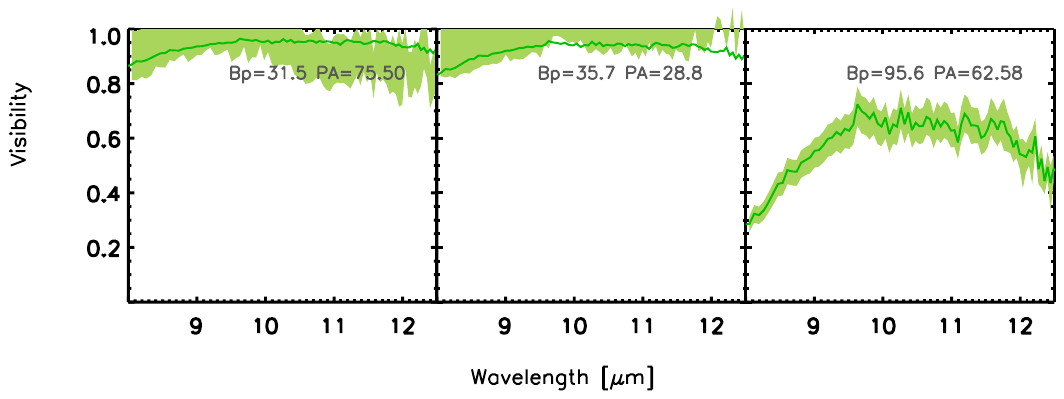}}
\caption{Best-fitting Gaussian profile (solid-line) for the MIDI visibilities of U~Ant.}
\label{uant-vis}
\end{center}
\end{figure*}

Even though both MIDI and IRAS spectra exhibit a weak SiC feature, the visibilities are typical of the
carbon stars without SiC (Fig.~\ref{fig:shape}). Indeed the typical visibility drop around 11.3~$\mu$m is not observed here.
A small decrease of the visibility is observed only for one of the observations with the 30~m baseline. 
On the other hand, the high level of visibility and, consequently, its large associated uncertainty, do not allow us to infer whether or not that decrease is real and due to SiC. 


\subsection{R~Lep}

R~Lep is one of the closest carbon-rich Miras in the southern hemisphere
showing intermediate mass-loss rates and very red colours \citep[][$J-K = 2.23$,]{whitelock2006}.
\cite{vanbelle1997} reported a $K$-band diameter of 11.50 mas, while \cite{chagnon02} measured 37.10 mas in the $L^{\prime}$ band.
Signatures of asymmetric structures were observed by \cite{ragland2006} with the IOTA interferometer in the near-IR.
The object is classified as non-detection by \cite{cox2012}. Nevertheless, the authors predict the presence 
of a bow shock at a distance < 1 pc from the stellar envelope.
The star was observed with MIDI in 2010, 2011, and 2012.

\subsubsection{Variability}
The MIDI spectra of R~Lep, shown in Fig.~\ref{spec.fig}, are very noisy. The flux also changes a lot within the MIDI data and 
between MIDI and IRAS. Excluding the data point from cycle 0 (yellow square in Fig.~\ref{flux-phase-rlep}), which has an unrealistic low flux, the variations within
the MIDI data are of the order of 0.2 magnitude. This is a reasonable value if compared with the variations\footnote{\cite{lebertre1992} observed a decrease in the amplitude of variability
of Mira stars towards the long wavelengths. There are no $N-$band measurements reported for R~Lep, but the variability in the $M$ band is 0.48 mag, i.e. the variability in the $N$ band must be smaller
than this value.} $\Delta m_N<0.48$~mag predicted by \cite{lebertre1992}.
The flux variation between the MIDI data from cycle 2 and the IRAS spectrum is of the order of 0.7 mag. 
Nevertheless, the MIDI spectra should be considered with caution. 
Regarding interferometric variability, the left panel of Fig.~\ref{rlep-interfvariab.fig} shows data sets that are too close in time to detect any intra-cycle variability.
On the other hand, in the right panel of Fig.~\ref{rlep-interfvariab.fig} there is a hint for a cycle-to-cycle variability: the visibility from cycle 0 is systematically lower than the others in the range where molecular opacities are at play ($8-10~\mu$m). Since the variability effect is small, the data are still combined together for the GEM-FIND fit.

\begin{figure*}[htbp]
\begin{center}
\resizebox{\hsize}{!}{
\includegraphics[width=\textwidth,angle=180,bb=48 86 769 311]{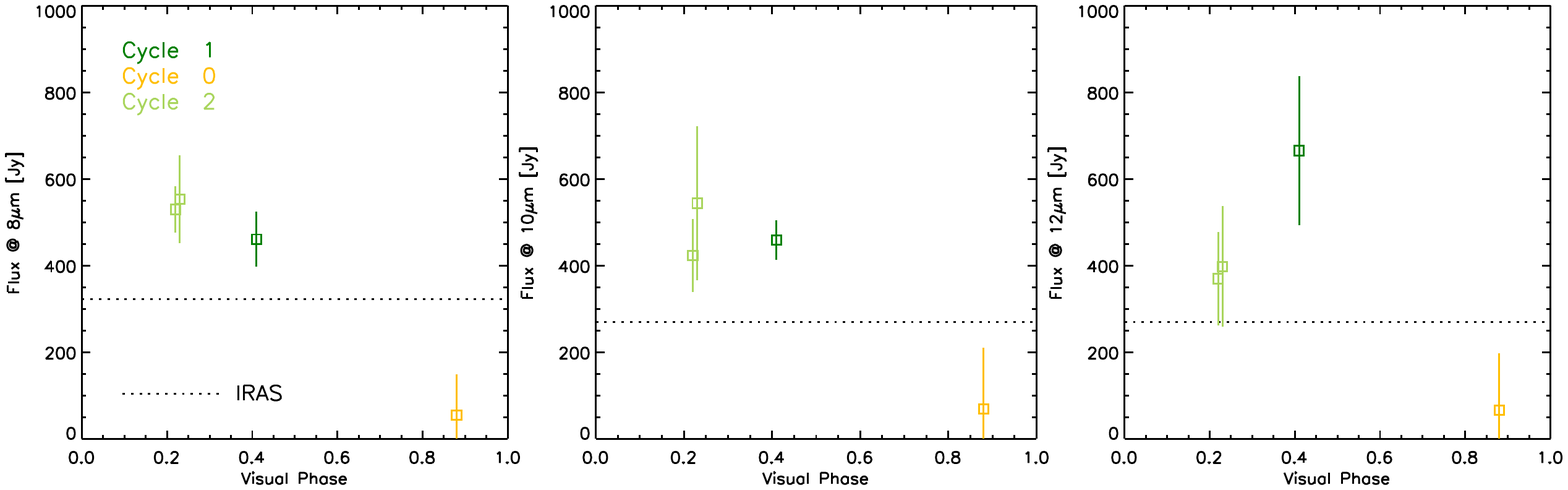}}
\caption{MIDI fluxes at 8, 10, and 12~$\mu$m for R~Lep. The dotted line represents the IRAS flux.}
\label{flux-phase-rlep}
\end{center}
\end{figure*}

\begin{figure*}[htbp]
\begin{center}
\resizebox{\hsize}{!}{
\includegraphics[width=\textwidth,angle=180,bb=535 74 766 313]{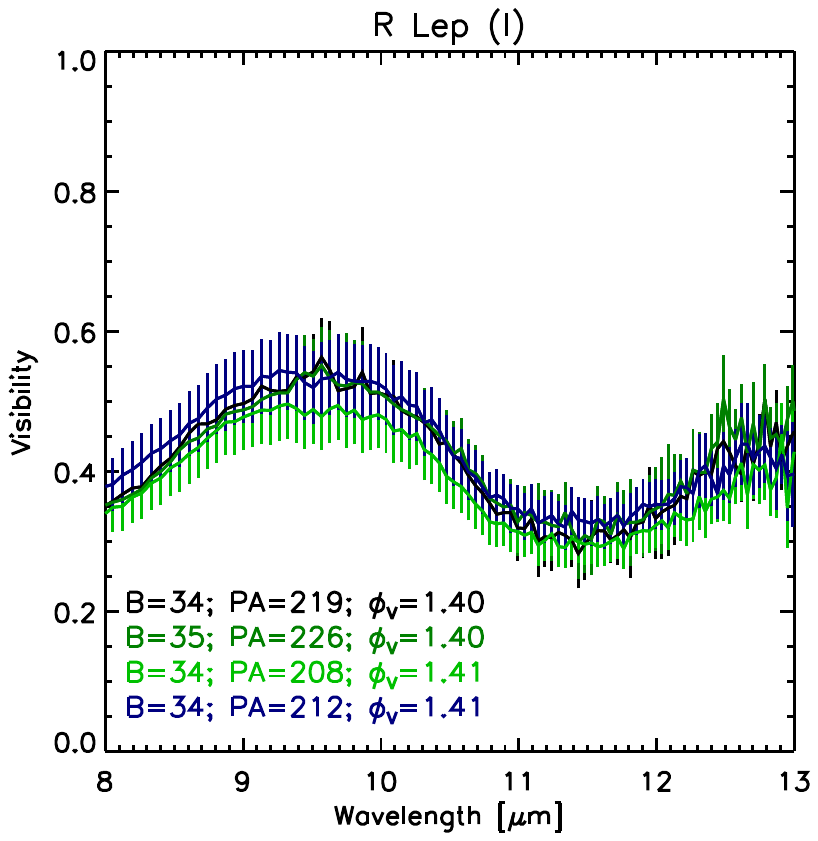}
\includegraphics[width=\textwidth,angle=180,bb=535 74 766 313]{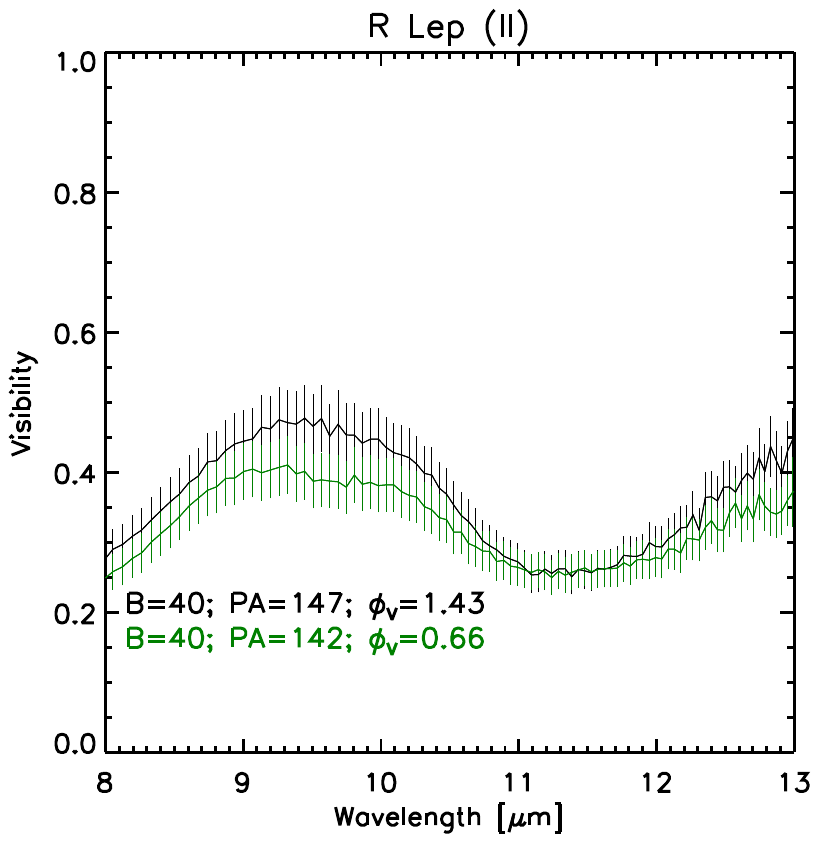}}
\caption{Interferometric variability for R~Lep.}
\label{rlep-interfvariab.fig}
\end{center}
\end{figure*}

\subsubsection{Morphology}
The visibility curve of R~Lep (Fig.~\ref{rlepvis}) is typical for a carbon Mira with a SiC feature, and it is described in Sect.~\ref{par:results}.
By using only the LP data, we obtain a better $\chi^2_{\rm{red}}$ for the Gaussian model, whereas we have a good fit of the data with a two-component model by adding all the observations together
 (UD+Gaussian; see Fig.~\ref{rlepvis}). 
We do not observe any asymmetric structure that might be related to the asymmetries observed at other spatial scales and wavelengths \citep[i.e. near-infrared by][]{ragland2006}. 
The SiC depression is always present in the visibility curve, but we note that at long baselines and long wavelengths the visibility increases again. 
This might be explained by the fact that at those spatial frequencies, there is some extra molecular opacity appearing to make the object smaller. 
SiC is clearly present in the IRAS spectrum of the star, and 
since we are sampling the spatial frequencies at 2 stellar radii \citep[using as a reference the photospheric diameter given by][]{vanbelle1997}, we can say that SiC is already observed at 2 stellar radii.

\begin{figure*}[htbp]
\begin{center}
\resizebox{\hsize}{!}{
\includegraphics[width=\textwidth,angle=0,bb=58 583 411 716]{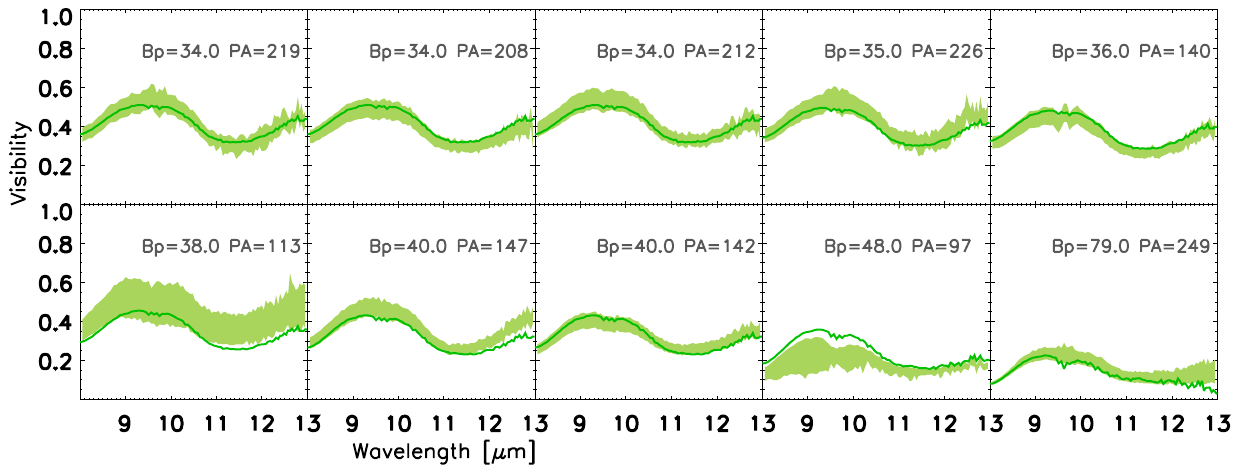}}
\caption{Best-fitting UD+Gaussian profile (solid line) for the MIDI visibilities of R~Lep.}
\label{rlepvis}
\end{center}
\end{figure*}
\subsection{Y~Pav}
Y~Pav is a semi-regular C-rich variable with period 232 days and distance 400 pc \citep{vanleeuwen2007}. 
The object is classified class X by \cite{cox2012}; no bow-shock nor detached shell were identified. 
\cite{sloan1998} classified the spectrum of this object as Br1 (Broad 1).
This class includes stars showing broad emission features extending from 8-9~$\mu$m to 12~$\mu$m, and
the authors are not able to identify the opacities contributing to these extended features. It is observed that the Br1 stars have similar colour distribution as the SiC class, but the central object
is cooler. 
According to the IRAS colour-colour diagram, this object is evolving towards a carbon Mira. 
Y~Pav was observed only in the framework of our programme in 2011, and we used four out of seven observations for the modelling ($u,v$-coverage shown in Fig.~\ref{uvcoverage.fig}). The data discarded suffered from poor weather conditions.

\subsubsection{Variability}
No interferometric variability can be assessed, as the data were taken at a single epoch. 
The difference between the IRAS and MIDI spectra is very likely due to calibration problems.
They are not mentioned further.

\subsubsection{Morphology}
The visibility versus wavelength plotted in Fig.~\ref{ypav.vis} shows a minimum in the visibility curve slightly shifted 
towards shorter wavelength than the SiC feature.
The difference is very small and can also be seen in Fig.~\ref{compypav:fig}.
It is not clear whether this difference is related to the Br1 spectral classification, mainly because this is the only Br1 of the sample. 
A detailed modelling of the visibilities and of the ISO spectrum
by means of model atmospheres clarifies what kind of opacities are contributing most (Rau et al., prep.). 

The visibility obtained are very high and can be well fitted by a Gaussian profile (Fig.~\ref{ypav.vis}). 

\begin{figure}[htbp]
\begin{center}
\resizebox{\hsize}{!}{
\includegraphics[width=\textwidth,angle=180,bb=74 66 699 528]{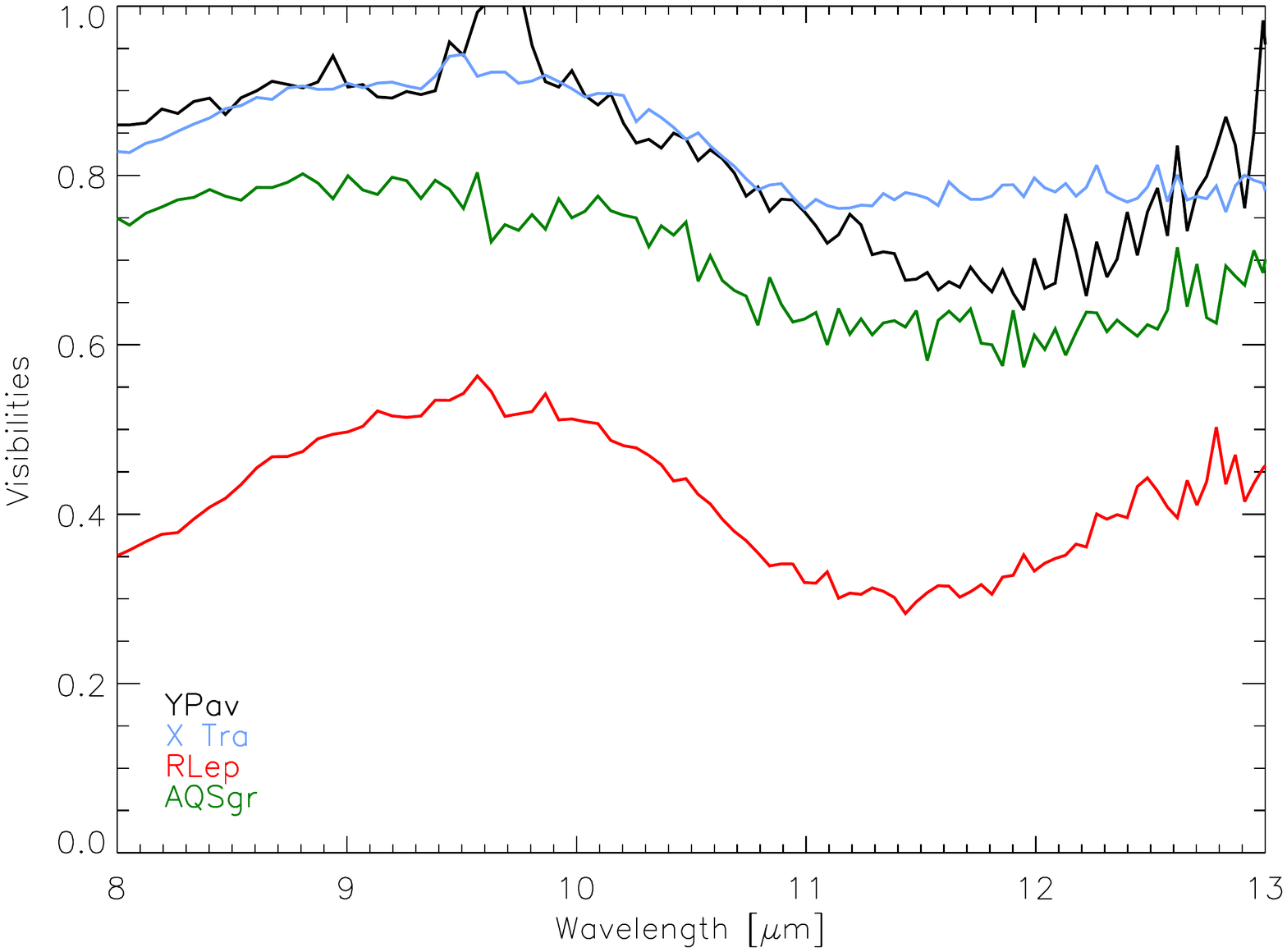}}
\caption{Comparison between the shape of the visibility of all the stars showing SiC.}
\label{compypav:fig}
\end{center}
\end{figure}

\begin{figure*}[htbp]
\begin{center}
\resizebox{\hsize}{!}{
\includegraphics[width=\textwidth,angle=0,bb=58 640 280 716]{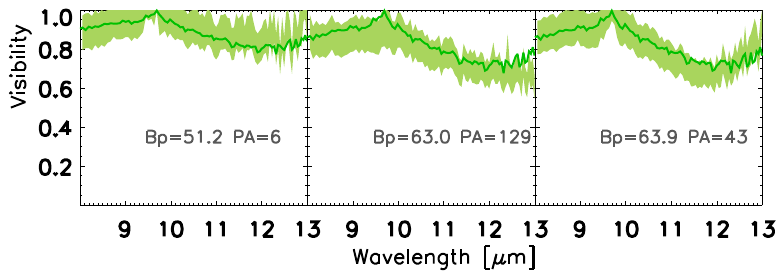}}
\caption{Best-fitting Gaussian profile (solid line) for the MIDI visibilities of Y~Pav.}
\label{ypav.vis}
\end{center}
\end{figure*}

\subsection{TX~Psc}
A detailed mid-infrared spectro-interferometric study of TX~Psc can be found in \cite{klotz2013}. 
In this section, we present the new visibility
data from the LP and discuss the morphology of the source.
As the spectral type of the calibrators used for the LP data are
MI and MIII, the MIDI spectra were not flux calibrated (see details in Sect.~\ref{par:datareduction}), and therefore
we do not discuss spectroscopic variability.

\subsubsection{Morphology}
The model fitting best both the archive and LP data is a UD with size between 12 and 9 mas. The star is larger
in the $8-9~\mu$m region where C$_2$H$_2$ and HCN opacities contribute, following the same 
wavelength dependency plotted in Fig.~3 (right panel) of \cite{klotz2013}.
We confirm that no asymmetric structures are observed, 
and that the object can be described by a single component. This result is in line with the finding of \cite{klotz2013},
however asymmetric structures are detected much further than 10~R$_{\star}$ by several authors \citep[see][and references therein]{hron2015}.
The object is classified in class N (naked) by \cite{sloan1998}. This class includes stars that do not exhibit dust
excess in the ISO spectrum. We confirm that the visibilities do not show signs of dust (Fig.~\ref{txpsc-uv}).

\begin{figure*}[htbp]
\begin{center}
\resizebox{\hsize}{!}{
\includegraphics[width=\textwidth,angle=0,bb=52 527 411 716]{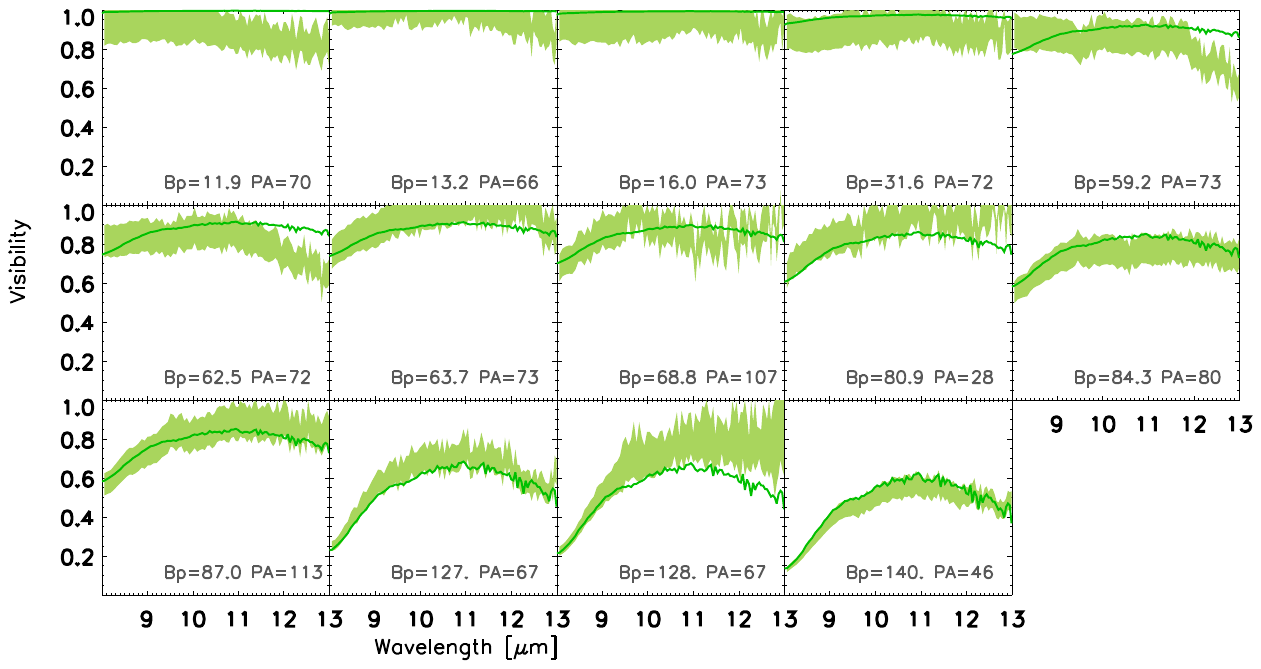}}
\caption{Best-fitting GEM-FIND model (solid line) for the MIDI visibilities of TX~Psc.}
\label{txpsc-uv}
\end{center}
\end{figure*}

\subsection{S~Sct}\label{ssct.sect}
S~Sct is a carbon-rich semi-regular pulsator with a variability amplitude of 1.5 mag in the $V $ band
and an uncertain distance \citep[between 367 pc and 580 pc according to][respectively]{vanleeuwen2007,Bergeat2005}.
The temperature of the star is 2755 K according to \cite{Bergeat2005} and 2895 K following \cite{lambert1986}. 
The mass-loss rate reported in literature is of the order of $10^{-6}~M_{\sun} $~yr$^{-1}$ (see Table~\ref{table:1}).
The star is one of the targets surrounded by a detached shell \citep{olofsson1992} that formed $\sim10^4$~yrs ago because of a superwind episode with a mass-loss rate of $\sim10^4~M_{\sun}$~yr$^{-1}$. 
This shell is also observed in the Herschel/PACS images
and was recently modelled by \cite{mecina2014}. The inner spatial scales probed by VLTI/PIONIER \citep[$H-$band 4-beam combiner interferometer;][]{lebouquin2009}
can be well fitted by a UD of diameter $6.22\pm 0.3$~mas  ($\chi^2 = 1.4$).
PIONIER did not detect any asymmetric structure (Le Bouquin, private communication). 

MIDI observed this object seven times between the 2011 May 28 and 30. 
For the modelling and data interpretation, we use
only two data sets, keeping in mind that the data set of May 30 is very good, while the data of May 28 are very noisy after 11.5~$\mu$m.
The other five observations were affected by poor weather conditions.

\subsubsection{Variability}
The MIDI spectrum agrees with the flux level of the ISO observations. However, ISO shows a small bump around 11.3~$\mu$m attributable to SiC,
while nothing similar is seen by MIDI.
Although we cannot exclude that the shape of the MIDI spectrum is still affected by calibration problems, it is an
interesting coincidence that neither is SiC observed in the visibilities (see following Section), nor that the shape of the MIDI spectrum resembles that from IRAS
(Fig.~\ref{ssct-spec}). 
Like TX~Psc, S~Sct is also classified as N (naked) in the \cite{sloan1998} classification based on
IRAS spectra.
This could mean that the amount of SiC was too small at the time of the IRAS observations, and that it increased afterwards.
But given the shape of the MIDI spectrum, one presumes that the feature recently disappeared again (Fig.~\ref{ssct-spec}).
A more likely possibility would be a photospheric variation, i.e. no change in the SiC, but rather in the molecular opacity at wavelengths $>10~\mu$m.

\begin{figure}[htbp]
  \begin{center}
    \resizebox{\hsize}{!}{
      \includegraphics[width=\textwidth,angle=0,bb=86 373 539 691]{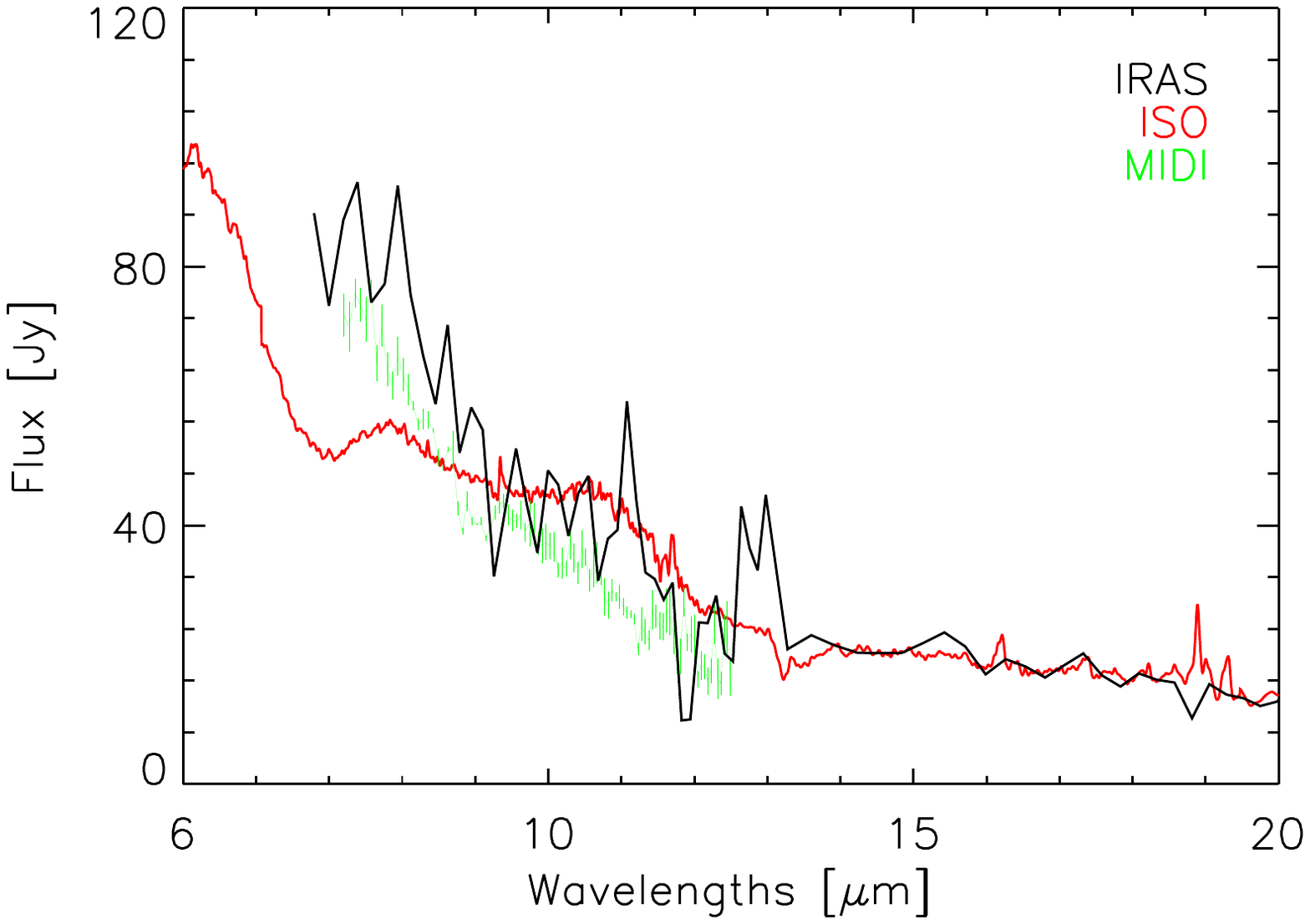}}
    \caption{Comparison of MIDI spectrum of S~Sct with those from IRAS and ISO. While the flux levels are very similar, the
MIDI and IRAS spectra do not show any signature of SiC, despite being clearly visible in the ISO spectrum.}
    \label{ssct-spec}
  \end{center}
\end{figure}

\subsubsection{Morphology}
The two visibility points shown in Fig.~\ref{ssct-vis} are easily fitted with circular models and
the fit shows a slight preference for the Gaussian shape. Nevertheless, the robustness of that conclusion should not be overestimated  
because of the high visibilities and small amount of data available. 
As in the case of U~Ant, we stress that the visibility of S~Sct does not show any signature of SiC.
By comparing the $N$-band UD diameter with the $H$-band UD diameter measured by PIONIER, we conclude that molecular and dust (amorphous carbon)
material enshrouds the object in the $N$ band. No SiC is detected either in the correlated or uncorrelated flux.
\begin{figure}[htbp]
  \begin{center}
    \resizebox{\hsize}{!}{
      \includegraphics[width=\textwidth,angle=0,bb=39 640 215 716]{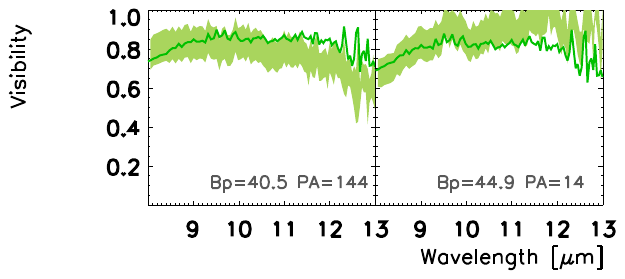}}
    \caption{Best-fitting GEM-FIND model for the MIDI visibilities of S~Sct.}
    \label{ssct-vis}
  \end{center}
\end{figure}

\subsection{AQ~Sgr}

AQ~Sgr is a carbon-enriched semi-regular variable. The period
of variability is 190~d and the distance is 330~pc \citep{vanleeuwen2007}.
The mass-loss estimates vary between $7.7$x$10^{-7}~M_\sun~$yr$^{-1}$ \citep{Bergeat2005}, and 10$^{-7}~M_\sun$ yr$^{-1}$ \citep{ramstedt2014}. 
The star is classified as fermata by \cite{cox2012}.
\cite{richichi05} report a K-band diameter of 6.13 mas. The star was already observed with MIDI in 2008,
but those data are very noisy and they were discarded during data reduction.
Only two points from the LP were used for the analysis. The archive data are corrupted by weather.

\subsubsection{Variability}
The MIDI spectrum of AQ~Sgr is compared with the spectrum from IRAS in Fig.~\ref{spec.fig}, and they fit within the uncertainties of IRAS.
The shape of the spectra is rather similar, and we observe a bump around 11.3~$\mu$m due to SiC.
The star was classified as SiC+ by \cite{sloan1998} and \cite{gupta2004}.
No interferometric variability can be determined from the available data set.

\subsubsection{Morphology}

The visibility curve presented in Fig.~\ref{aqsgr-vis} matches the shape of the spectrum, showing the SiC signature.
The observations are well fitted by a UD profile with a diameter between 17 and 33 mas (depending on wavelength). 
The level of visibility at 11.3~$\mu$m is comparable to that at 12~$\mu$m, where the UD diameter measures 32 mas.
By comparing the latter value with the $K$-band diameter,
we can conclude that SiC is being detected at 5~R$_{\star}$.

\begin{figure}[htbp]
  \begin{center}
    \resizebox{\hsize}{!}{
      \includegraphics[width=\textwidth,angle=0,bb=39 640 215 717]{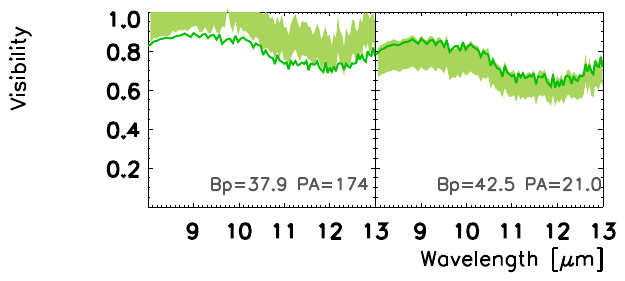}}
    \caption{Best-fitting GEM-FIND model (solid line) for the MIDI visibilities of AQ~Sgr.}
    \label{aqsgr-vis}
  \end{center}
\end{figure}

\subsection{X~TrA} 
X~TrA is a carbon-rich irregular variable.
A detailed near-infrared spectroscopic study with line identification was presented for this object by \cite{lebzelter2012}.
The star is a single object, as no indication of binarity was reported so far.
\cite{izumiura1995} detected a detached shell at a distance of 1.3$^{\prime}$.
In \cite{cox2012} the star is classified among the rings, but the ring detected by Herschel is faint with 
only a bright arc to the east. These authors conclude that more observations are needed.
X~TrA was observed for the first time with MIDI within the frame of our LP. Four points out of seven
are used for the geometric modelling. The other points were affected by weather conditions.

\subsubsection{Variability}
Because of the irregular nature of the light curve, 
time variability cannot be studied in detail for this star. However a comparison between
the MIDI and IRAS spectra (Fig.~\ref{spec.fig}) shows that the flux level is unchanged and SiC is observed.
The object is classified as SiC+ in the \cite{sloan1998} classification.
The presence of SiC is already detected in the IRAS spectrum shown in Fig.~\ref{spec.fig} and in the MIDI spectra as well.
\subsubsection{Morphology}
The X~Tra MIDI data can be reproduced with a UD of size 22 - 39 mas. The $\chi^2_{\rm{red}}$ of the
Gaussian (Table~\ref{tab:gem-find}) is similar for the already mentioned reason: the data sample the upper part of the visibility curve, where
both Gaussian and UD profiles are very similar. The visibility curve versus wavelength shown in Fig.~\ref{xtra-vis}
exhibits the shape typical of stars with SiC. The level of visibility at 11.3~$\mu$m is comparable to the level of visibility at 12 $\mu$m, as do the diameters.
Therefore we can conclude that for X~Tra, SiC is detected at $\sim$4~R$_{\star}$.

\begin{figure*}[htbp]
  \begin{center}
    \resizebox{\hsize}{!}{
      \includegraphics[width=\textwidth,angle=0,bb=58 640 411 716]{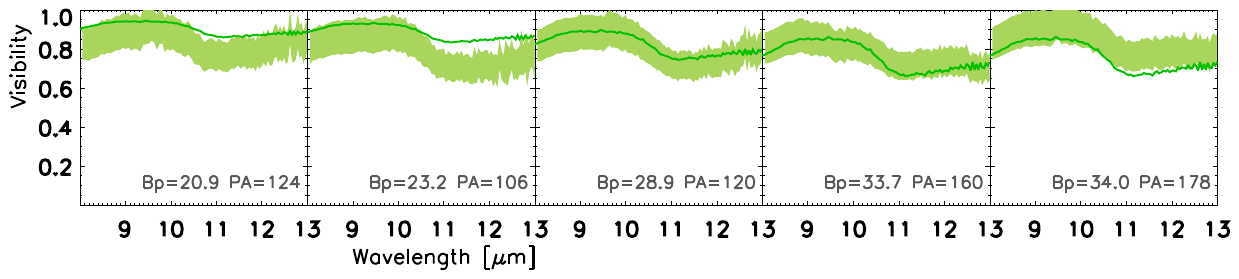}}
    \caption{Best-fitting GEM-FIND model (solid line) for the MIDI visibilities of X~Tra.}
    \label{xtra-vis}
  \end{center}
\end{figure*}

\newpage

\section{Journal of observations}
\label{journal.appendix}
This section presents the journal of observations for the LP dataset and the archive data used in this work.
Where calibrated fluxes are available, points are marked with `F'. 
Calibrators used to calibrate the visibilities are given below the science observation. If no calibrator is listed, the science point was not 
used in the astrophysical interpretation.

\begin{small}

\begin{table*}[]
\caption{\label{journal-tetaps}Journal of the MIDI Auxiliary Telescopes observations of $\theta$\,Aps.}
\centering
\vspace{0.5cm}
\begin{tabular}{llllllllll}
\hline
\hline
Target & UT date \& time &      $\phi_V$& Config. & $B_p$ & PA & Seeing & Airmass & Mode \\
& & && [m] & [$^\circ$]&['']\\
\hline
\textbf{LP data}\\
\hline
$\theta$ Aps    &       2011-06-11 00:28        &       9.27    &       A1-B2         &       10.4    &       92              &       1.01    &       1.651   &       SCIPHOT/PRISM   \\
HD167618        &       2011-06-11 06:24        &       \ldots          &         \ldots  &       \ldots  &       \ldots  &       0.69    &       1.036   &       SCIPHOT/PRISM   \\ 
HD167618        &       2011-06-11 08:20        &       \ldots          &         \ldots  &       \ldots  &       \ldots  &       0.63    &       1.101   &       SCIPHOT/PRISM   \\ 
\hline
$\theta$ Aps    &       2011-06-11 01:21        &       9.27    &       A1-C1         &       15.5    &       76              &       1.66    &       1.630         &       SCIPHOT/PRISM   \\
HD167618        &       2011-06-11 07:23        &       \ldots          &         \ldots  &       \ldots  &       \ldots  &       0.67    &       1.231   &       SCIPHOT/PRISM   \\
\hline
\hline
\end{tabular}
\end{table*}

\begin{table*}[]
\caption{\label{journal-rcrt}Journal of the MIDI Auxiliary Telescopes observations of R\,Crt.}
\centering
\vspace{0.5cm}
\begin{tabular}{llllllllll}
\hline
\hline
Target & UT date \& time &      $\phi_V$& Config. & $B_p$ & PA & Seeing & Airmass & Mode \\
& & && [m] & [$^\circ$]&['']\\
\hline
\textbf{LP data}\\
\hline
R Crt$^\mathrm{F}$      &       2012-01-24 04:57        &       6.66    &       C1-A1         &       13.1    &       52              &       1.60    &       1.242   &       SCIPHOT/PRISM         \\
HD81797         &       2012-01-24 05:16        &       \ldots          &       \ldots  &       \ldots  &       \ldots         &       1.08    &       1.057   &       SCIPHOT/PRISM   \\
HD123139        &       2012-01-24 07:53        &       \ldots          &       \ldots  &       \ldots  &       \ldots         &       1.16    &       1.252   &       SCIPHOT/PRISM   \\
HD112142        &       2012-01-24 08:25        &       \ldots          &       \ldots  &       \ldots  &       \ldots         &       0.81    &       1.069   &       SCIPHOT/PRISM   \\
HD123139        &       2012-01-24 08:59        &       \ldots          &       \ldots  &       \ldots  &       \ldots         &       1.05    &       1.097   &       SCIPHOT/PRISM   \\
\hline
\hline
\end{tabular}
\end{table*}

\begin{table*}[]
\caption{\label{journal-rleo}Journal of the MIDI Auxiliary Telescopes observations of R\,Leo.}
\centering
\vspace{0.5cm}
\begin{tabular}{llllllllll}
\hline
\hline
Target & UT date \& time &      $\phi_V$& Config. & $B_p$ & PA & Seeing & Airmass & Mode \\
& & && [m] & [$^\circ$]&['']\\
\hline
\textbf{LP data}\\
\hline
R Leo                   &       2012-01-24 06:18        &7.78&  C1-A1   &       15.7         &       75              &       0.77    &       1.235   &       SCIPHOT/PRISM   \\
\hline
R Leo$^\mathrm{F}$      &       2012-01-25 07:18        & 7.78& C1-A1   &       16.0         &       72              &       0.75    &       1.289   &       SCIPHOT/PRISM   \\
HD81797                 &       2012-01-25 05:57        &\ldots&        \ldots  &       \ldots  &       \ldots         &       0.71    &       1.040   &       SCIPHOT/PRISM   \\
HD81797                 &       2012-01-25 06:38        &\ldots&        \ldots  &       \ldots  &       \ldots         &       0.69    &       1.057   &       SCIPHOT/PRISM   \\
HD81797                 &       2012-01-25 07:01        &\ldots&        \ldots  &       \ldots  &       \ldots         &       0.67    &       1.083   &       SCIPHOT/PRISM   \\
\hline
\hline
\end{tabular}
\end{table*}

\begin{table*}[]
\caption{\label{journal-rleo}Journal of R\,Leo continued.}
\centering
\vspace{0.5cm}
\begin{tabular}{llllllllll}
\hline
\hline
Target & UT date \& time &      $\phi$ &Config. & $B_p$ & PA & Seeing & Airmass & Mode \\
& & && [m] & [$^\circ$]&['']\\
\hline
R Leo$^{\mathrm{F}}$    &       2006-12-19 06:01        &1.77&  K0-G0   &         42.5    &       77    &   0.39  &       1.63            &       SCIPHOT/PRISM   \\ 
$\alpha$Hya             &       2006-12-19 05:49        &\ldots&        \ldots  &       \ldots  &       \ldots         &   0.50        &       1.28            &       SCIPHOT/PRISM   \\
$\alpha$Hya             &       2006-12-19 07:09        &\ldots&        \ldots  &       \ldots  &       \ldots         &   0.65        &       1.08            &       SCIPHOT/PRISM   \\
$\alpha$Hya             &       2006-12-19 07:46        &\ldots&        \ldots  &       \ldots  &       \ldots         &   0.68        &       1.049   &       SCIPHOT/PRISM   \\
\hline  
R Leo$^{\mathrm{F}}$    &       2006-12-21 07:07        &1.78&  H0-G0   &         27.5    &       77              &       0.91    &       1.33            &         SCIPHOT/PRISM   \\   
$\alpha$Hya             &       2006-12-21 07:26        &\ldots&        \ldots  &       \ldots  &       \ldots         &   1.20                &       1.057           &       SCIPHOT/PRISM   \\
\hline
\hline
\end{tabular}
\end{table*}

\begin{table*}[]
\caption{\label{journal-tmic}Journal of the MIDI Auxiliary Telescopes observations of T\,Mic.}
\centering
\vspace{0.5cm}
\begin{tabular}{llllllllll}
\hline
\hline
Target & UT date \& time &      $\phi_V$ &Config. & $B_p$ & PA & Seeing & Airmass & Mode \\
& & && [m] & [$^\circ$]&['']\\
\hline
\textbf{LP data}\\
\hline
T Mic   &       2011-06-06 07:02        &       0.36    &       B2-C1   &       11.3                 &       18              &       1.33    &       1.040   &       SCIPHOT/PRISM   \\
\hline
T Mic   &       2011-06-06 07:57        &       0.36    &       B2-C1   &       11.3                 &       18              &       1.33    &       1.004   &       SCIPHOT/PRISM   \\
HD167618&       2011-06-06 07:39        &       \ldots          &       \ldots         &       \ldots          &       \ldots  &       1.60    &       1.096   &       SCIPHOT/PRISM   \\
\hline
T Mic   &       2011-06-11 06:44        &       0.37    &       B2-A1   &       10.9                 &       109     &       0.64    &       1.039   &       SCIPHOT/PRISM   \\
\hline
T Mic   &       2011-06-11 07:40        &       0.37    &       C1-A1   &       16.0                 &       70              &       0.73    &       1.003   &       SCIPHOT/PRISM   \\
\hline
\hline
\end{tabular}
\end{table*}


\begin{table*}[]
\caption{\label{journal-rtvir}Journal of the MIDI Auxiliary Telescopes observations of RT\,Vir.}
\centering
\vspace{0.5cm}
\begin{tabular}{llllllllll}
\hline
\hline
Target & UT date \& time &      $\phi_V$ &Config. & $B_p$ & PA & Seeing & Airmass & Mode \\
& & && [m] & [$^\circ$]&['']\\
\hline
\textbf{LP data}\\
\hline
RT Vir$^\mathrm{F}$     &       2012-01-24 07:33        &       0.76    &         C1-A1   &       12.6    &       72              &       0.86    &       1.338         &       SCIPHOT/PRISM   \\
HD81797&        2012-01-24 05:16        &       \ldots          &       \ldots  &       \ldots  &       \ldots         &       1.08    &       1.057   &       SCIPHOT/PRISM   \\
HD112142&       2012-01-24 07:17        &       \ldots          &       \ldots  &       \ldots  &       \ldots         &       0.86    &       1.207   &       SCIPHOT/PRISM   \\
HD123139&       2012-01-24 07:53        &       \ldots          &       \ldots  &       \ldots  &       \ldots         &       1.16    &       1.252   &       SCIPHOT/PRISM   \\
HD112142&       2012-01-24 08:25        &       \ldots          &       \ldots  &       \ldots  &       \ldots         &       0.81    &       1.069   &       SCIPHOT/PRISM   \\
\hline
\hline
\end{tabular}
\tablefoot{{$^\ast$} These observations have been published in \cite{sacuto2013}.}
\end{table*}

\begin{table*}[]
\caption{\label{journal-pigru}Journal of the MIDI Auxiliary Telescopes observations of $\pi^1$\,Gru.}
\centering
\vspace{0.5cm}
\begin{tabular}{llllllllll}
\hline
\hline
Target & UT date \& time &      $\phi_V$& Config. & $B_p$ & PA & Seeing & Airmass & Mode \\
& & && [m] & [$^\circ$]&['']\\
\hline
\textbf{LP data}\\
\hline
$\pi^1$\,Gru    &       2011-05-09 09:03        &\ldots &       B2-C1   &       11.1    &       4               &       0.80    &       1.30    &       SCI-PHOT   \\
\hline
$\pi^1$\,Gru    &       2011-05-09 09:34        &\ldots &       B2-C1   &       11.1    &       8               &       0.80    &       1.22    &       SCI-PHOT   \\
\hline
$\pi^1$\,Gru    &       2011-06-12 07:52        &\ldots &       A1-B2   &       10.0    &       96              &       1.45    &       1.16    &       SCI-PHOT   \\
HD206778                &       2011-06-12 07:36        &\ldots &       \ldots  &       \ldots  &       \ldots  &       1.08    &       1.32    &       SCI-PHOT        \\
HD206778                &       2011-06-12 08:50        &\ldots &       \ldots  &       \ldots  &       \ldots  &       0.62    &       1.22    &       SCI-PHOT        \\
\hline
\hline
\end{tabular}
\tablefoot{{$^\ast$} These observations have been published in \cite{sacuto2008}.}
\end{table*}

%
\begin{table*}[]
\caption{\label{journal-omiori}Journal of observations of omi\,Ori. }
\centering
\vspace{0.5cm}
\begin{tabular}{llllllllll}
\hline
\hline
Target & UT date \& time &      $\phi_V$& Config. & $B_p$ & PA & Seeing & Airmass & Mode \\
& & && [m] & [$^\circ$]&['']\\
\hline
\textbf{LP data}\\
\hline
$o$\,Ori        &       2011-11-01 04:08        &        \ldots         &         G1-I1   &       26.8    &       27              &       1.42    &         1.794   &       HIGHSENS/PRISM  \\
\hline
$o$\,Ori        &       2011-11-01 05:04        &               \ldots  &         K0-I1   &       34.9    &       0               &       0.77    &         1.474   &       HIGHSENS/PRISM  \\
HD18884 &       2011-11-01 04:50        &               \ldots  &       \ldots  &       \ldots  &       \ldots         &       0.83    &       1.143   &       HIGHSENS/PRISM  \\
HD18884 &       2011-11-01 06:50        &               \ldots  &       \ldots  &       \ldots  &       \ldots         &       1.45    &       1.277   &       HIGHSENS/PRISM  \\
\hline
$o$\,Ori        &       2011-11-01 06:04        &               \ldots  &         G1-I1   &       36.9    &       45              &       1.04    &         1.321   &       HIGHSENS/PRISM  \\
\hline
\hline
\end{tabular}
\end{table*}

\begin{table*}[]
\caption{\label{journal-rlep}Journal of the MIDI Auxiliary Telescopes observations of U~Ant.}
\centering
\vspace{0.5cm}
\begin{tabular}{llllllllll}
\hline
\hline
Target & UT date \& time &      $\phi_V$ &Config. & $B_p$ & PA & Seeing & Airmass & Mode \\
& & && [m] & [$^\circ$]&['']\\
\hline
\textbf{LP data}\\
\hline
U~Ant   &       2012-01-25 04:03        &       \ldots          &       D0-A1   &       35.7    &       28              &       1.26    &       1.3     &       HIGHSENS/PRISM  \\
\hline
U~Ant   &       2012-01-25 05:10        &       \ldots          &       D0-A1   &       35.7    &       28.8            &       1.08    &       1.12    &       HIGHSENS/PRISM  \\
HD29139 &       2012-01-25 02:12     &  \ldots          &       \ldots  &       \ldots  &       \ldots          &       1.11    &       1.40    &       HIGHSENS/PRISM  \\
\hline
U~Ant   &       2012-01-29 04:46        &       \ldots          &       B2-D0   &       33.8    &       11              &       1.00    &       1.144&  HIGHSENS/PRISM  \\
\hline
U~Ant   &       2011-11-04 08:38        &       \ldots          &       H0-I1   &       30.2    &       129             &       1.54    &       1.54    &       HIGHSENS/PRISM  \\
\hline

\end{tabular}
\tablefoot{{$^\ast$} These observations have been published in \cite{ladjal2011}, but we performed a new data reduction.}
\end{table*}

\begin{table*}[]
\caption{\label{journal-rlep}Journal of the MIDI Auxiliary Telescopes observations of R\,Lep.}
\centering
\vspace{0.5cm}
\begin{tabular}{llllllllll}
\hline
\hline
Target & UT date \& time &      $\phi_V$ &Config. & $B_p$ & PA & Seeing & Airmass & Mode \\
& & && [m] & [$^\circ$]&['']\\
\hline
\textbf{LP data}\\
\hline
R Lep$^\mathrm{F}$&     2011-10-24 06:36        &       1.40            &       D0-A1         &       34      &       219     &       0.82    &       1.042   &       SCIPHOT/PRISM         \\
HD20720         &       2011-10-24 06:16        &       \ldots          &       \ldots  &       \ldots  &       \ldots         &       0.69    &       1.00    &       SCIPHOT/PRISM   \\
HD48915         &       2011-10-24 06:54        &       \ldots          &       \ldots  &       \ldots  &       \ldots         &       0.56    &       1.208   &       SCIPHOT/PRISM   \\
HD20720         &       2011-10-24 07:25        &       \ldots          &       \ldots  &       \ldots  &       \ldots         &       0.41    &       1.077   &       SCIPHOT/PRISM   \\
HD48915         &       2011-10-24 08:02        &       \ldots          &       \ldots  &       \ldots  &       \ldots         &       0.62    &       1.059   &       SCIPHOT/PRISM   \\
\hline
\hline
\end{tabular}
\end{table*}

\begin{table*}[]
\caption{\label{journal-ypav}Journal of the MIDI Auxiliary Telescopes observations of Y Pav}
\centering
\vspace{0.5cm}
\begin{tabular}{llllllllll}
\hline
\hline
Target & UT date \& time &      $\phi_V$& Config. & $B_p$ & PA & Seeing & Airmass & Mode \\
& & && [m] & [$^\circ$]&['']\\
\hline
\textbf{LP data}\\
\hline
Y Pav$^\mathrm{F}$  & 2011-05-30 07:12:53 &  0.57  &    D0-H0 & 63.9 &  43    & 0.63 &     1.55 &    HIGH-SENS     \\
HD 150798         & 2011-05-30 06:55:42 &  \ldots&      \ldots&\ldots& \ldots & 0.84 &        1.48 &    \ldots     \\
\hline
Y Pav             & 2011-05-30 09:07:32 &  0.57  & H0-G1 & 51.2   &  6     &     0.78 &  1.42 &  HIGH-SENS \\
HD 150798         & 2011-05-30 08:50:24 &  \ldots&\ldots & \ldots & \ldots &       0.75 &  1.77 &  \ldots  \\
HIP 82363         & 2011-05-30 09:22:51 &  \ldots&\ldots & \ldots & \ldots &       0.82 &  1.83 &  \ldots  \\
\hline
Y Pav             & 2011-05-30 09:59:51 &  0.57  & D0-G1 & 63 &129    &         0.82 &       1.42 &   HIGH-SENS \\
HIP 82363         & 2011-05-30 09:42:36 & \ldots &\ldots&\ldots &\ldots &       0.78 & 2.03  & \ldots\\
HIP 82363         & 2011-05-30 10:15:22 & \ldots &\ldots& \ldots&\ldots &       0.87 &       2.12 & \ldots\\
\hline
Y Pav             & 2011-05-30 08:04:00 &   0.57        &       D0-H0 & 63.6 &       55 &    0.95 &  1.47  & HIGH-SENS \\
\hline
Y Pav              & 2011-05-31 07:54:30 &   0.57   & H0-G1 &51.3 &     -4 &       0.77 &  1.47 &  HIGH-SENS \\
\hline
Y Pav             & 2011-05-31 08:50:28 &  0.57    &    D0-G1 & 60.1 & 116 &       0.63 &  1.42 &  HIGH-SENS \\
\hline
Y Pav             & 2011-07-02 04:26:06 &    0.71        &      H0-I1 & 51.0 &       85 & 0.53  &    1.83 &  HIGH-SENS\\
\hline
\hline
\end{tabular}
\end{table*}

    \begin{table*}[htp]
            \caption{\label{journal}Journal of MIDI observations of TX\,Psc }
            \centering
            \vspace{0.5cm}
            \begin{tabular}{llllllllll}
            \hline
            \hline
            Target & UT date \& time &  $\phi$& Config. & $B_p$ & PA & Seeing & Airmass & Mode \\
            & & && [m] & [$^\circ$]&['']\\
            \hline
           \textbf{LP data}\\   
            \hline  
            TX~Psc & 2011-10-30 01:42 & \ldots & K0-G1 & 80.9 & 28 & 1.26 & 1.136 & HIGH-SENS \\
            HIP 1170&2011-10-30 01:55 &    & \ldots    & \ldots   & \ldots & 1.31 & 1.012 & HIGH-SENS\\
            \hline
            TX~Psc & 2011-10-30 02:29 & \ldots & K0-A1 & 128 & 67 & 1.44 & 1.147 & HIGH-SENS \\
            HIP~154& 2011-10-30 02:16 & \ldots & \ldots    & \ldots   & \ldots & 1.29 & 1.056 & HIGH-SENS \\
            HIP~1170&2011-10-30 02:41 & \ldots & \ldots    & \ldots   & \ldots & 1.30 & 1.008 & HIGH-SENS \\
\hline                                                                                                         
                \end{tabular}
                \tablefoot{{$^\ast$} These observations have been published in \cite{klotz2013}.}
        \end{table*}
        %

\begin{table*}[]
\caption{\label{journal-ssct}Journal of the MIDI Auxiliary Telescopes observations of S Sct}
\centering
\vspace{0.5cm}
\begin{tabular}{llllllllll}
\hline
\hline
Target & UT date \& time &      $\phi_V$& Config. & $B_p$ & PA & Seeing & Airmass & Mode \\
& & && [m] & [$^\circ$]&['']\\
\hline
\textbf{LP data}\\
\hline
S Sct & 2011-05-02 08:18:31 &   0.84 &  H0-I1 & 40 &    146 &   0.37 &  1.05 &        HIGH-SENS   \\
S Sct & 2011-05-28 05:34:10 &  0.02      &      K0-I1 & 44 &    1       &1.23 &       1.13 &   HIGH-SENS    \\ 
S Sct & 2011-05-28 06:34:05 & 0.02      &       K0-I1 & 44.3 &  8 &     1.24 &       1.05 &  HIGH-SENS     \\ 
S Sct & 2011-05-28 08:28:37&    0.02 &  G1-I1   & 46.6  &       45 &    0.55 &       1.11 &  HIGH-SENS\\
S Sct & 2011-05-28 09:18:32&     0.02 & G1-I1 & 45.9  & 46  &   0.90 &  1.23 &  HIGH-SENS\\
\hline
S Sct$^\mathrm{F}$ & 2011-05-28 07:27:11 & 0.02 &   K0-I1    &  44.9     &    14  &    0.96  &   1.05  &  HIGH-SENS   \\
HIP89931  &     2011-05-28 06:15:23 &\ldots     &       K0-I1  & 46.5 & 9         &1.01 & 1.01     &   HIGH-SENS \\
HIP89931  &     2011-05-28 07:08:03 & \ldots  &         K0-I1 & 46.3 & 14 &       1.41  & 1.01 & HIGH-SENS \\
\hline
S Sct$^\mathrm{F}$ & 2011-05-30 06:06:29 &      0.03 &  H0-I1   & 40.5 &        144 &       1.01 & 1.07 &  HIGH-SENS\\
HIP89931  &  2011-05-30 05:44:39 &      \ldots          &       H0-I1 & 40.1 & 144   &0.94 &         1.02  &  HIGH-SENS\\
HIP90496  &  2011-05-30 06:23:06 &      \ldots          &H0-I1  & 40.7   & 148   & 0.86 &        1.00  & HIGH-SENS\\
\hline
\hline
\end{tabular}
\end{table*}

\begin{table*}[]
\caption{\label{journal-aqsgr}Journal of the MIDI Auxiliary Telescopes observations of AQ Sgr}
\centering
\vspace{0.5cm}
\begin{tabular}{llllllllll}
\hline
\hline
Target & UT date \& time &      $\phi_V$ &Config. & $B_p$ & PA & Seeing & Airmass & Mode \\
& & && [m] & [$^\circ$]&['']\\
\hline
\textbf{LP data}\\
\hline
AQ Sgr$^\mathrm{F}$ &2011-05-02 07:14:32  &  0.86       &       G1-I1 & 42.5 &       21 & 0.46 &     1.20 &  HIGH-SENS  \\
HIP89931  &     2011-05-02 07:00:03     & \ldots        &       G1-I1 & 46.3 & 29 & 0.47 &   1.06 &   HIGH-SENS  \\
HIP89931  &     2011-05-02 09:01:28     & \ldots        &       H0-I1 & 40.7 & 154 & 0.49 &  1.02  &  HIGH-SENS  \\
HIP90496  &     2011-05-02 07:31:27     & \ldots         &      G1-I1 & 46.2 & 33 & 0.42 &   1.02 &  HIGH-SENS  \\
HIP90496  &     2011-05-02 08:01:01     & \ldots         &      H0-I1 & 40.6 & 146 & 0.37 &  1.01 &  HIGH-SENS  \\
\hline
\hline
\end{tabular}
\end{table*}

\begin{table*}[]
\caption{\label{journal-xtra}Journal of the MIDI Auxiliary Telescopes observations of X~TrA}
\centering
\vspace{0.5cm}
\begin{tabular}{llllllllll}
\hline
\hline
Target & UT date \& time &      $\phi_V$ & Config. & $B_p$ & PA & Seeing & Airmass & Mode \\
& & && [m] & [$^\circ$]&['']\\
\hline
\textbf{LP data}\\
\hline
X TrA$^\mathrm{F}$ & 2011-04-23 03:47:06        &\ldots &       K0-I1 &34.0 & 178 & 0.82 &  1.52 &     HIGH-SENS   \\
X TrA$^\mathrm{F}$ & 2011-04-23 05:11:55        &\ldots         &       \ldots &33.7 & 160 & 0.67 &    1.43 &     HIGH-SENS    \\
HD 150798 & 2011-04-23 03:24:38         & \ldots      &       \ldots &\ldots      &\ldots      & 0.85 &      1.81 &   HIGH-SENS   \\
HD 150798 & 2011-04-23 04:02:21         &  \ldots     &       \ldots &\ldots     &   \ldots   &0.71 &        1.68 &     HIGH-SENS\\
HD 150798 & 2011-04-23 04:56:30         &  \ldots     &       \ldots &\ldots  &\ldots  &0.74 &1.54 &   HIGH-SENS\\
\hline
\hline
\end{tabular}
\end{table*}  

\end{small}

\end{document}